\documentclass[preprint,showpacs,showkeys,aps]{revtex4}
\usepackage{graphicx}
\usepackage{dcolumn}
\usepackage{bm}
\usepackage{graphicx}
\usepackage{latexsym,color}
\usepackage{amssymb}
\begin{document}
\newcommand{\ti}[1]{\mbox{\tiny{#1}}}
\newcommand{\im}{\mathop{\mathrm{Im}}}
\def\be{\begin{equation}}
\def\ee{\end{equation}}
\def\bea{\begin{eqnarray}}
\def\eea{\end{eqnarray}}
\newcommand{\tb}[1]{\textbf{\texttt{#1}}}
\newcommand{\rtb}[1]{\textcolor[rgb]{1.00,0.00,0.00}{\tb{#1}}}
\newcommand{\il}{~}
\title{Motion of charged test particles  in Reissner--Nordstr\"om spacetime}

\author{Daniela Pugliese$^{1,2}$, Hernando Quevedo$^{1,3}$, and Remo Ruffini$^1$}
\email{d.pugliese.physics@gmail.com, quevedo@nucleares.unam.mx, ruffini@icra.it}
\affiliation{$^1$Dipartimento di Fisica, Universit\`a di Roma "La Sapienza", Piazzale Aldo Moro 5, I-00185 Roma, Italy\\
ICRANet, Piazzale della Repubblica 10, I-65122 Pescara, Italy \\
$^2$ School of Mathematical Sciences, Queen Mary, University of London,
Mile End Road, London E1 4NS, United Kingdom\\
$^3$Instituto de Ciencias Nucleares, Universidad Nacional Aut\'onoma de M\'exico,
AP 70543, M\'exico, DF 04510, Mexico
}

\date{\today}

\begin{abstract}
We investigate the circular motion of charged test particles in the gravitational field of
a charged mass described by the Reissner-Nordstr\"om (RN) spacetime. We study in detail
all the spatial regions where circular motion is allowed around either black holes or naked
singularities. The effects of repulsive gravity are discussed by finding all the
circles at which a particle can have vanishing angular momentum. We show that the
geometric structure of stable accretion disks, made of only test particles moving along circular
orbits around the central body, allows us to clearly distinguish between black holes and
naked singularities.
\end{abstract}

\maketitle

%
\section{Introduction}\label{xanes}

Let us consider the background of a static gravitational source of mass $M$ and
charge $Q$, described by the Reissner--Nordstr\"om (RN) line element in
standard spherical coordinates
\begin{equation}\label{11metrica}
ds^2=-\frac{\Delta}{r^2}dt^2+\frac{r^2}{\Delta}dr^2
+r^2\left(d\theta^2+\sin^2\theta d\phi^2\right)\ ,
\end{equation}
where $\Delta = (r-r_+)(r-r_-)$ and $r_\pm = M\pm\sqrt{M^2 -Q^2}$ are the radii of the outer and inner horizon, respectively.
Furthermore, the associated electromagnetic potential and field are
\begin{equation}\label{EMF}
A=\frac{Q}{r}dt,\quad F=dA=-\frac{Q}{r^2}dt\wedge dr \ ,
\end{equation}
respectively.

The motion of a test particle of charge $q$ and mass $\mu$ moving in a
RN   background (\ref{11metrica}) is described by the
following Lagrangian density:
\begin{equation}\label{LagrangianaRN}
\mathcal{L}=\frac{1}{2} g_{\alpha\beta}\dot{x}^{\alpha}\dot{x}^{\beta}+\epsilon A_\alpha x^\alpha,
\end{equation}
where $A_{\alpha}$ are the components of the electromagnetic 4--potential, the dot
represents differentiation with respect to the proper
time, and the parameter $\epsilon = q/\mu$ is the specific charge of the test particle.
The equations of motion of the test particle  can be derived from Eq.\il(\ref{LagrangianaRN}) by using
the Euler--Lagrange equation. Then,
\begin{equation}\label{COv}
\dot{x}^{\alpha}\nabla_{\alpha}\dot{x}^{\beta}=\epsilon F^{\beta}_{\ \gamma}\dot{x}^{\gamma},
\end{equation}
where $F_{\alpha\beta}\equiv A_{\alpha,\beta}-A_{\beta,\alpha}$.

Since the Lagrangian  density (\ref{LagrangianaRN})  does not depend explicitly on the variables $t$ and $\phi$, the
following two conserved quantities exist
\begin{eqnarray}
\label{10000} p_t&\equiv& \frac{\partial\mathcal{L}}{\partial
\dot{t}}=-\left(\frac{\Delta}{r^2}\dot{t}+\frac{\epsilon Q}{r}\right)=-\frac{E}{\mu},\\
\label{100000}  p_{\phi}&=&\frac{\partial\mathcal{L}}{\partial
\dot{\phi}}=r^2\sin^2\theta \dot{\phi}=\frac{L}{\mu},
\end{eqnarray}
where $L$ and  $E$ are respectively the angular momentum and energy
of the particle as measured by an observer at rest at infinity.
Moreover, to study the motion of charged test particles in the RN spacetime
it is convenient to use the fact  if the initial position and the tangent
vector of the trajectory of the particle lie on a plane that contains the center of the body,
then the entire trajectory must lie on this plane.
Without loss of generality we may therefore restrict ourselves to the study of
equatorial trajectories with $\theta =\pi/2$.

On the equatorial plane $\theta=\pi/2$, the motion equations can be reduced to the form $\dot r^2 + V^2 = E^2/\mu^2$ which
describes the motion inside an effective potential $V$. Then, we define the potential
\begin{equation}\label{9}
V_{\pm}=\frac{E^{\pm}}{\mu}=\frac{\epsilon Q}{r}\pm
\sqrt{\left(1+\frac{L^2}{\mu^2r^2}\right)\left(1-\frac{2M}{r}+\frac{Q^2}{r^2}\right)}
\end{equation}
as the value of $E/\mu$ that makes $r$ into a ``turning point''
$(V=E/\mu)$; in other words, the  value of $E/\mu$ at which the
(radial) kinetic energy of the particle vanishes \cite{RuRR,Chandra,Levin:2008mq,Bilic:2006bh}.
The effective potential with positive (negative) sign corresponds to
the solution with
$$
\lim_{r\rightarrow\infty} E^{+}=+\mu;\quad
\left(\lim_{r\rightarrow\infty} E^{-}=-\mu\right),
$$
where
\be
E^{+}(L,\epsilon,r)\geq E^{-}(L,\epsilon,r),
\ee
and  the following  relation holds:
\begin{equation}\label{IR}
E^{+}(L,\epsilon,r)=-E^{-}(L,-\epsilon,r).
\end{equation}
The behavior of the effective potential strongly depends on the sign
of $\epsilon Q$; in particular in the case of $\epsilon Q<0$, negative
energy states for the solution $E^{+}$ can exist (see also \cite{Grunau:2010gd,[11],[12],[13],[14],Belbruno:2011nn,Barack:2011ed}).

The problem of finding exact solutions of  the motion equations of test particles moving  in  a
RN spacetime has been widely studied in literature in many contexts and ways.
For a recent discussion we mention
the works \cite{Grunau:2010gd,[11],[12],[13],[14],Belbruno:2011nn,Barack:2011ed}.
In particular, in a recent paper \cite{Grunau:2010gd}
the full set of analytical solutions of the motion equations for electrically and magnetically charged
test particles is  discussed in terms of the Weierstrass ($\gamma$, $\sigma$ and $\zeta$)
functions. The general structure of the geodesics was discussed and  a  classification of their types was proposed.
Remarkably, analytical solutions are found in the case of a central RN source not only with constant
electric charge, but also with constant magnetic charge. It is interesting to notice
that if either the test particle or the central body possesses
both types of charge, it turns out that the motion is no longer confined to a plane.
In the present work, we consider only equatorial circular orbits  around a central RN source with constant electric charge.
Instead of solving directly the equations of motion, we explore the properties of the effective potential function associated
to the motion. Thus, we discuss and propose  a  classification of the equatorial  orbits in terms of the two constants of motion: the energy $E/\mu$ and the orbital angular momentum $L/(\mu M)$.
In fact, we focus our attention on some peculiar features of the circular motion and the physics around
black holes and naked singularities. In particular, we are interested in exploring the possibility of distinguishing between black holes and naked singularities by studying the motion of circular test particles. In this sense, the present work complements and is different from previous studies \cite{Grunau:2010gd,[11],[12],[13],[14],Belbruno:2011nn,Barack:2011ed}.

In a previous work \cite{Pugliese:2010ps,Pugliese:2010he}, we analyzed the dynamics of the RN spacetime by studying the motion of neutral test particles for which the effective
potential turns out to coincide with $V_+$ as given in Eq.(\ref{9}) with $\epsilon=0$. We will see that in the case of charged test particles the term
$\epsilon Q/r$  drastically changes the behavior of the effective potential, and leads to several possibilities which must analyzed in the case of
black holes and naked singularities. In particular, we will show that for particles moving along circular orbits there exist stability regions
whose geometric structure clearly distinguishes naked singularities from black holes (see also \cite{Virbhadra:2002ju, Virbhadra:2007kw} and \cite{do1,Dabrowski:2002iw}).
The plan of this  paper is the  following: In Sec.   \ref{sec:veff} we investigate the behavior of the effective potential and the conditions
for the motion
of positive and negative charged test particles moving on circular orbits around the central charged mass.
This section also contains a brief analysis of the Coulomb approximation of the effective potential.
In   Sec.   \ref{BHBHTR}, we will consider the black hole case  while in  Sec.   \ref{NSNSTRE} we shall focus on  the motion around naked singularities.
The conclusions are in Sec.  \ref{milka}.

\section{Circular motion}
\label{sec:veff}

The circular motion of charged test particles is governed by the behavior of the effective potential (\ref{9}).
In this work, we will mainly consider the special case of a positive solution $V_+$ for the potential in order
to be able to compare our results with those obtained in the case of neutral test particles analyzed in \cite{Pugliese:2010ps,Pugliese:2010he}.
Thus, the radius of circular orbits and the corresponding values of the energy  $E$
and the angular momentum $L$ are given by the extrema of the  function $V_{+}$.
Therefore, the conditions for the occurrence of circular orbits are:
\begin{equation}\label{fg2}
\frac{d V_{+}}{d r}=0, \quad V_+=\frac{E^+}{\mu}.
\end{equation}
When possible, to simplify the notation we will drop the subindex $(+)$ so that, for example, $V=E/\mu$ will denote the
positive effective potential solution.
Solving Eq.\il(\ref{fg2}) with respect to $L$,  we find the specific angular momentum

\be
\frac{(L^\pm)^2}{\mu^2} = \frac{r^2}{2\Sigma^2}\left[ 2(Mr-Q^2)\Sigma
+ \epsilon^2 Q^2 \Delta\pm Q\Delta\sqrt{\epsilon^2 \left(4\Sigma+\epsilon^2  Q^2\right)}\right],
\label{ECHEEUNAL}
\ee
where $\Sigma\equiv r^2-3Mr+2Q^2$,
%
of the test particle on a circular orbit of radius $r$. The corresponding energy can be obtained
by introducing the expression for the angular momentum into Eq.\il(\ref{9}). Then, we obtain
\begin{equation}
\label{Lagesp}
\frac{E^{\pm}}{\mu}=\frac{\epsilon Q}{r}
+\frac{\Delta\sqrt{2\Sigma+\epsilon^2Q^2\pm Q\sqrt{\epsilon^2(4\Sigma+\epsilon^2Q^2)}}}
{\sqrt{2}r|\Sigma|}\  .
\end{equation}
The sign in front of the square root should be chosen in accordance with the physical situation.
This point will be clarified below by using the formalism of orthonormal frames.

An interesting particular orbit is the one in which the particle is located at rest as seen by an observer at infinity, i.e.,
$L=0$. These ``orbits'' are therefore characterized by the following conditions
\be\label{Nolia}
L=0,\quad \frac{d V}{dr}=0.
\ee
\cite{Interessantissimo}. Solving Eq.\il(\ref{Nolia}) for $Q\neq0$ and $\epsilon\neq0$, we find the following radius
%
\be r_{s}^{\pm}\equiv\frac{\left(\epsilon ^2-1\right)Q^2M }{\epsilon ^2Q^2 -M^2}\pm\sqrt{\frac{\epsilon^2Q^4 \left(\epsilon ^2-1\right) \left(M^2-Q^2\right) }{\left(\epsilon ^2Q^2 -M^2\right)^2}}\label{rE}\ .
\ee

Table\il\ref{TabL0} shows the explicit values of all possible radii for different values of the ratio $Q/M$.
\begin{table}
\begin{center}
\resizebox{.9\textwidth}{!}{%
\begin{ruledtabular}
\begin{tabular}{ll|ll|ll}
$0<Q<M$&
&$Q=M$&
&$Q>M$
\\
\toprule
$\epsilon$&Radius
&$\epsilon$&Radius
&$\epsilon$&Radius
\\
$\epsilon>M/Q$&$ r=r_s^+$
&$\epsilon=1$&$ r>M$
&$-M/Q<\epsilon<0$&$ r=r_s^-$
\\
$$&$$
&$$&$$
&$\epsilon=-M/Q$&$r=Q^2/(2M)$
\\
$$&$$
&$$&$$
&$-1<\epsilon\leq-M/Q$&$r=r_s^+$
\\
$$&$$
&$$&$$
&$\epsilon=0$&$r=Q^2/M$
\\
$$&$$
&$$&$$
&$0<\epsilon<M/Q$&$r=r_s^+$
\\
\end{tabular}
\end{ruledtabular}}
\caption[font={footnotesize,it}]{\footnotesize{Radii of the ``orbits'' characterized by the  conditions $L=0$ and $dV/dr=0$.}}
\label{TabL0}
\end{center}
\end{table}
A particle located at $r=r_{s}$ with angular momentum $L=0$ will have the energy (see also
\cite{CONS,Bonnor,Interessantissimo,Bini:2006dp,Bini:2008zza,Bini:2006pk,Bini:2005xg})
%
%
\be
\frac{E_s^{\pm}}{\mu}\equiv\frac{1}{Q}\left(\sqrt{\frac{M^2-Q^2}
{\epsilon^2-1}}+\frac{\epsilon}{\frac{\epsilon^2-1}{\epsilon^2Q^2-M^2}\pm\sqrt{\frac{\epsilon^2(M^2-Q^2)(\epsilon^2-1)}
{(\epsilon^2Q^2-M^2)^2}}}\right)\ .
\ee
%

The minimum radius for a stable circular orbit occurs at the
inflection points  of the effective potential function; thus, we must
solve the equation
\begin{equation}\label{Lagespep}
\frac{d^2 V}{d r^2}=0,
\end{equation}
for the orbit radius $r$, using the expression
(\ref{ECHEEUNAL}) for the angular momentum $L$.
From Eq.\il(\ref{fg2}) and Eq.\il(\ref{Lagespep}) we find that the radius of the last stable circular orbit and the angular momentum of this orbit are
related by the following equations
\bea\nonumber
\left(L^2+Q^2-1\right) r^6-6 L^2 r^5+6 L^2\left(1+ Q^2\right) r^4-2L^2\left(2 L^2+5Q^2\right) r^3\\
\nonumber+ L^2\left(3 L^2+3 L^2 Q^2+3 Q^4\right) r^2-6 L^4 Q^2 r+2 L^4 Q^4=0  \ ,\label{nube}
\eea
and
\be\label{nuvola}
Q^2 r^2-r^3+L^2 \left(2 Q^2-3 r+r^2\right)+Q r^3 \sqrt{\frac{\left(L^2+r^2\right) \left(Q^2-2 r+r^2\right)}{r^4}} \epsilon =0\ ,
\ee
where in order to simplify the notation we introduced the normalized quantities $L\rightarrow L/(M/\mu)$,  $r\rightarrow r/M$, and $Q\rightarrow Q/M$.
Equation (\ref{nube}) depends on the test particle specific charge $\epsilon$ via the function $L$ as given in Eq.\il (\ref{ECHEEUNAL}).
It is possible to solve Eq.\il(\ref{nube}) for the last stable circular orbit radius as a function of the free parameter $L$. We find the expression
\bea
\nonumber\frac{(L^\pm_{\ti{lsco}})^2}{\mu^2}&=&\frac{r^2}{2\left[2 Q^4+3 Q^2 r(r-2M) -(2 r-3M) r^2\right]} \left[2 Q^2 (5M-3 r) r-3 Q^4-r^2 [6M^2+(r-6M) r]\right.\\\label{Carmara}
\nonumber&&\left.\pm\sqrt{9 Q^2+(r-6M) r} \left(Q^2+(r-2M) r\right)^{3/2}\right]
\\
\eea
for the angular momentum of last stable circular orbit.
Eq.\il(\ref{Carmara}) can be substituted in Eq.\il(\ref{nuvola})
to find the radius of the last stable circular orbit.

\subsection{Coulomb potential approximation}
Consider the case of a charged particle moving in the  Coulomb potential
$$
U(r)=\frac{ Q}{r} \ .
$$
This means that we are considering the motion described by the following effective
potential
\begin{equation}\label{vecchiaRR}
V_+= \frac{E^{+}}{\mu}=\frac{\epsilon Q}{r} +\sqrt{1+\frac{L^2}{\mu^2
r^2}}\ ,
\end{equation}
where $\epsilon Q<0$. The Coulomb approximation is interesting for our further analysis
because it corresponds to the limiting case for large values of the radial coordinate $r$ [cf. Eq.(\ref{9})].

Circular orbits are therefore situated at  $r=r_c$ with
%
%
%
\begin{equation}\label{mai più}
r_c = \sqrt{\frac{L^2}{\mu^{2}}\left(\frac{L^2}{\epsilon^2 Q^{2}}-1\right)}\quad\mbox{and}\quad
\frac{L^2}{\mu^2}\geq \epsilon^2 Q^2,
\end{equation}
and in the case  $\epsilon=0$ with $Q>0$, circular orbits exist in all $r>0$ for $L=0$.
We conclude that in this approximation circular orbits always exist with orbital radius $r_c$ and angular
momentum satisfying the condition $|L|/\mu \geq |\epsilon Q|$.
For the last stable circular orbit situated at  $r=r_{\ti{lsco}}$ we find
\begin{equation}
\label{mascalzone}
r_{\ti{lsco}}=0\quad\mbox{with}\quad \frac{E^+(r_{\ti{lsco}})}{\mu}= 0
\quad\mbox{and}\quad \frac{|L|}{\mu}=|\epsilon Q|\ .
\end{equation}
This means that, in the approximation of the Coulomb potential, all the circular orbits are stable, including the limiting case of
a particle at rest on the origin of coordinates.

\begin{figure}
\centering
\begin{tabular}{c}
\includegraphics[scale=1]{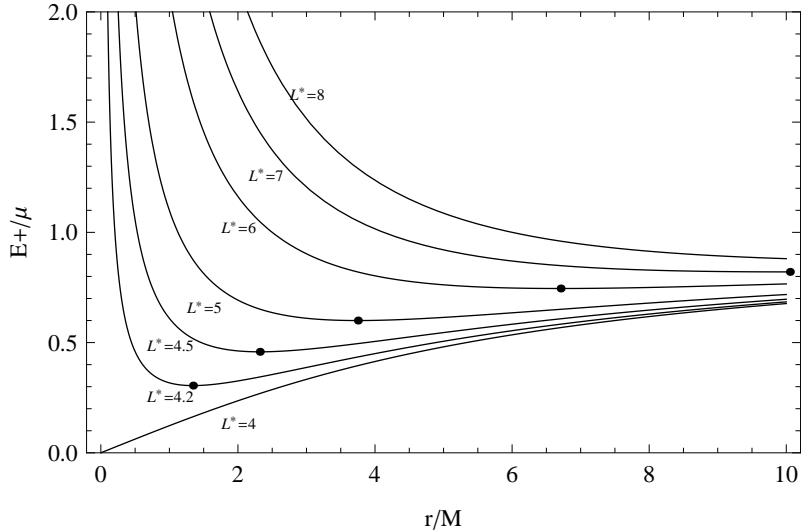}
\end{tabular}
\caption[font={footnotesize,it}]{\footnotesize{Effective potential
for a charged test particle with $\epsilon=-2$ moving  in a Coulomb potential with
$Q/M=2$ for different values of the momentum $L^{*}\equiv L/(\mu
M)$. The points indicate the minima of the potential. In particular, for
$L^{*}=|\epsilon Q|/M$ the potential vanishes on the origin $r=0$ (see text).}}
\label{Plotfxue}
\end{figure}

Furthermore, Eqs.\il(\ref{mai più}--\ref{mascalzone}) show that, in contrast with the general RN case, for a
charged particle moving in a Coulomb potential only positive or null energy
solutions can exist.
See Fig. \ref{Plotfxue} where the potential (\ref{vecchiaRR}) is plotted as a function of the orbital radius for different
values of the angular momentum.
\section{Black holes}
\label{BHBHTR}

\begin{figure}
\centering
\begin{tabular}{c}
\includegraphics[scale=1]{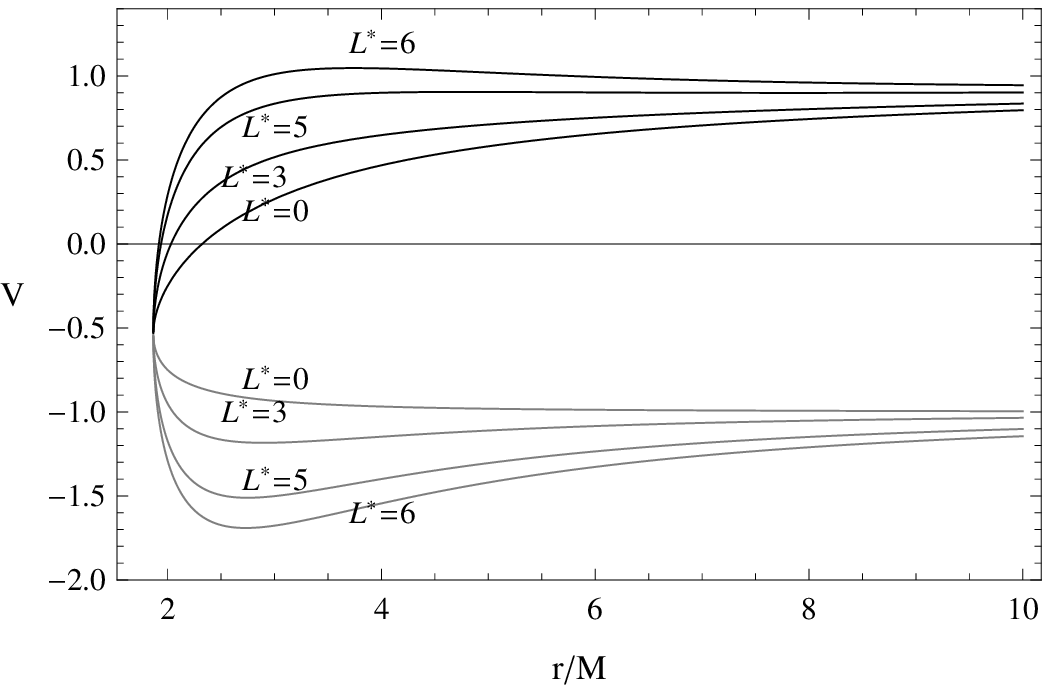}
\end{tabular}
\caption[font={footnotesize,it}]{\footnotesize{The
effective potential as a function of $r/M$  for a charged particle of
charge--to--mass ratio  $\epsilon\equiv q/\mu$ moving in a  Reissner--Nordstr\"om
black hole of charge $Q$ and mass $M$.  The graphic shows
the positive $E^{+}/\mu$ (black curves) and negative roots $E^{-}/\mu$
(gray curves) of the effective potential for $Q/M=0.5$, $\epsilon=-2$, and
 different
values of the momentum $L^{*}\equiv L/(M\mu)$.
The outer horizon is
located at $r_{+}\equiv M+\sqrt{M^2-Q^2}\approx1.87M$. Note the
presence of negative energy states for the positive roots.}
}
\label{Pcomparativa2u}
\end{figure}

In the case of a black hole $(M^2>Q^2)$
the two roots $V_{\pm}$ of the effective potential  are plotted as a
function of the ratio
$r/M$ in Fig. \il\ref{Pcomparativa2u} for a fixed value of the charge--to--mass
ratio of the test particle and different values of the angular momentum
$L/(M \mu)$ (see also \cite{Pradhan:2010ws,Gladush:2011cz,Olivares:2011xb,Zaslavskii:2010aw,Gad,Dotti:2010uc}).
Notice the presence of negative energy states for the  positive solution  $V_+=E^+/\mu$  of the effective potential function.
Negative energy states for $V_+$ are possible only in the case $\epsilon Q<0$.
In particular, the largest region in which the
$V_{+}$ solution has negative energy states is
\begin{equation}
\label{assurdo!}
M+\sqrt{M^2-Q^2}<r\leq M+\sqrt{M^2-Q^2\left(1-\epsilon^2\right)}
\end{equation}
and corresponds to the limiting case of vanishing angular momentum ($L=0$).
%
For $L\neq 0$ this region becomes smaller and decreases as $L$ increases.
For a given value of the orbit radius, say $r_0$, such that $r_0<M+\sqrt{M^2-Q^2\left(1-\epsilon^2\right)} $, the
angular momentum of the test particle must be chosen within the interval
\be
0<\frac{L^2}{\mu^2}<{r_0^2}{} \left(\frac{ \epsilon ^2 Q^2}{r_0^2-2Mr_0 +Q^2 }-1\right)
\ee
for a region with negative energy states to exist. This behavior is illustrated in Fig. \ref{Pcomparativa2u}.

\begin{figure}
\centering
\begin{tabular}{cc}
\includegraphics[scale=.7]{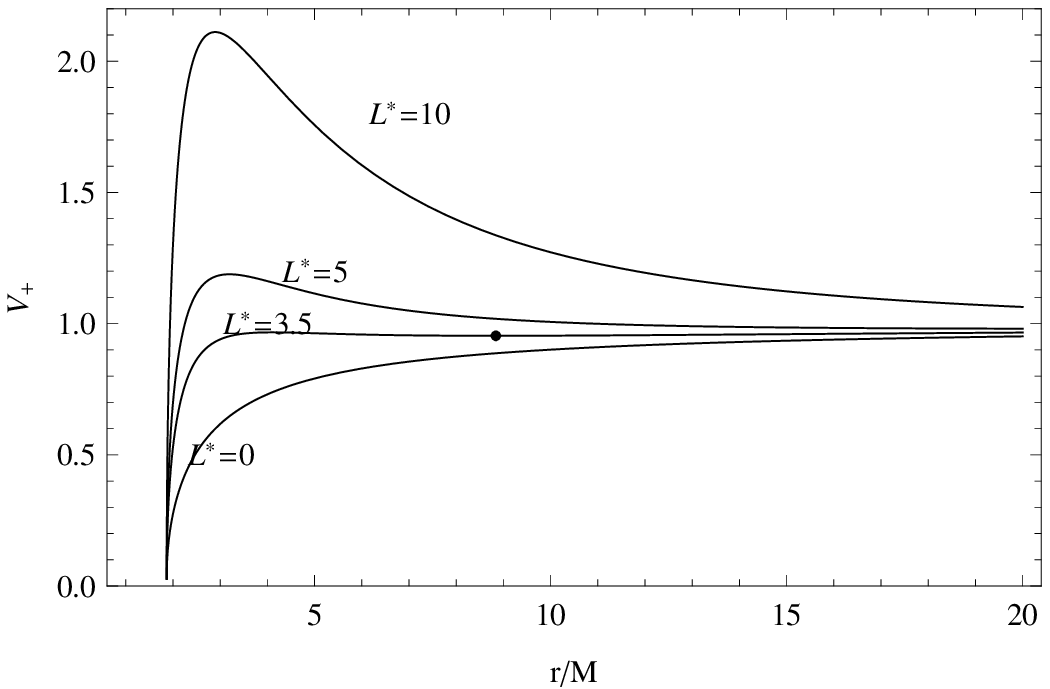}
\includegraphics[scale=.7]{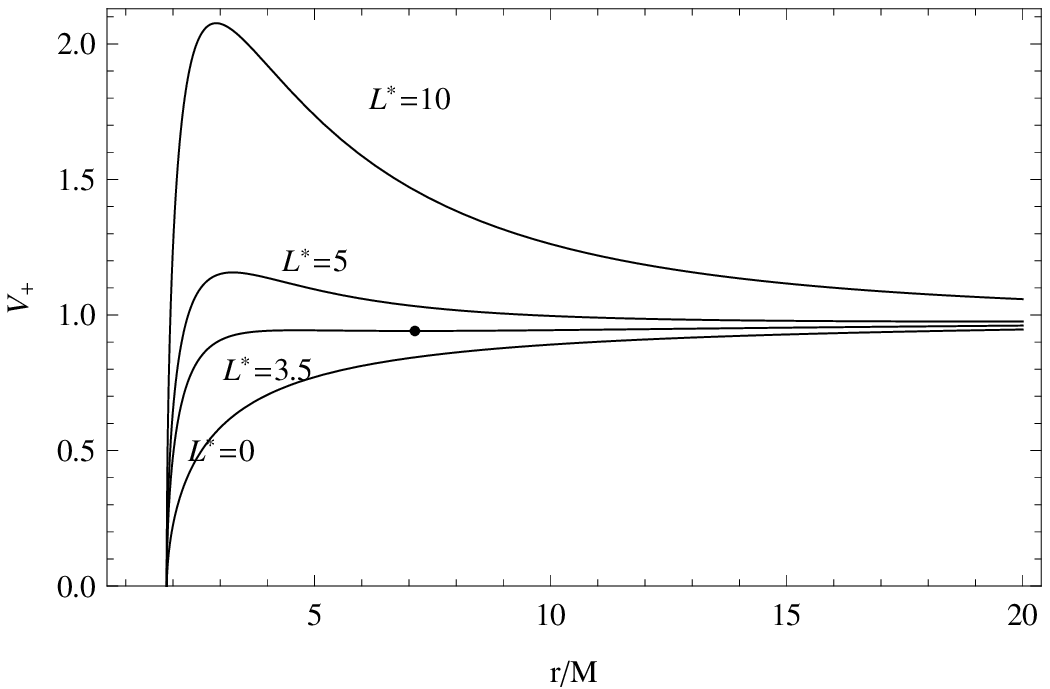}
\end{tabular}
\caption[font={footnotesize,it}]{\footnotesize{The effective
potential $V_{+}$ for a charged particle of charge--to-mass ratio,
$\epsilon=q/\mu$, moving in a Reissner-Nordstr\"om spacetime of charge $Q$
and mass $M$ with charge--to--mass ratio $Q/M=0.5$ is plotted as a function
of the radial coordinate $r/M$ for different values of the angular momentum
$L^{*}\equiv L/(M\mu)$. The outer horizon is located at
$r_{+}\approx1.87M$. In the graphic on the left with $\epsilon=0.1$, the effective potential for
$L^{*}\approx3.5$ has a minimum $V_{min}\approx0.954$  at $r_{min}\approx8.84M$. In the graphic on the right with
$\epsilon=-0.1$, the minimum $V_{min}\approx0.94$ is located at $r_{min}\approx7.13M$ for $L^{*}\approx3.5$. } }
\label{Plot0501}
\end{figure}

\begin{figure}
\centering
\begin{tabular}{c}
\includegraphics[scale=1]{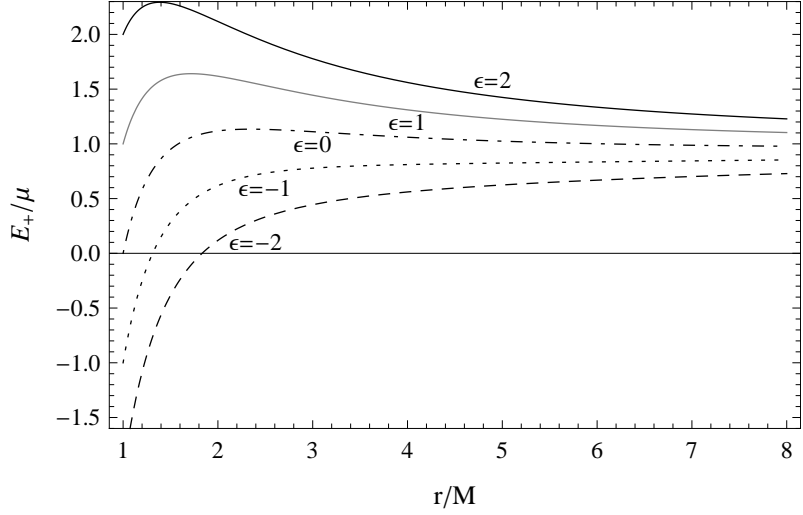}\\
\end{tabular}
\caption[font={footnotesize,it}]{\footnotesize{The
effective potential $V_{+}$ is plotted as
 a function of $r/M$  for a charged
test particle with specific charge $\epsilon=q/\mu$ moving in the field
of a Reissner-Nordstr\"om extreme black hole ($Q=M$). Here  $L/(M\mu)=4$, and the effective potential
is plotted for different values of $\epsilon$. The outer horizon is
located at $r_{+}\equiv M+\sqrt{M^2-Q^2}=M$. Note the presence of
negative energy states for particles with negative $\epsilon$. } }
\label{Pcomparativa2a}
\end{figure}

Fig. \il\ref{Plot0501}  shows the positive solution $V_+$ of
the effective potential for different values  of the momentum and for
positive  and negative    charged
particles. In particular, we note that, at fixed $Q/M$ for a particle with $|\epsilon|<1$,
 in the case $\epsilon Q>0$ the  stable orbit radius is larger than in the case of attractive
 electromagnetic interaction, i. e., $\epsilon Q<0$.
In Fig. \il\ref{Pcomparativa2a}, the potential   $V_+$ of an extreme black hole is plotted for
different, positive and negative values of the test particle with charge--to--mass ratio $\epsilon$.
In this case, it is clear that the magnitude of the energy increases as the magnitude of the specific charge of the particle $\epsilon$ increases.

As mentioned in Sec.  \ref{sec:veff}, in the case of the positive solution for the effective potential
the conditions for the existence of circular orbits
\be
\dot{r}=0, \quad V=\frac{E}{\mu},\quad \frac{d V}{d r} =0.
\ee
lead to Eqs.(\ref{ECHEEUNAL}) and (\ref{Lagesp}) in which the selection of the $(\pm)$ sign inside the square root
should be done properly.
To clarify this point we consider explicitly the equation of motion for a charged particle
in the gravitational field of a RN black hole.
\begin{equation}\label{100}
a(U)^{\alpha}=\epsilon F^{\alpha}_{\phantom\ \beta}U^{\beta},
\end{equation}
where $a(U) = \nabla_{U} U$ is the particle's 4--acceleration.
Introducing the orthonormal frame
\be
e_{\hat{t}}=\frac{r}{\Delta^{1/2}} \partial_{t},\quad
e_{\hat{r}}=\frac{\Delta^{1/2}}{r}\partial_{r},\quad
e_{\hat{\theta}}=\frac{1}{r}\partial_{\theta},\quad
e_{\hat{\phi}}=\frac{1}{r\sin\theta}\partial_{\phi} ,
\ee
with dual
\be
\omega^{\hat{t}}=\frac{\Delta^{1/2}}{r}dt,\quad
\omega^{\hat{r}}=\frac{r}{\Delta^{1/2}}dr,\quad
\omega^{\hat{\theta}}=rd\theta,\quad \omega^{\hat{\phi}}=r\sin\theta
d\phi\ ,
\ee
the tangent to a (timelike) spatially circular orbit $u^{\alpha}$ can be expressed as
$$
u=\Gamma(\partial_t+\zeta\partial_{\phi})=\gamma\left(e_{\hat{t}}+\nu
e_{\hat{\phi}}\right)\ ,
$$
where $\Gamma$ and $\gamma$ are normalization factors
$$
\Gamma^2=(-g_{tt}-\zeta^2g_{\phi\phi})^{-1}\quad\mbox{and}\quad\gamma^2=(1-\nu^2)^{-1},
$$
which guarantees that $u_{\alpha}u^{\alpha} = -1$. Here $\zeta$ is the angular
velocity with respect to infinity and $\nu$ is the ``local proper
linear velocity" as measured by an observer associated with the orthonormal frame. The angular
velocity $\zeta$ is related to the local proper linear velocity by
$$
\zeta=\sqrt{-\frac{g_{tt}}{g_{\phi\phi}}}\nu\ .
$$
Since only the radial component of the 4--velocity is non-vanishing, Eq.\il(\ref{100}) can be
written explicitly as
\begin{equation}
\label{11}
0=\gamma (\nu^{2}-\nu_{g} ^{2})+\frac{\nu_{g} }{\zeta_{g}}\frac{\epsilon Q}{r^{2}}\ ,
\end{equation}
where
\begin{equation}\label{RR}
\zeta_{g}= \pm\frac{\sqrt{Mr-Q^2}}{r^2}\ ,\quad\nu_{g} =\sqrt{\frac{Mr-Q^2}{\Delta}}\ .
\end{equation}
This equation gives the values of the particle linear velocity $\nu=\pm
\nu_{\epsilon}^{\pm}$  which are compatible with a given value of $\epsilon Q$ on a
circular orbit of radius $r$, i. e.,
\begin{equation}\label{12}
\nu_{\epsilon}^{\pm}=\nu_g  \sqrt{1-\frac{Q^{2} \epsilon ^{2}}{2 r^4
\zeta_g^{2}}\pm \frac{Q }{r^{2} \zeta_g \nu_g } \sqrt{\frac{\epsilon
^{2}}{\gamma_g^{2}}+\frac{Q^{2} \epsilon ^4 \nu_g ^{2}}{4 r^4
\zeta_g^{2}}}},
\end{equation}
where
$$
\gamma_{g}=\left(\frac{\Delta}{r^2-3Mr+2Q^2}\right)^{1/2},
$$
%
and
\be
\gamma_{\epsilon}^{\pm}=(1-\nu_{\epsilon}^{\pm}\-^{2})
^{-1/2}.
\ee
In the limiting case of a neutral particle ($\epsilon =0)$, Eq.(\ref{11}) implies that the linear velocity of the particle
is $\nu_g $.

We  introduce the limiting value of the parameter
$\epsilon$ corresponding to a particle at rest, $\nu=0$, in
Eq.\il(\ref{11}), i. e.,
\begin{equation}\label{13}
\epsilon_{0}=\nu_{g}  \zeta_g \frac{r^{2}}{Q}=\frac{M r-Q^{2}}{Q
\sqrt{\Delta}}.
\end{equation}
By introducing this quantity into Eq.\il(\ref{11}), one gets the following
equivalent relation
\begin{equation}\label{14}
\frac{\epsilon}{\epsilon_{0}}=\gamma
\left(1-\frac{\nu^{2}}{\nu_{g} ^{2}}\right),
\end{equation}
whose solution (\ref{12}) can be conveniently rewritten as
\begin{equation}\label{15}
\nu_{\epsilon}^{\pm}=\nu_{g} \left[\Lambda\pm\sqrt{\Lambda^{2}-1+(\epsilon/\epsilon_0)^{2}}\right]^{1/2}\ ,
\end{equation}
where
\begin{equation}\label{16}
\Lambda=1-\frac{\nu_{g} ^{2}}{2}\left(\frac{\epsilon}{\epsilon_0}\right)^{2}.
\end{equation}
Moreover, from Eq.\il(\ref{14}) it follows that $\epsilon<0$ implies
that $\nu^{2}> \nu_{g} ^{2}$ (because $\epsilon_{0}$ is always positive for
$r > r_+$), so that the allowed solutions for $\nu$ can exist only for
$r \geq r^+_\gamma$, where
\be
r_\gamma^+ \equiv\frac{1}{2}\left(3M+\sqrt{9M^2-8Q^2}\right)\ ,
\ee
the equality corresponding to $\nu_{g} = 1$. In this
case,  the solutions of Eq.\il(\ref{11}) are given by
$\nu=\pm\nu_{\epsilon}^{+}$.

For $ \epsilon> 0$, instead, solutions can exist also for $r_+ < r <
r^+_\gamma$. The situation strongly depends on the considered range of
values of $\epsilon$ and is summarized below.

Equation \il(\ref{15}) gives the following conditions for the existence of
velocities
\begin{eqnarray}
\label{170} \Lambda^{2}-1+(\epsilon/\epsilon_{0})^{2} &\geq& 0\ , \\
\label{180}
\Lambda\pm\sqrt{\Lambda^{2}-1+(\epsilon/\epsilon_{0})^{2}}&\geq& 0\ .
\end{eqnarray}
%
%
The second condition, Eq.\il(\ref{180}), is satisfied by \
\begin{equation}\label{19}
r\geq r_l \equiv\frac{3M}{2}+\frac{1}{2}\sqrt{9M^{2}-8Q^{2}-\epsilon^{2}Q^{2}}.
\end{equation}
Moreover for $Q=M$ and $\epsilon=1$ it is
$\Lambda+\sqrt{\Lambda^{2}-1+(\epsilon/\epsilon_{0})^{2}}\geq0$ when $M<r<(3/2)M$. However it is also possible to show that condition Eq.\il(\ref{fg2}) is satisfied for $0<Q<M$ and $\epsilon>0$ only in the range $r\geq r_l$.

Requiring that the argument of the square root be nonnegative
implies
\begin{equation}\label{20}
\epsilon\le\epsilon _{l}\equiv \frac{\sqrt{9M^{2}-8Q^{2}}}{Q}.
\end{equation}
The condition (\ref{180}) will be discussed later.

From the equation of motion (\ref{14}) it follows that the velocity
vanishes for $\epsilon/\epsilon_0 = 1$, i. e., for [cf. Eq.(\ref{rE})]
\begin{equation}\label{21}
r=r_{s}\equiv
\frac{Q^{2}}{\epsilon^{2}Q^{2}-M^{2}}
\left[M(\epsilon^{2}-1)+\sqrt{\epsilon^2(\epsilon^2-1)(M^{2}-Q^{2})}\right]\ ,
\end{equation}
which exists only for $\epsilon> M/Q$.
We thus have that
\be
\frac{\epsilon}{\epsilon_0}> 1 \quad\mbox{for}\quad r > r_{s},
\ee
whereas
\be
\frac{\epsilon}{\epsilon_0}< 1  \quad\mbox{for}\quad r_+ < r < r_{s}\ .
\ee
On the other hand, the condition $\nu = 0$ in Eq.\il(\ref{15}) implies that
\begin{eqnarray}
\label{22}
\left[\Lambda\pm\sqrt{\Lambda^{2}-1+(\epsilon/\epsilon_{0})^{2}}\right]_{\epsilon/\epsilon_0=1}&=&
0,
\\
\label{230}\mbox{i. e.} \quad
\left[\Lambda\pm\sqrt{\Lambda^{2}-1}\right]_{r=r_{s}}&=& 0,
\end{eqnarray}
thus $\nu_{\epsilon}^{-}$ is identically zero whereas
$\nu_{\epsilon}^{+}= {2}\Lambda(r_{s})=0$ only for
\begin{equation}\label{24}
\epsilon=\tilde{\epsilon}\equiv
\frac{1}{\sqrt{2}Q}\sqrt{5M^{2}-4Q^{2}+\sqrt{25M^{2}-24Q^{2}}}
.
\end{equation}
Finally, the lightlike condition $\nu=1$ is reached only at $r =
r_\gamma^+$, where $\nu_g  = 1 = \nu$.

The behavior of charged test particles depends very strongly on their location with respect to the special radii
$r_+$, $r_l$, $r_\gamma^+$, and $r_{s}$. In Sec.  \ref{sec:stb} the behavior of these radii will be analyzed in connection
with the problem of stability of circular orbits.


On the other hand, the particle's 4--momentum is given by $P = m U-qA$. Then, the conserved quantities associated with the temporal and
azimuthal Killing vectors $\xi=\partial_{t}$ and
$\eta=\partial_{\phi}$ are respectively
\bea\label{250}
P \cdot \xi&=&-\frac{\epsilon
Q}{r}-\gamma\frac{\sqrt{\Delta}}{r}=-\frac{E}{\mu},\\
P \cdot
\eta&=&\frac{r}{M}\gamma \nu=\frac{L}{M\mu},
\eea
where $E/\mu$ and $L/\mu$ are the particle's energy and angular
momentum per unit mass, respectively (see also Eqs.(\ref{ECHEEUNAL}) and
(\ref{Lagesp}) ).

Let us summarize the results.
%
\subsubsection{Case $\epsilon<0$}
\begin{figure}
\centering
\begin{tabular}{c}
\includegraphics[scale=1]{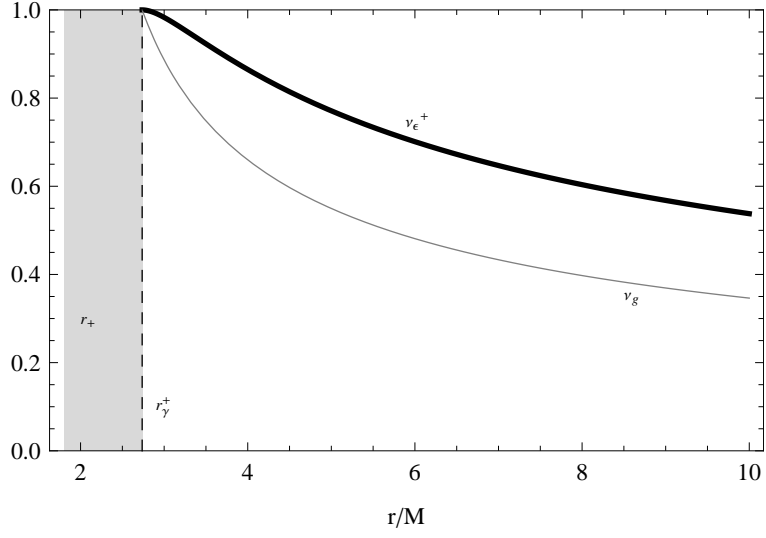}
\end{tabular}
\caption[font={footnotesize,it}]{\footnotesize{The positive solution
of the linear velocity $\nu^+_{\epsilon}$ is plotted as a function of the radial
distance $r/M$ for the parameter choice $Q/M = 0.6$ and $\epsilon =-
3$ so that $r_\gamma^+/M \approx 2.74$ and the outer horizon is located at $r_+/M =1.8$.
The geodesic velocity $\nu_g $ is also shown (gray curve).
The shaded region $(r<r_\gamma^+)$ is forbidden.
}} \label{Figure1a}
\end{figure}

\begin{figure}
\centering
\begin{tabular}{c}
\includegraphics[scale=1]{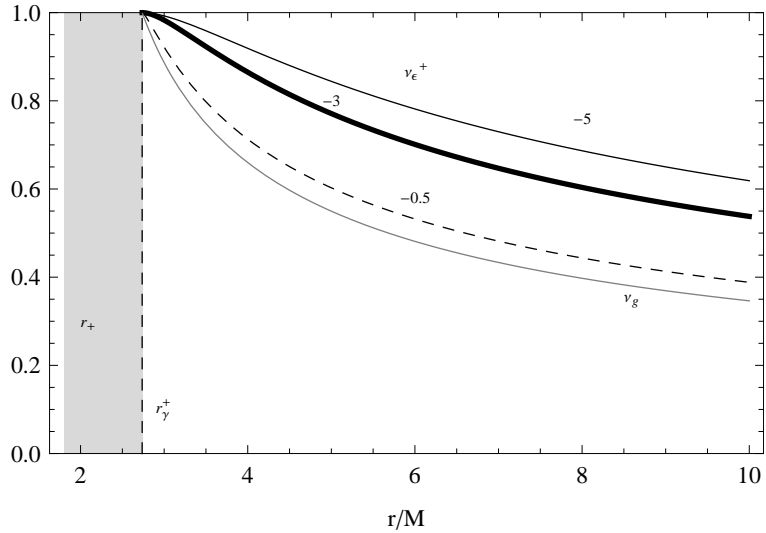}
\end{tabular}
\caption[font={footnotesize,it}]{\footnotesize{The positive solution
of the linear velocity $\nu^+_{\epsilon}$ is plotted as a function of the radial
distance $r/M$ for the parameter choice $Q/M = 0.6$ and different values of
$\epsilon =-5$ (black curve), $\epsilon =-3$ (thick black curve), and $\epsilon =-0.5$ (dashed curve).
The geodesic velocity $\nu_g $  for $\epsilon=0$ is also shown (gray curve). The choice of parameters implies
that $r_\gamma^+/M \approx 2.74$ and the outer horizon is located at $r_+/M = 1.8$.
The shaded region is forbidden. For $r>r_\gamma^+$ it holds that  $\nu^+_{\epsilon}>\nu_g $.
}} \label{Figure1an}
\end{figure}

\begin{figure}
\centering
\begin{tabular}{cc}
\includegraphics[scale=1]{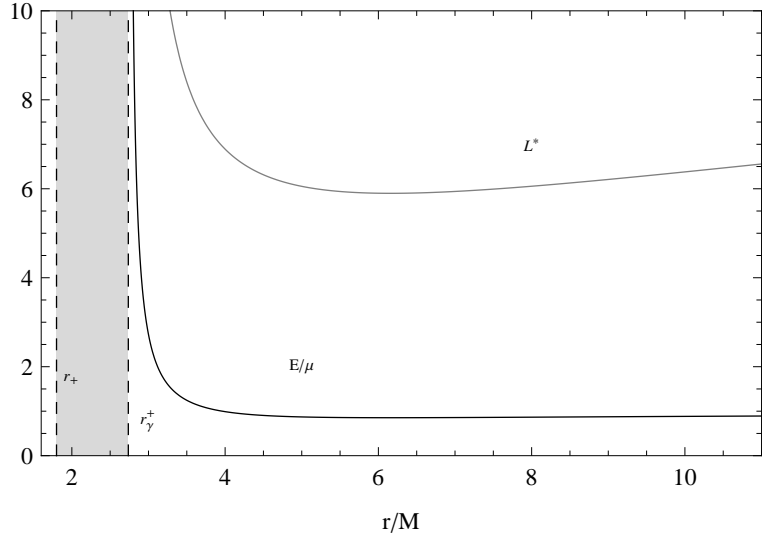}
\end{tabular}
\caption[font={footnotesize,it}]{\footnotesize{The energy
$E/\mu$ and angular momentum $L^*\equiv L/(\mu M)$ of a charged particle of charge--to--mass ratio $\epsilon$ moving
in the field of a RN black hole with charge $Q$ and mass $M$ are plotted as
functions of the radial distance $r/M$ for the parameter choice $Q/M =
0.6$ and $\epsilon= -3$, with $r_\gamma^+/M \approx 2.74$ and the outer
horizon located at $r_+/M = 1.8$. The shaded region is
forbidden.
}} \label{Figure1ab}
\end{figure}

\begin{figure}
\centering
\begin{tabular}{cc}
\includegraphics[scale=1]{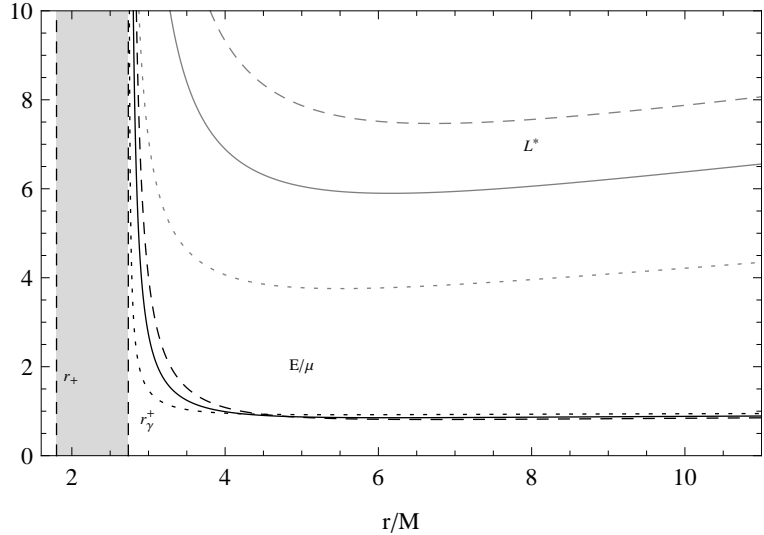}
\end{tabular}
\caption[font={footnotesize,it}]{\footnotesize{
The energy
$E/\mu$ and angular momentum $L^*\equiv L/(\mu M)$ of a charged particle of charge--to--mass ratio $\epsilon$ moving
in the field of a RN black hole with charge $Q$ and mass $M$ are plotted as
functions of the radial distance $r/M$ for the parameter choice $Q/M
0.6$ and $\epsilon= -3$ (solid curves), $\epsilon= -5$ (dashed curves), $\epsilon= -0.5$ (dotted curves). Here  $r_\gamma^+/M \approx 2.74$
and the outer
horizon is located at $r_+/M = 1.8$. The shaded region is
forbidden. The energy and angular momentum decrease as $|\epsilon|$ decreases.
}} \label{Figure1absil}
\end{figure}

The solutions are the geodesic velocities $\nu=\pm\nu_{\epsilon}^+$ in the
range $r\geq r_\gamma^+$ as illustrated in  Fig. \il\ref{Figure1a}.
Orbits with radius $r= r_\gamma^+$ are lightlike.
We can also compare the velocity of charged test particles  with the geodesic velocity $\nu_g $ for neutral particles.
For $r> r_\gamma^+$ we see that  $\nu_{\epsilon}^+>\nu_g $ always. This means that, at fixed orbital radius,
charged test particles acquire a larger orbital velocity compared to that of neutral test particles
in the same orbit.
As it is possible to see from Eq.\il(\ref{12}) and also in Fig. \il\ref{Figure1an},
an increase in  the particle charge  $\epsilon<0$ corresponds to an  increase in
the velocity $\nu_{\epsilon}^+$.
As the orbital radius decreases, the velocity increases until it reaches the limiting value $\nu_{\epsilon}^+=1$ which corresponds to the velocity
of a photon.
This fact can be seen also in Fig. \il\ref{Figure1ab}, where the energy and angular momentum for circular orbits are plotted in terms
of the distance $r$.
Clearly, this graphic shows that to reach the photon orbit at $r=r_\gamma^+$, the particles must acquire and infinity amount of energy
and angular momentum.
In Fig. \il\ref{Figure1absil} we analyze the behavior of the particle's energy and angular momentum in terms of
the specific charge $\epsilon$. It follows that both quantities decrease as the value of $|\epsilon|$ decreases.

%
%
\subsubsection{Case $\epsilon=0$}
The solutions are the geodesic velocities $\nu=\pm\nu_g $ in the
range $r\geq r_\gamma^+$. This case has been studied in detail in \cite{Pugliese:2010ps}.
%
\subsubsection{Case $\epsilon>0$}
Depending on the explicit values of the parameters $Q$ and $\epsilon$ and the radial coordinate $r$, it is necessary to analyze several subcases.
\begin{description}
\item[a)] $\epsilon < M/Q$ and  $r \geq r_l$.

There are two different branches for both signs of the linear velocity:
$\nu=\pm\nu_{\epsilon}^+$ in the range $r_l \leq r\leq r_\gamma^+$, and
$\nu=\pm\nu_{\epsilon}^-$ in the whole range $r\geq r_l$. The two
branches join at $r=r_l$, where $\nu^+_{\epsilon}
=\nu^{-}_{\epsilon}=\nu_{g}  \sqrt{\Lambda}$, as shown in Fig. \il\ref{Figure2abcd}. First we note that in this case for $r> r_\gamma^+$ it always holds that $\nu_{\epsilon}^{-}<\nu_g $. This means that, at fixed orbital radius,  charged test particles possess a smaller orbital velocity than that of neutral test particles in the same orbit.  This is in accordance to the fact that in this case, a black hole with  $\epsilon Q>0$, the attractive gravitational force is balanced by the repulsive electromagnetic force. In the region $r>r_\gamma^+$, the orbital velocity increases as the radius approaches the value $r_{\gamma}^+$ (see Fig. \il\ref{Figure2abcd}). The interval $r_l \leq r\leq r_\gamma^+$ presents a much more complex dynamical structure.
First we note that, due to the Coulomb repulsive force, charged particle orbits are  allowed in a region which is forbidden for neutral test particles.
This is an interesting result leading to the possibility of accretion disks in which the innermost part forms a ring of charged particles only.
Indeed, suppose that an accretion disk around a RN black hole is made of neutral and charged test particles. Then, the accretion disk can exist only
in the region $r\geq r_l$ with a ring of charged particles in the interval $[r_l,r_\gamma^+)$. Outside the exterior radius of the ring $(r>r_\gamma^+)$, the disk
can be composed of neutral and charged particles.
\begin{figure}
\centering
\begin{tabular}{cc}
\includegraphics[scale=.7]{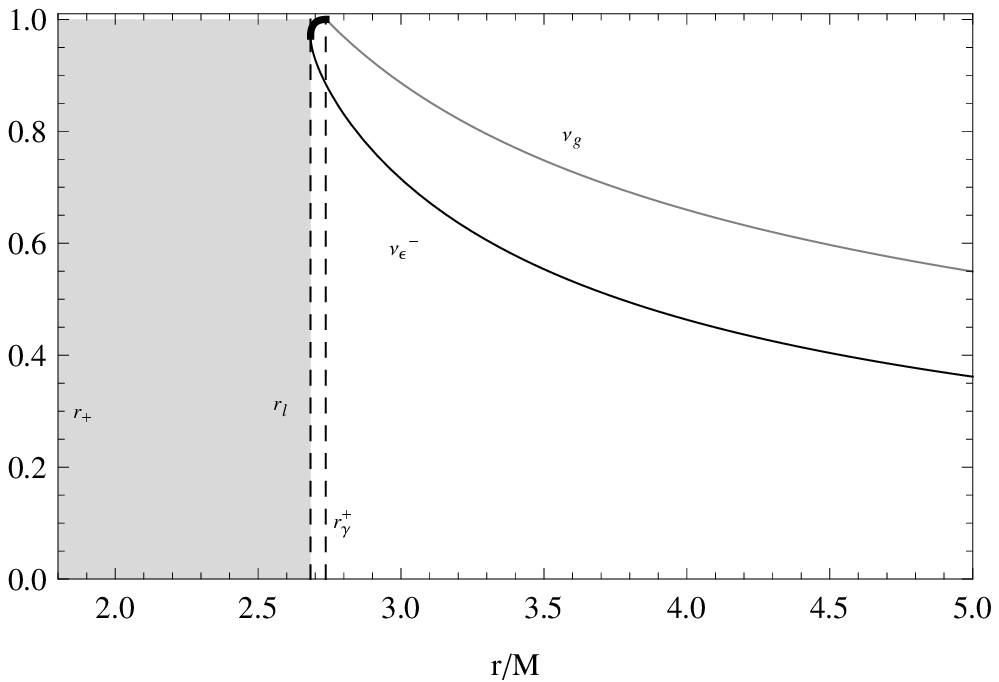}
\includegraphics[scale=.7]{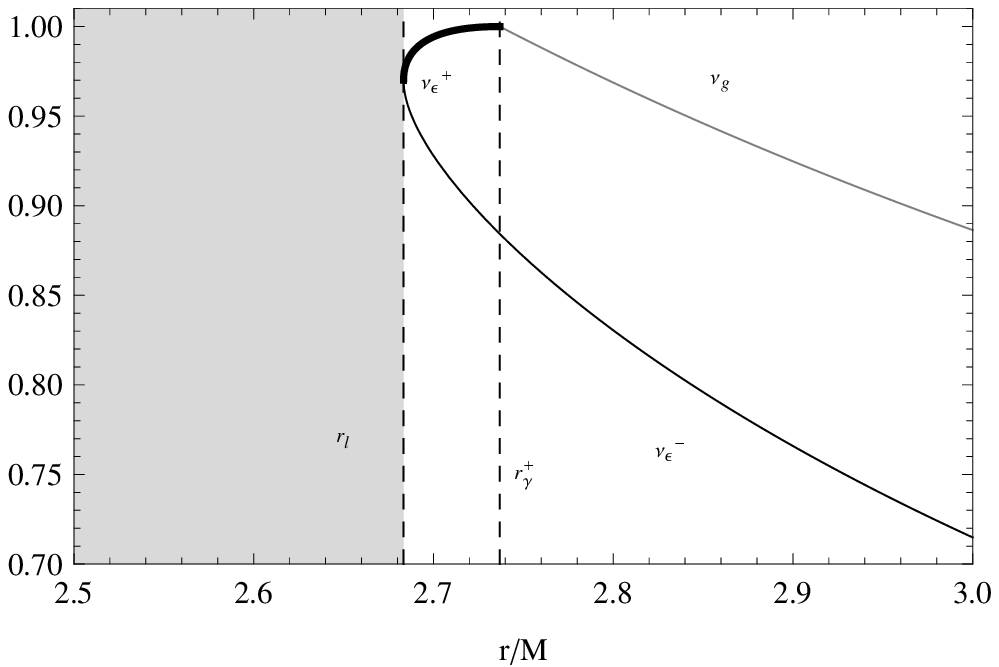}
\end{tabular}
\caption[font={footnotesize,it}]{\footnotesize{The positive solution of the linear velocity $\nu_{\epsilon}^{\pm}$ is plotted as a function of the radial
distance $r/M$ in the region $[1.8,5]$ (left graphic) and $[2.5,3]$ (right graphic).  Here
$Q/M = 0.6$ and $\epsilon = 1.2$, so that  $r_l/M =2.68$ and $r_\gamma^+=2.737M$.  For the chosen
parameters we have that  $\tilde{\epsilon}=3.25$ and $\epsilon_{l} = 4.12$. The region within
the interval $[r_l,r_\gamma^+]$ is forbidden for neutral particles.}} \label{Figure2abcd}
\end{figure}
This situation can also be  read from  Fig. \il\ref{Figure2a} where the energy and the angular momentum are plotted as functions of the radial distance
$r/M$.
\begin{figure}
\centering
\begin{tabular}{cc}
\includegraphics[scale=.7]{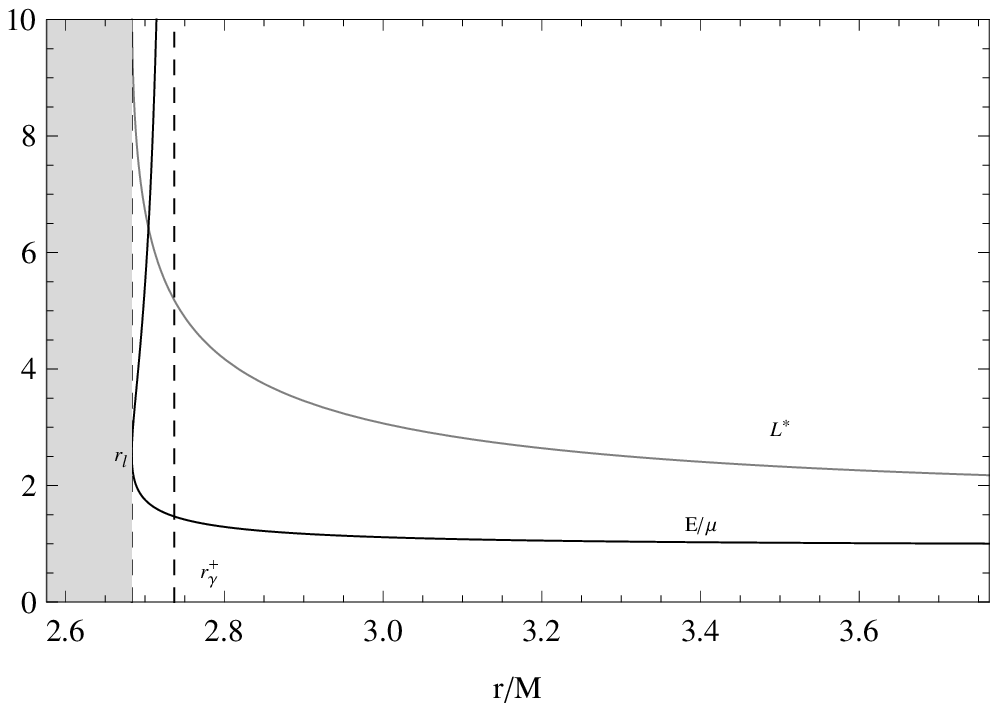}
&\includegraphics[scale=.7]{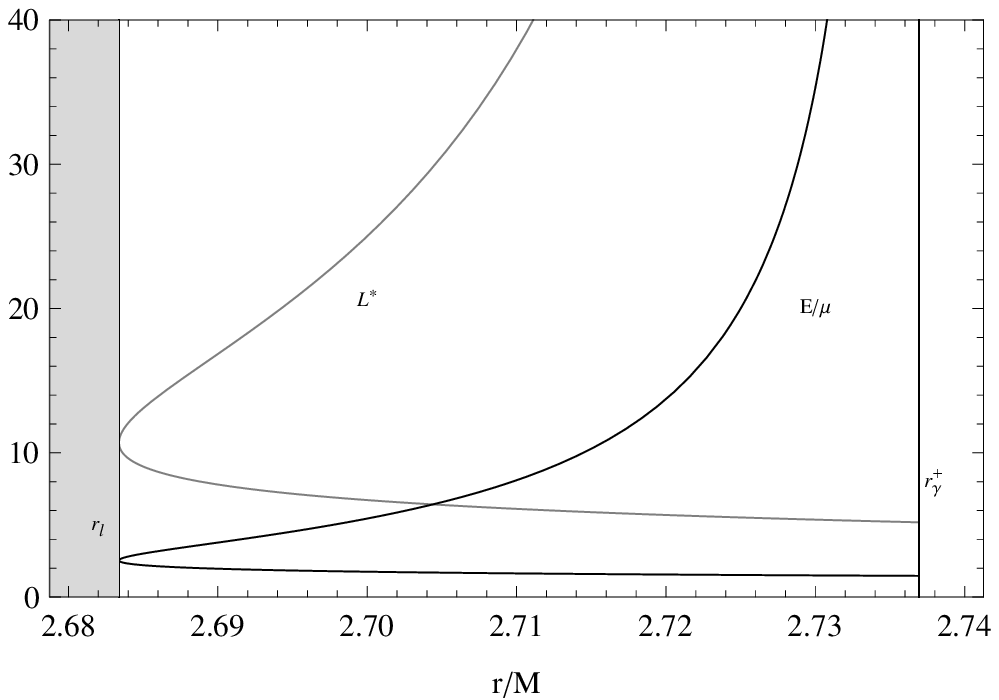}
\end{tabular}
\caption[font={footnotesize,it}]{\footnotesize{The energy
$E/\mu$ and angular momentum $L^*\equiv L/(\mu M)$  of a charged particle of charge--to--mass ratio $\epsilon$ moving
along circular orbits in a
Reissner--Nordstr\"om black hole of charge $Q$ and mass $M$ are plotted in terms of
the radial distance $r/M$ in the range $[2.6,3.8]$ (left graphic)  $[2.68,2.74]$ (right graphic). Here $Q/M= 0.6$
and $\epsilon = 1.2$, so that  $r_l/M =2.68$ and $r_\gamma^+/M=2.737$.  For the
chosen parameters we have that $\tilde{\varepsilon}=3.25$
$\epsilon_{l} = 4.12$. The shaded region is forbidden for any particles.
}}
\label{Figure2a}
\end{figure}
\\
\item[b)] $M/Q <\epsilon < \tilde{\epsilon}$ and $r_l \leq r\leq r_{s}$.

Since $r < r_{s}$, one has that $\epsilon/\epsilon_0 < 1$, implying that
both solutions $\nu^{+}_{\epsilon}$ and $\nu^{-}_{\epsilon}$ can exist.
There are two different branches for both signs:
$\nu=\pm\nu^+_{\epsilon}$ in the range $r_l \leq r\leq r_{\gamma}^{+}$,
and
 $\nu^{-}_{\epsilon}$ in the entire range $r_l\leq r\leq r_{s}$. The two branches join at $r= r_l$.
 Note that for increasing values of $\epsilon$, the radius
$r_{s}$ decreases and approaches $r_l$, reaching it at $\epsilon
=\tilde{\epsilon}$, and as $\epsilon$ tends to infinity $r_s$ tends to the outer horizon $r_+$ (see Fig. \il\ref{PlotRsRt}).
\begin{figure}
\centering
\begin{tabular}{c}
\includegraphics[scale=.7]{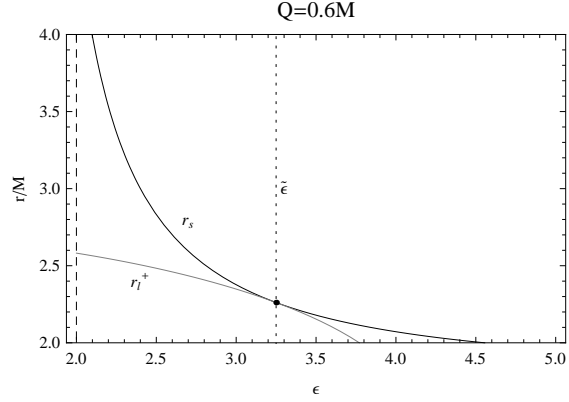}
\end{tabular}
\caption[font={footnotesize,it}]{\footnotesize{Radius $r_s=r_s^+$ (black curve) and $r_l=r_l^+$ (gray curve), are plotted as function of $\epsilon$ for $Q=0.6M$. $r_s^+=r_l^+$ for $\epsilon=\tilde{\epsilon}\approx3.25$. }} \label{PlotRsRt}
\end{figure}
\begin{figure}
\centering
\begin{tabular}{cc}
\includegraphics[scale=.7]{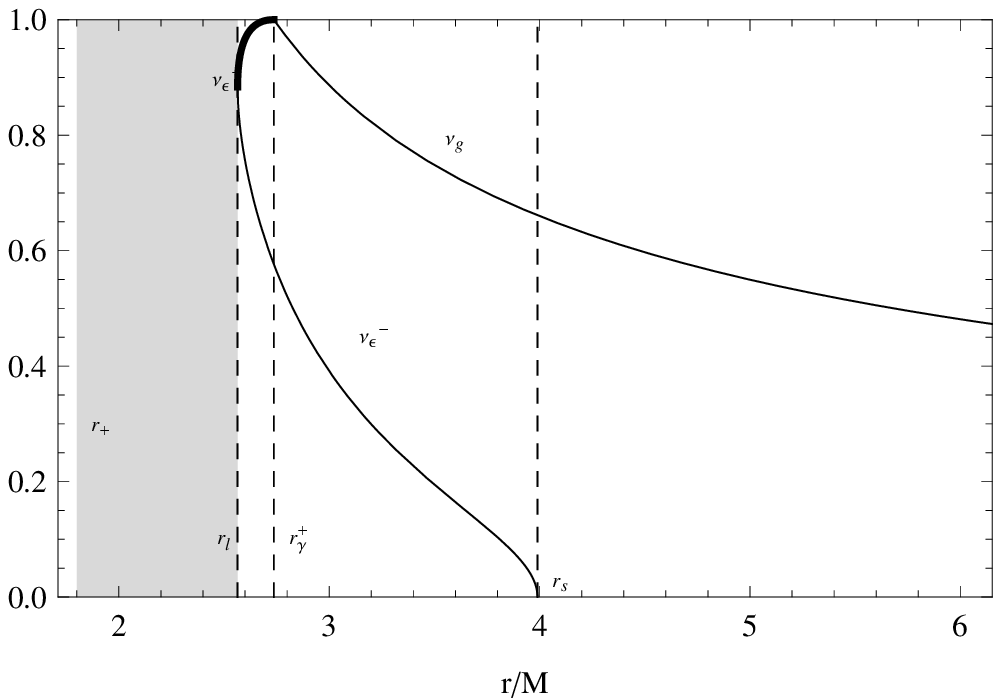}
\includegraphics[scale=.7]{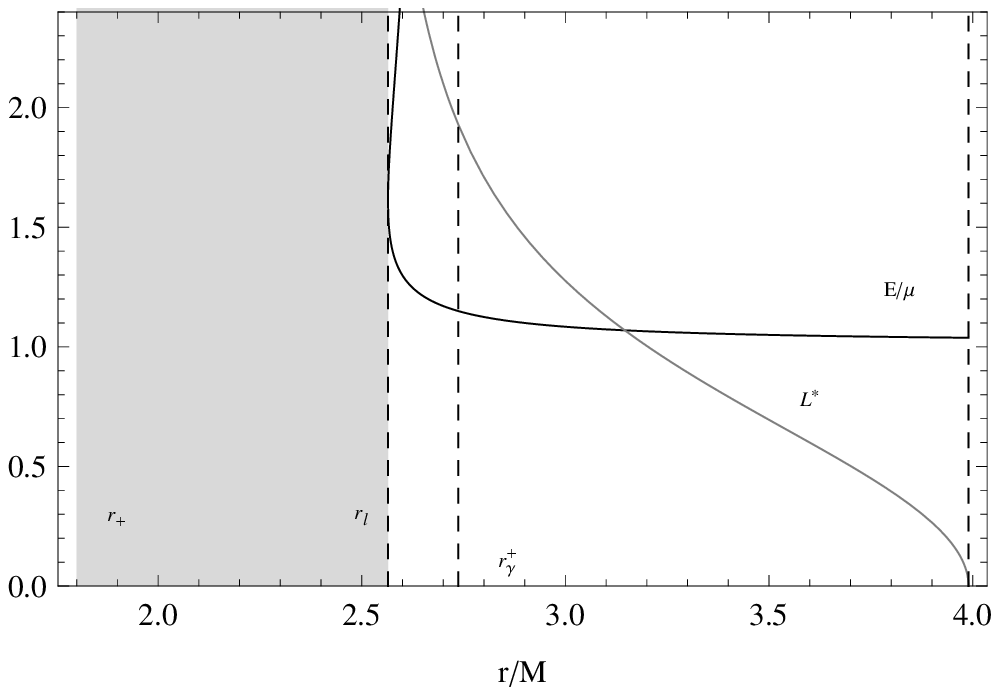}
\end{tabular}
\caption[font={footnotesize,it}]{\footnotesize{Left graphic: The positive solution
of the linear velocity $\nu$ is plotted as a function of the radial
distance $r/M$. Right graphic: The energy
$E/\mu$ and angular momentum $L^*\equiv L/(\mu M)$
of a charged particle of charge--to--mass ratio $\epsilon$  are plotted in terms of $r/M$.
The parameter choice is $Q/M = 0.6$ and $\epsilon = 2.1$. Then, $r_l/M =2.56$, $r_\gamma^+/M =2.737$,
and $r_{s}/M=3.99$. Moreover, for this choice $\tilde{\epsilon}=3.25$ and $\epsilon_{l} = 4.12$.
The shaded region is
forbidden.}} \label{Figure2bc}
\end{figure}
In particular, the interaction between the attractive gravitational force and the Coulomb force generates a zone
$r_l \leq r\leq r_\gamma^+$  in which only charged test particles can move along circular trajectories while neutral
particles are allowed in the region $r>r_\gamma^+$ (see Fig. \ref{Figure2bc}).
This result again could be used to construct around black holes accretion disks with rings made of charged particles.
\\
\item[c)]
$\tilde{\epsilon}<\epsilon <\epsilon_{l} $ and $r_{s} < r < r_\gamma^{+}$.

The solution $\nu^{-}_{\epsilon}$ for the linear velocity is not allowed whereas the
solution $\nu^+_{\epsilon}$ is valid in the entire range. In fact, the condition $r > r_{s}$
implies that $\epsilon/\epsilon_{0} > 1$, and therefore
$\Lambda^{2}-1+(\epsilon/\epsilon_{0})^{2} > \Lambda^{2}$, so that
the condition (\ref{180}) for the existence of velocities is
satisfied for the plus sign only. Therefore, the solutions are given
by $\nu=\pm\nu^+_{\epsilon}$ in the entire range as shown in Fig. \il\ref{Figure2abcdz}.
At the radius orbit $r=r_{s}$, the angular momentum and the velocity of the test particle vanish,
indicating that the particle remains at rest with respect to static observers located at
infinity. In the region $r_{s} < r < r_\gamma^{+}$ only charged particles can move along circular trajectories.
\begin{figure}
\centering
\begin{tabular}{cc}
\includegraphics[scale=.7]{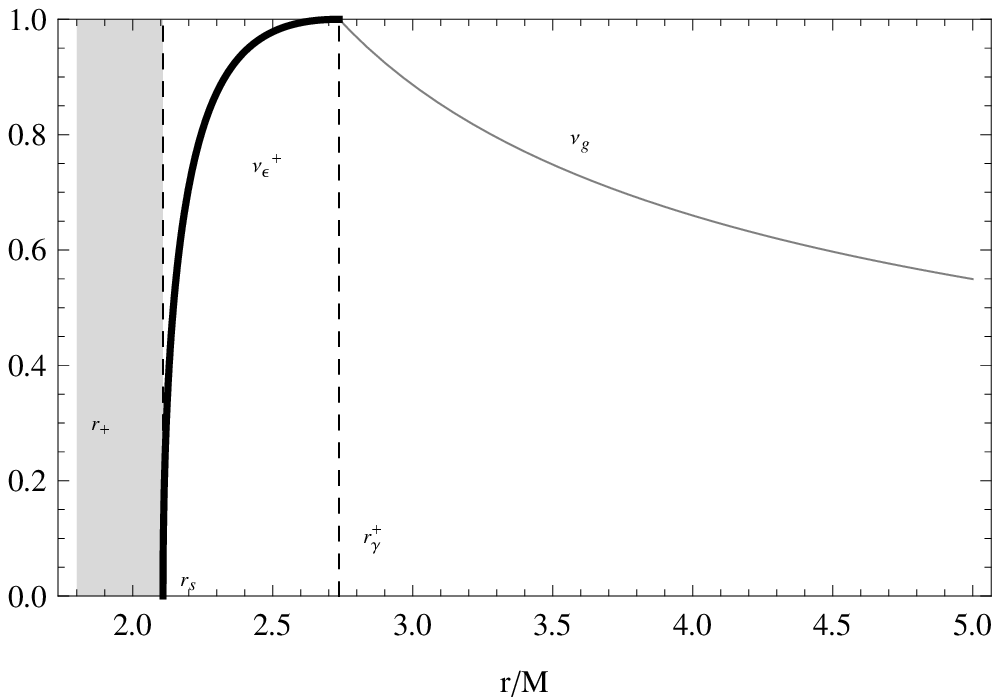}
\includegraphics[scale=.7]{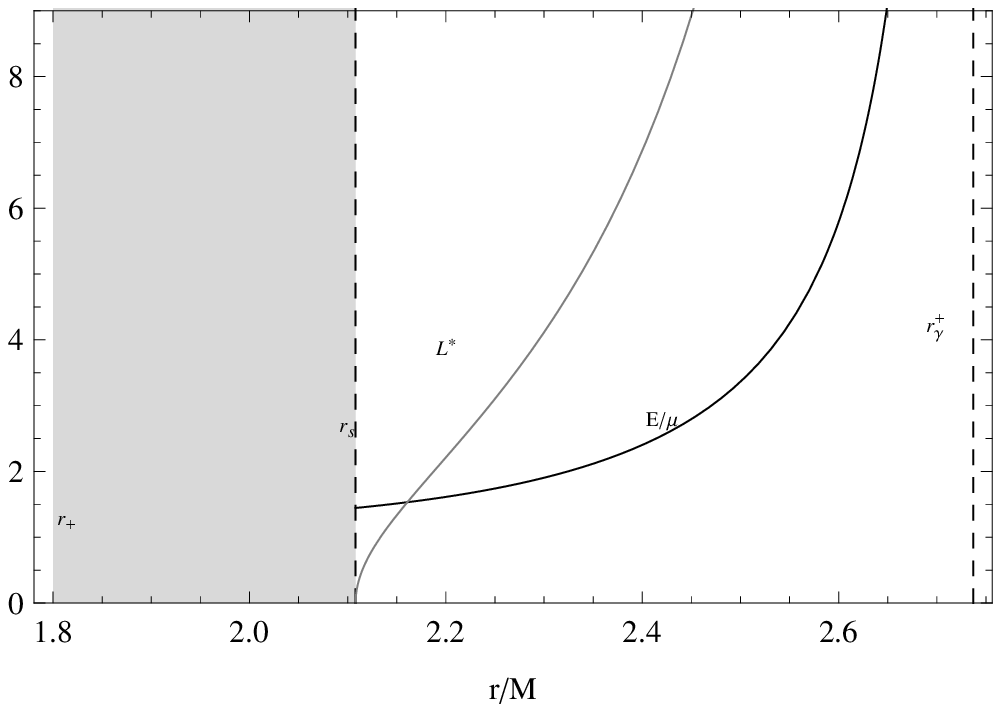}
\end{tabular}
\caption[font={footnotesize,it}]{\footnotesize{Left graphic: The positive solution
of the linear velocity $\nu$ is plotted as a function of the radial
distance $r/M$. Right graphic: The energy
$E/\mu$ and angular momentum $L^*\equiv L/(\mu M)$
of a charged particle of charge--to-mass ratio $\epsilon=3.8$ moving in
a RN spacetime with $Q/M=0.6$ are plotted in terms of
the radial distance $r/M$. For this choice of parameters the radii
are $r_l/M =1.98$, $r_{s}/M =2.11$, and $r_\gamma^+/M=2.737$ whereas the
charge parameters are $\tilde{\epsilon}=3.25$ and $\epsilon_{l} = 4.12$.
 }} \label{Figure2abcdz}
\end{figure}
\\
\item[d)] $ \epsilon > \epsilon_l$ and $r_{s}<r<r^{+}_\gamma$.

In this case the radius $r_l$ does not exist.
The solutions are the velocities $\nu=\pm\nu^+_{\epsilon}$ in the entire range.
Note that for
$\epsilon\rightarrow\infty$ one has that $r_{s}\rightarrow r_+$.
Also in this case we note that neutral particles can stay in circular orbits with a velocity $\nu_g $ only in the region
$r>r^{+}_\gamma$ whereas charged test particles are allowed within the interval $r_{s}<r<r^{+}_\gamma$, as shown in Fig. \il\ref{Figure2abcdx}.
Clearly, for charged and neutral test particles the circular orbit at $r=r^{+}_\gamma$ corresponds to a limiting orbit.
\end{description}
\begin{figure}
\centering
\begin{tabular}{cc}
\includegraphics[scale=.7]{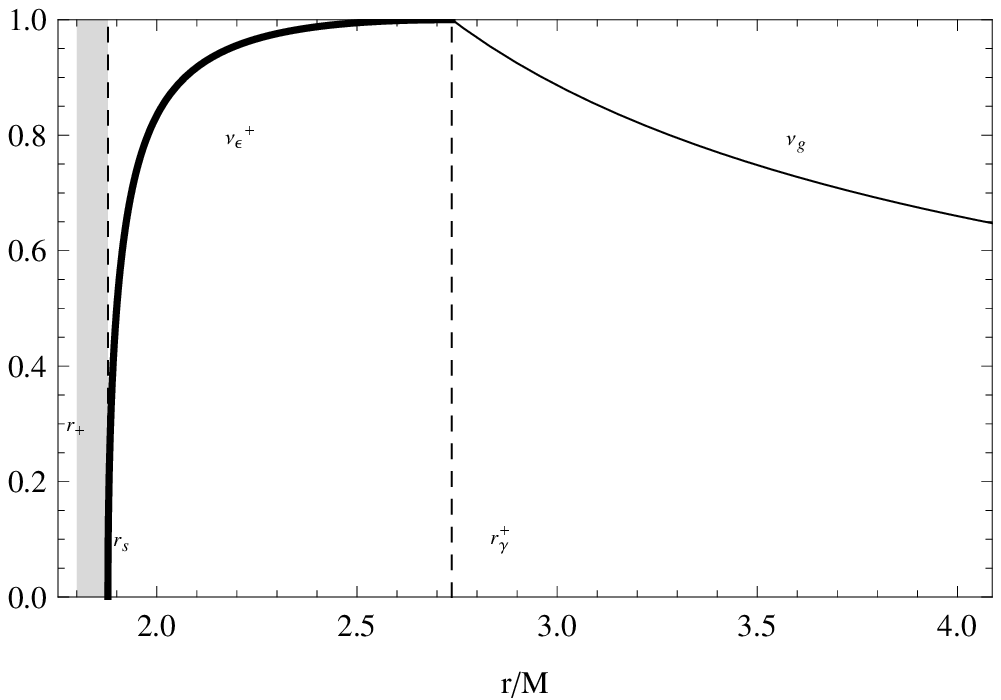}
\includegraphics[scale=.7]{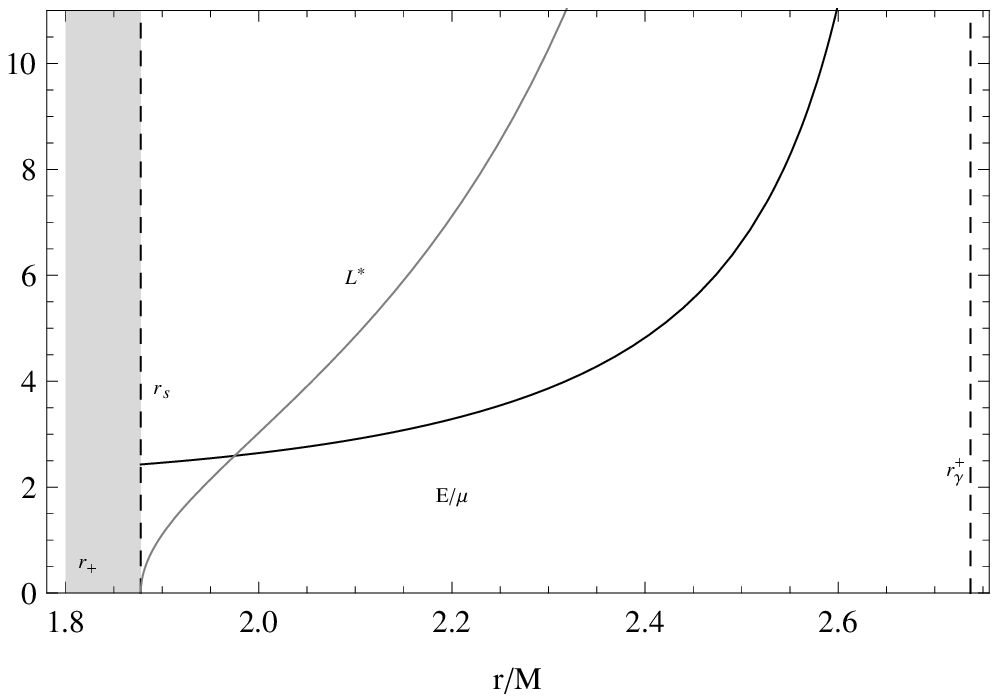}
\end{tabular}
\caption[font={footnotesize,it}]{\footnotesize{Left graphic: The positive solution
of the linear velocity $\nu$ is plotted as a function of the radial
distance $r/M$. Right graphic: The energy
$E/\mu$ and angular momentum $L^*\equiv L/(\mu M)$
of a charged particle of charge--to--mass ratio $\epsilon=7$, moving in the field of a
RN black hole with $Q/M=0.6$, are plotted in terms of the radial distance $r/M$.
For this parameter choice $r_{s}/M =1.88$, $\tilde{\epsilon}=3.25$, and
$\epsilon_{l} = 4.12$.
}} \label{Figure2abcdx}
\end{figure}
%
\subsection{Stability}
\label{sec:stb}
\begin{figure}
\centering
\begin{tabular}{cc}
\includegraphics[scale=.71]{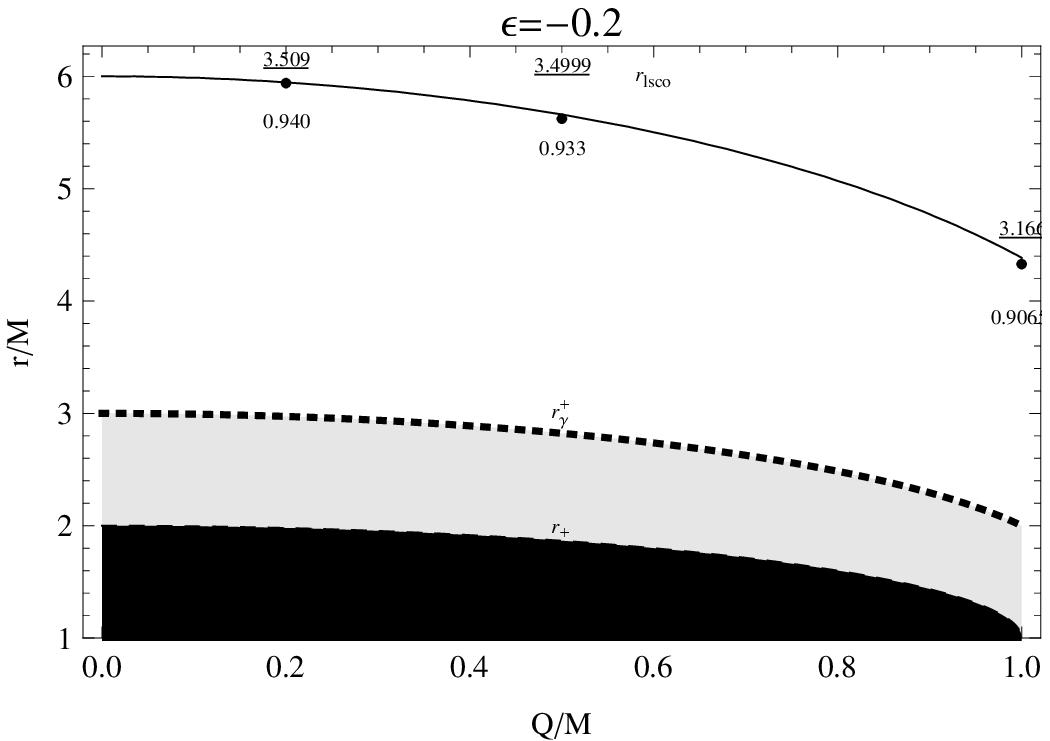}
\includegraphics[scale=.71]{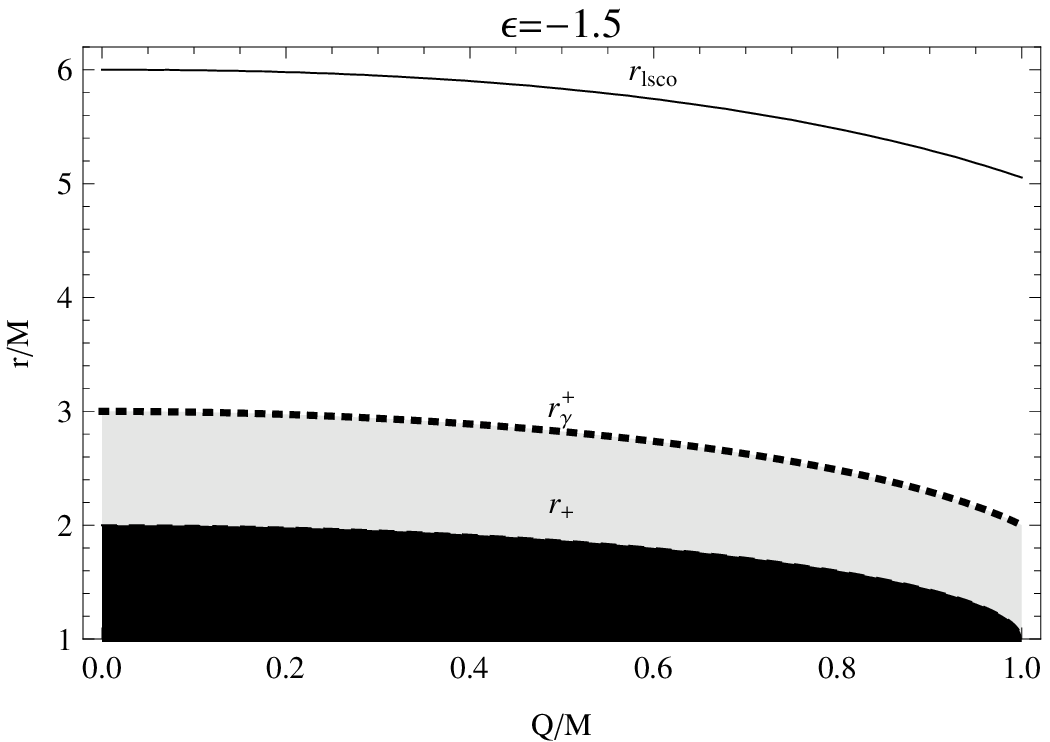}
\end{tabular}
\caption[font={footnotesize,it}]{\footnotesize{The
radius  of the last stable circular orbit $r_{\ti{lsco}}$ in
a RN black hole of mass $M$ and charge $Q$ for
a  particle with ratio $\epsilon=-0.2$ (left plot) and $\epsilon=-1.5$ (right plot).
Numbers close to the point represent the energy $E/\mu$ of the last stable circular orbits at that point.
Underlined numbers represent the corresponding angular momentum $L/(M\mu)$.
Stable orbits are possible only for $r>r_{\ti{lsco}} $. For comparison we also include the
curves for the radii $r_+$ and $r_\gamma^+$.
}}
\label{ultimaepsilonEpm02}
\end{figure}

\begin{figure}
\centering
\begin{tabular}{c}
\includegraphics[scale=1]{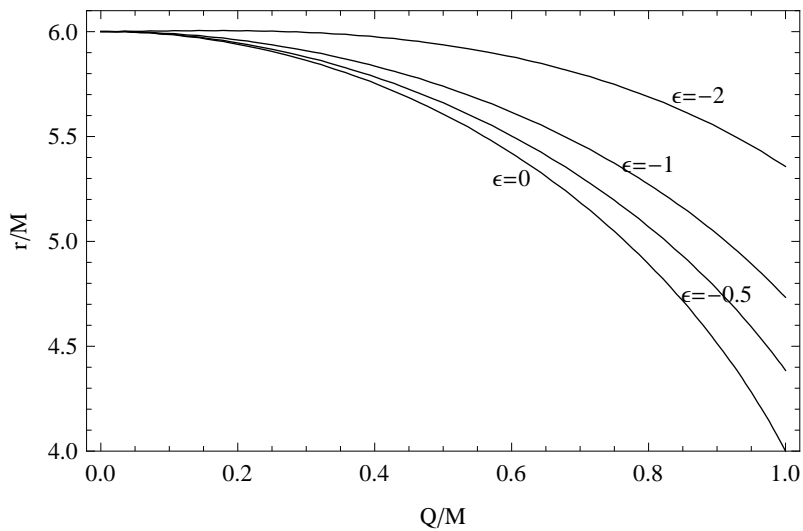}
\end{tabular}
\caption[font={footnotesize,it}]{\footnotesize{The
radius $r_{\ti{lsco}}$ of the last stable circular orbit in
a RN black with charge--to--mass ratio $Q/M$ for
selected values of the charge--to--mass ratio $\epsilon$ of the test particle.
Only the case $\epsilon Q\leq0$ is illustrated.
Stable orbits are possible only for $r>r_{\ti{lsco}}$}}
\label{ultimaepsilon}
\end{figure}

To analyze the stability of circular orbits for charged test particles  in
a RN black hole we must consider the condition (\ref{Lagespep}) which leads
to the Eqs.(\ref{nube}), (\ref{nuvola}), and (\ref{Carmara}).
So the stability of circular orbits   strongly depends on the
sign of $(\epsilon Q)$.
The case  $\epsilon Q\leq0$ is illustrated in
Fig. \il\ref{ultimaepsilonEpm02} where the radius of the last stable circular orbit
$r_{\ti{lsco}}$ is plotted for two different values of $\epsilon$ as a function of
$Q/M$. It can be seen that the energy and angular momentum of the particles decrease
as the value of $Q/M$ increases.
These graphics also include the radius of the outer horizon $r_+$ and the
radius $r_\gamma^+$ which  determines the last (unstable) circular orbit
of neutral particles. In Sec.  \ref{BHBHTR}, we found that circular orbits for charged particles
are allowed also inside the radius $r_\gamma^+$ for certain values of the parameters; however, since
$r_\gamma^+<  r_{\ti{lsco}}$, we conclude that all those orbits must be unstable.
From Fig. \il\ref{ultimaepsilon} we see that for  $Q=0$ and $\epsilon=0$, the
well-known result for the Schwarzschild case, $r_{\ti{lsco}}= 6M$, is recovered.
Also in the limiting case $Q=M$ and $\epsilon=0$, we recover the value of
$r_{\ti{lsco}}= 4M$ for neutral particles moving along circular orbits in
an extreme BN black hole.
In general, as the value of $|\epsilon|$ increases we see that the
value of $ r_{\ti{lsco}}$ increases as well. This behavior resembles
the case of the radius of the last stable orbit for neutral test particles \cite{Pugliese:2010ps,Pugliese:2010he}.
Indeed, in the case $\epsilon Q<0$ the attractive Coulomb force reinforces the attractive gravitational force
so that the general structure remains unchanged. We also can expect that an increase in the charge of the particle $|\epsilon|$
produces an increase in  the velocity of the stable circular orbits. In fact, this can be seen explicitly from Eq.\il(\ref{12}) and Fig. \il\ref{Figure1an}.
It then follows that the energy and angular momentum of the charged test particle increases as the value of $|\epsilon|$ increases.

\begin{figure}
\centering
\begin{tabular}{cc}
\includegraphics[scale=1]{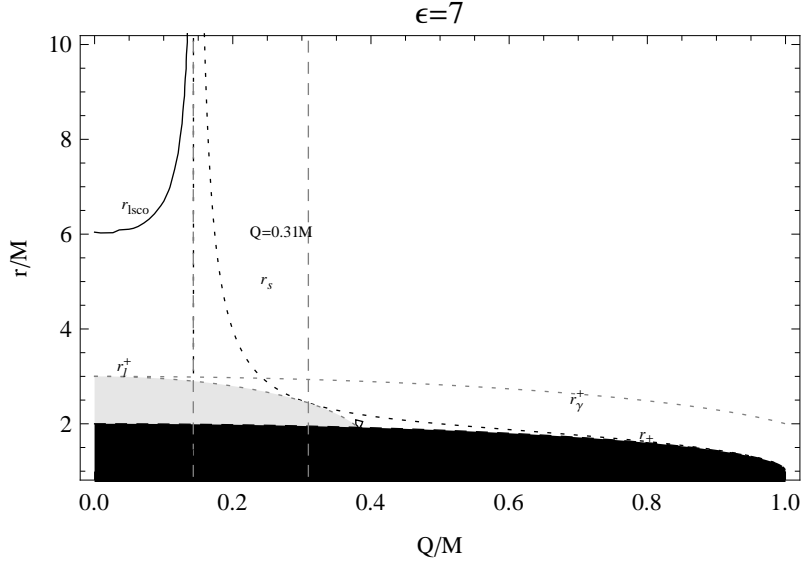}
\end{tabular}
\caption[font={footnotesize,it}]{\footnotesize{The radius of
the last stable circular orbit $r_{\ti{lsco}}$ (solid curve) for a charged
test particle with $\epsilon=7$, in a RN black hole with charge $Q$ and mass $M$,
is plotted as a function of the ratio $Q/M$. Other curves are
the outer horizon radius $r_{+}=M+\sqrt{M^2-Q^2}$ and the radii
$r_\gamma^{+}\equiv [3M+\sqrt{(9M^2-8Q^2)}]/2$, $r_{s}\equiv\frac{Q^{2}}{\epsilon^{2}Q^{2}-M^{2}}\left[\epsilon\sqrt{M^{2}-Q^{2}}\sqrt{\epsilon^{2}-1}+M(\epsilon^{2}-1)\right]$,
$r_l \equiv\frac{3 M}{2}+\frac{1}{2}
\sqrt{9 M^{2}-8 Q^{2}-Q^{2} \epsilon ^{2}} $. Shaded and dark regions are
forbidden for timelike particles.  Stable orbits are possible
only for $r>r_{\ti{lsco}}$.}} \label{BHP7}
\end{figure}

\begin{figure}
\centering
\begin{tabular}{cc}
\includegraphics[scale=1]{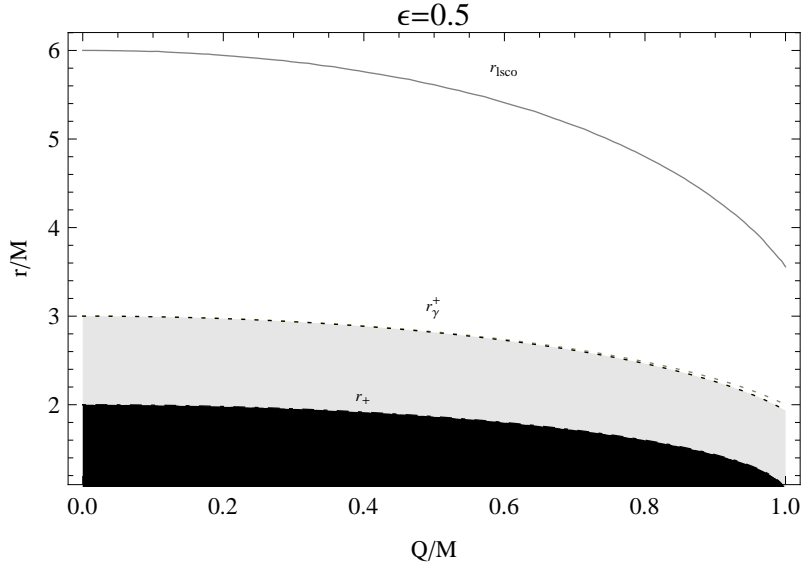}
\end{tabular}
\caption[font={footnotesize,it}]{\footnotesize{The radius of
the last stable circular orbit $r_{\ti{lsco}}$ (solid curve) for a charged
test particle with $\epsilon=0.5$, in a RN black hole with charge $Q$ and mass $M$,
is plotted as a function of the ratio $Q/M$. Other curves are
the outer horizon radius $r_{+}$ and the radius  $r_\gamma^{+}$.
Shaded and dark regions are
forbidden for timelike particles.  Stable orbits are possible
only for $r>r_{\ti{lsco}}$.
}} \label{BHP05}
\end{figure}

The case  of $\epsilon Q>0$ is illustrated in Figs.\il\ref{BHP7} and \ref{BHP05}.
The situation is  very different from the case of neutral particles or charged particles with $\epsilon Q<0$.
Indeed, in this case the Coulomb force is repulsive and leads to a non trivial interaction with the attractive gravitational
force, see also \cite{Joshi,Belinski:2008zz,Luongo:2010we,Pizzi:2008ti,Belinski:2008bn,Pizzi:2008zz,Paolino:2008qi,Manko:2007hi,Alekseev:2007re,Preti:2008zz}.
It is necessary to analyze two different subcases. The first subcase for $\epsilon>1$ is illustrated
in Fig. \il\ref{BHP7} while the second one for  $0<\epsilon<1$ is depicted in Fig. \il\ref{BHP05}.
We can see that in the case   $0<\epsilon<1$ the stability  regions   are similar  to those found in the case $\epsilon<0$ (cf.
Figs.\il\ref{ultimaepsilonEpm02} and \ref{BHP05}). This means that for weakly--charged test particles, $0<\epsilon  <1$,
it always exists a stable circular orbit and  $r_{\ti{lsco}}\geq 4M$, where the equality holds for an extreme black hole.
On the contrary, in the case  $\epsilon >1$ there are regions  of $Q$ and $\epsilon$ in which stable circular orbits cannot exist at all.
As can be seen from Fig. \ref{BHP7}, charged particles moving along circular orbits with radii
located within the region $r<r_\gamma^+$ or $r<r_{s}$ must be unstable.

We conclude that the ring structure of the hypothetical accretion disks around a RN black hole mentioned in Sec.  \ref{BHBHTR}
must be unstable.

\section{Naked singularities}
\label{NSNSTRE}
\begin{figure}
\centering
\begin{tabular}{cc}
\includegraphics[scale=.7]{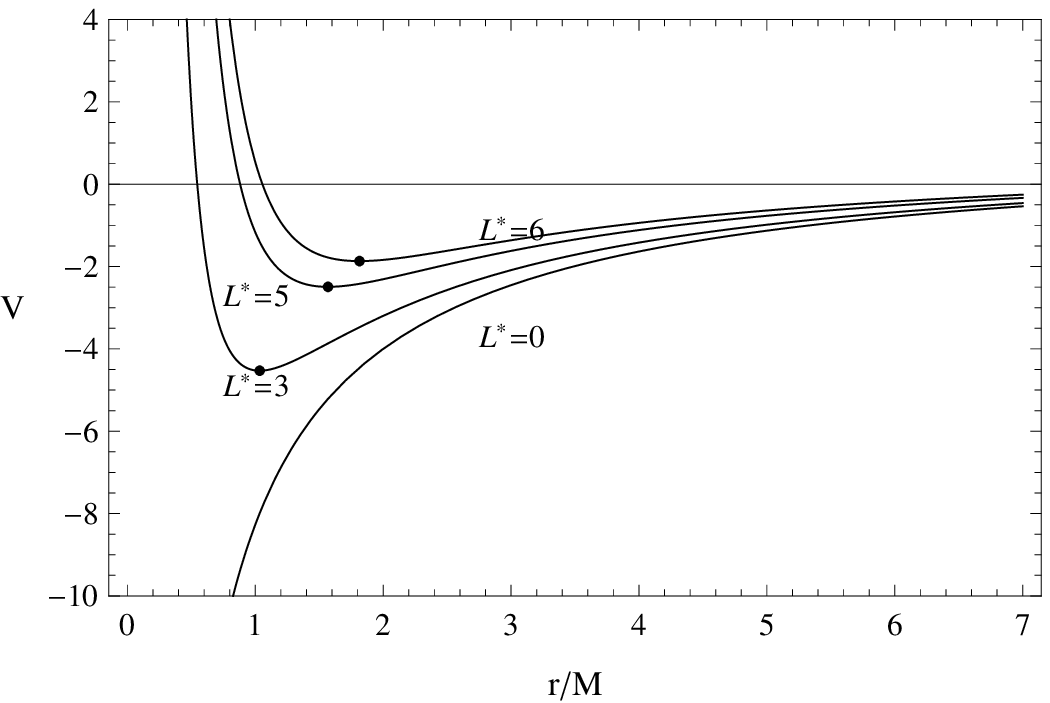}
\includegraphics[scale=.7]{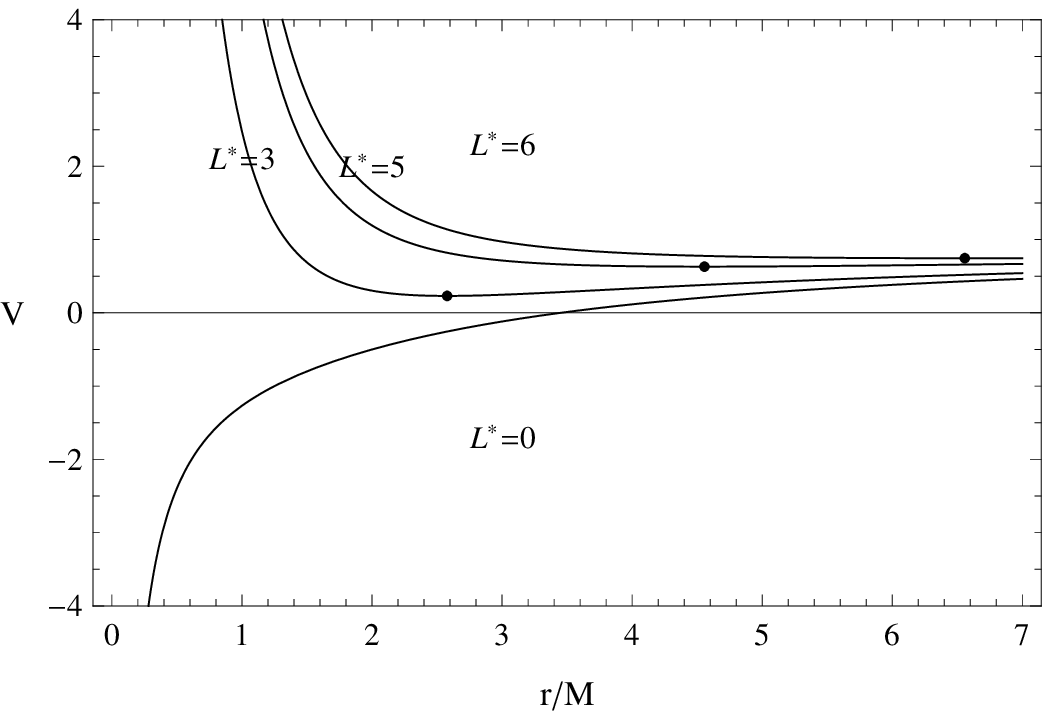}\\
\includegraphics[scale=.7]{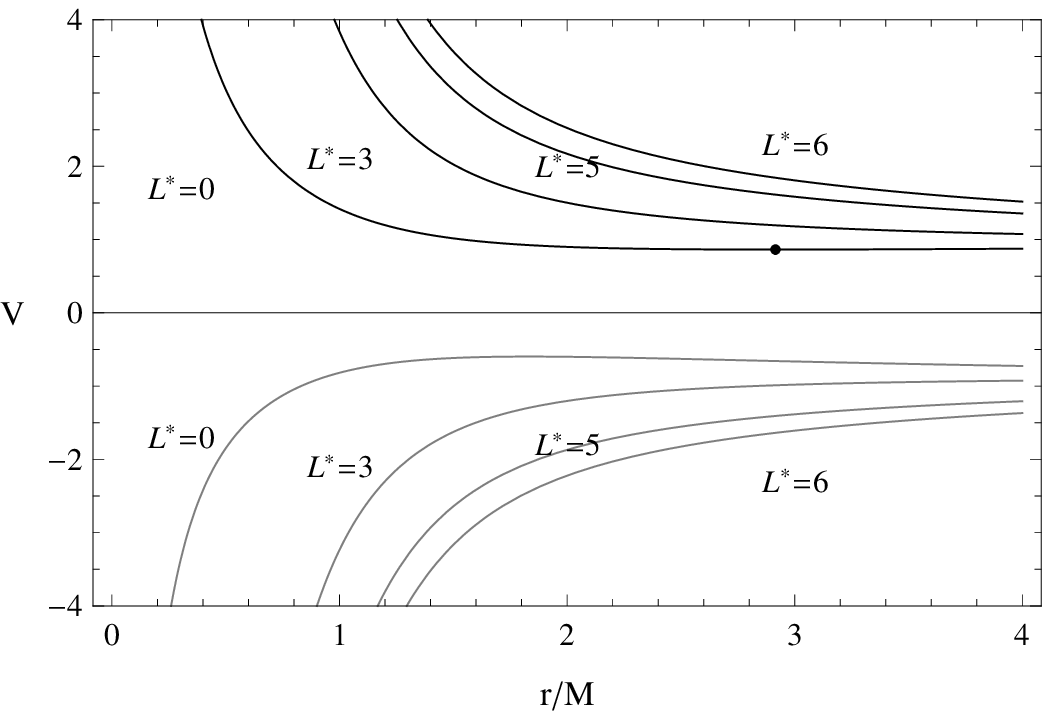}
\end{tabular}
\caption[font={footnotesize,it}]{\footnotesize{
The effective potential for a charged particle with charge--to--mass  ratio
$\epsilon$ in a RN naked singularity of charge $Q$ and mass
$M$ is plotted as a function of the radius $r/M$  for fixed values of the
angular momentum $L^*\equiv L/(\mu M)$. Black curves represent the
positive solution $V_+$ while gray curves correspond to $V_-$.
The boldfaced points denote the minima of the potentials. In upper left plot,
the parameter choice is $Q/M=2$ and $\epsilon=-1.5$; the upper right plot is for
$Q/M=2$ and $\epsilon=-5$ while the bottom plot corresponds to the choice
$Q/M=1.5$ and $\epsilon=-0.2$.} } \label{NKS}
\end{figure}

\begin{figure}
\centering
\begin{tabular}{c}
\includegraphics[scale=1]{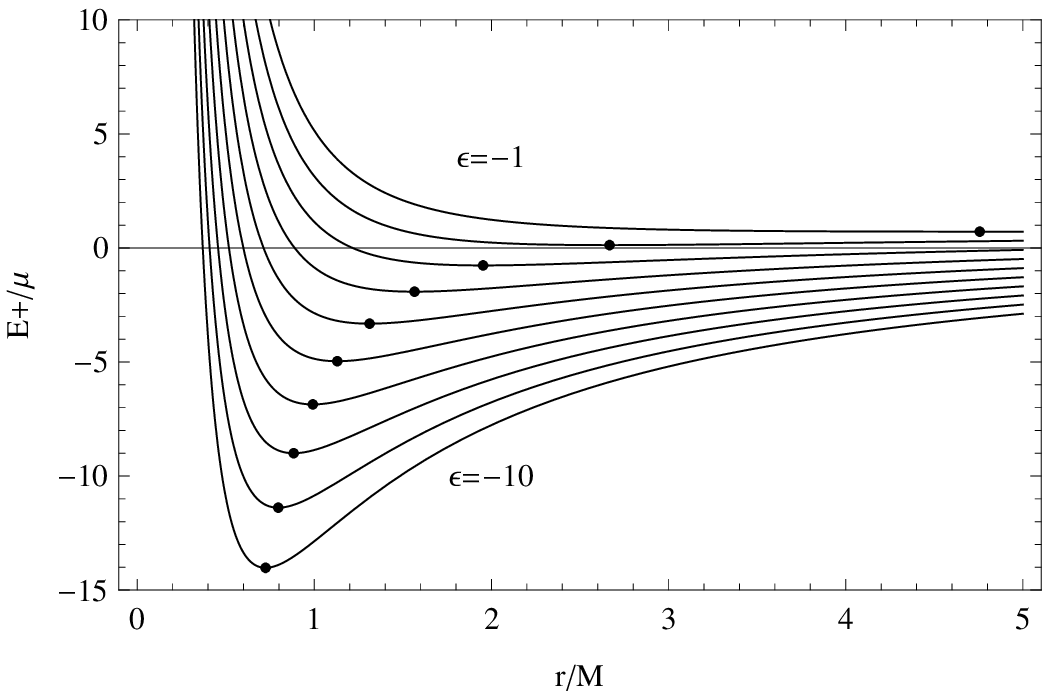}
\end{tabular}
\caption[font={footnotesize,it}]{\footnotesize{The effective potential
of a RN naked singularity with $Q/M=2$ for a particle with  charge--to--mass ratio
$\epsilon$ in the range $[-10,-1]$ and angular momentum $L/(M\mu)=4$. }}
\label{11:00M}
\end{figure}

\begin{figure}
\centering
\begin{tabular}{cc}
\includegraphics[scale=.7]{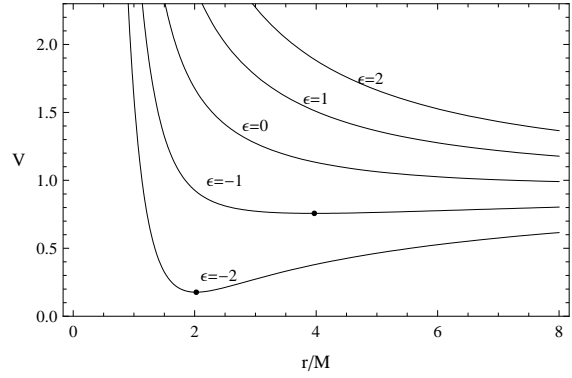}
\end{tabular}
\caption[font={footnotesize,it}]{\footnotesize{
The effective potential
of a RN naked singularity with $Q/M=3/2$ for a particle with  charge--to--mass ratio
$\epsilon$ in the range $[-2,+2]$ and angular momentum $L/(M\mu)=4$.}}
\label{NKSepsilon}
\end{figure}

\begin{figure}
\centering
\begin{tabular}{cc}
\includegraphics[scale=.7]{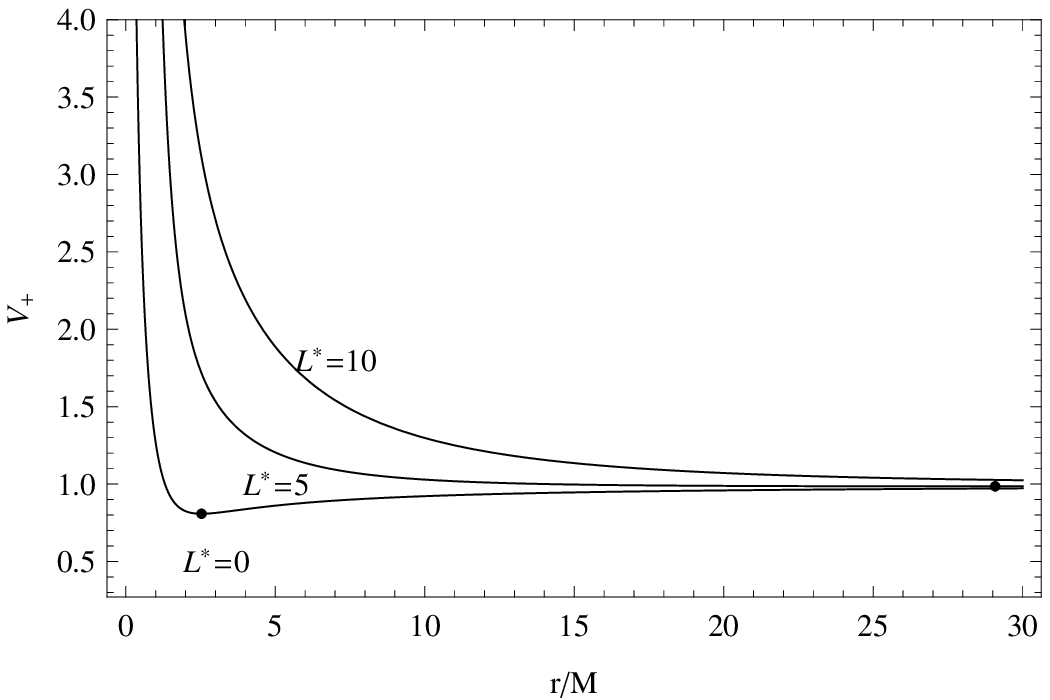}
\includegraphics[scale=.7]{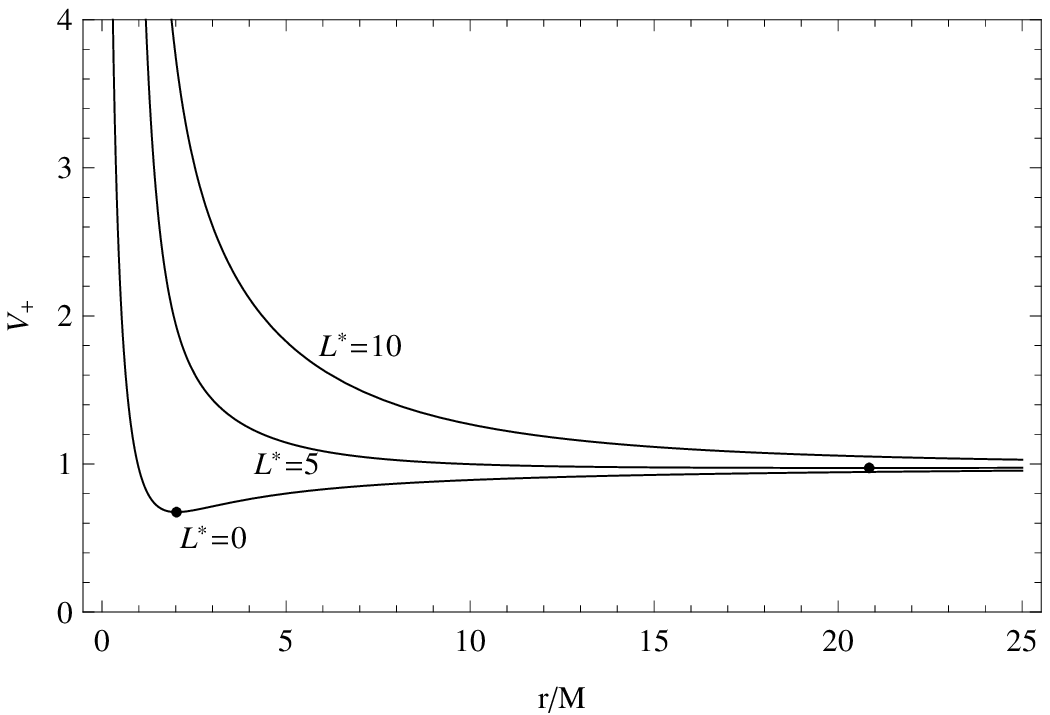}
\end{tabular}
\caption[font={footnotesize,it}]{\footnotesize{
The effective potential $V_+$
of a RN naked singularity with $Q/M=3/2$ for a charged particle
is plotted for different values of
the angular momentum $L^{*}\equiv L/(M\mu)$.
The left plot corresponds to the ratio $\epsilon=0.1$
while the right one is for $\epsilon=-0.1$.
For $\epsilon=0.1$ there is a minimum, $V_{min}\approx0.81$, at $r_{min}\approx2.52M$  for $L^{*}=0$,
and a minimum, $V_{min}\approx0.96$, at $r_{min}\approx29M$  for $L^{*}=5$.
For $\epsilon=-0.1$ the minimum, $V_{min}\approx0.67$, is located at $r_{min}\approx2.02M$  for $L^{*}=0$, and
at $r_{min}\approx20.8M$ with $V_{min}\approx0.97$ for $L^{*}=5$.
} }
\label{Plot15501}
\end{figure}

The effective potential $V_{\pm}$ given in Eq.\il({9})
in the case of naked singularities $(M^2<Q^2)$
is plotted in Figs.\il(\ref{NKS}--\ref{Plot15501}) in terms of the radial coordinate $r/M$
for selected values of the ratio $\epsilon$ and the angular momentum
$L/(M \mu)$ of the test particle, see also \cite{dfl,062,Cohen:1979zzb,primorg,Patil:2010nt,Patil:2011aw,Pradhan:2010ws,F1}.
The effective potential profile strongly depends on the sign of $\epsilon Q$. Moreover,
the cases with  $|\epsilon|\leq1$ and with $|\epsilon|>1$  must be explored separately.

Fig. \il\ref{11:00M} shows the effective potential for a particle of charge--to--mass $\epsilon$ in the range $[-10,-1]$.
The presence of minima (stable circular orbits) in the effective potential with negative energy states is evident.
Moreover, we note that the minimum  of each potential  decreases as $|\epsilon|$ increases.
This fact is due to the attractive and repulsive  effects of the gravitational and electric forces \cite{Joshi,Belinski:2008zz,Luongo:2010we,Pizzi:2008ti,Belinski:2008bn,Pizzi:2008zz,Paolino:2008qi,Manko:2007hi,Alekseev:2007re,Preti:2008zz}
.
In Fig. \il\ref{NKSepsilon} the
effective potential is plotted for negative and positive  values of the charge--to--mass  ratio $\epsilon$.
We see that for a fixed value of the radial coordinate and the angular momentum of the particle, the value of the potential $V$ increases as the value
of $\epsilon$ increases.
In the Fig. \il\ref{Plot15501}  we plot the effective potential for a fixed $Q/M$ as function  of the radial coordinate and the angular momentum
for two different cases, $\epsilon=0.1$ and $\epsilon=-0.1$.  We can see that in the first case the presence of a repulsive Coulomb force
reduces the value of the radius of the last stable circular orbit for a fixed angular momentum.
We note the existence of stable ``circular" orbits with $L=0$ at which the particle is at rest with respect to
static observers located at infinity.

Negative energy states are possible only in the case $\epsilon Q<0$.
The region in which the
solution $V_{+}$ has negative energy states is
%
%
\begin{equation}\label{assurdonk!}
 0<r<M+\sqrt{M^2-Q^2 \left(1-\epsilon^2\right)} \quad \mbox{for}\quad  \epsilon \leq-1 \ ,
\end{equation}
and
\begin{equation}\label{assurdonk!w}
 0<r<r_l^+ \quad \mbox{for} \quad 0\leq L<L_q\ , \quad  \epsilon \leq-1 \ ,
\end{equation}
\begin{equation}\label{assurdonk!3}
  r_l^-<r<r_l^+ \quad \mbox{for}\quad 0\leq L<L_q\ ,  \quad  -1<\epsilon \leq-\sqrt{1-\frac{M^2}{Q^2}} \ ,
\end{equation}
where
\be
\frac{L_q}{\mu}\equiv{r} \sqrt{\frac{\epsilon^2 Q^2 }{r^2-2Mr +Q^2 } -1}\ .
\ee

In general, for a particle in circular motion with radius $r_0$ and charge--to--mass ratio $\epsilon$, around
a RN naked singularity with charge $Q$ and mass $M$,
the corresponding angular momentum must be chosen as
\be
\frac{L^2}{\mu^2 }<{r_0^2} \left(\frac{\epsilon^2 Q^2 }{r_0^2-2Mr_0 +Q^2 } -1\right),
\ee
in order for negative energy states to exist.

The conditions for circular motion around a RN naked singularity
are determined by Eq.\il(\ref{fg2}) which can be used to find
the energy and angular momentum of the test particle.
Indeed, Eqs.\il(\ref{ECHEEUNAL}) and (\ref{Lagesp}) define the
angular momentum $L^\pm$  and the energy $E^{\pm}$, respectively,
in terms of $r/M$, $Q/M$, and $\epsilon$. The explicit dependence
of these parameters makes it necessary to investigate several
intervals of values. To this end, it is useful to introduce  the following
notation
\begin{eqnarray}
r^{\pm}_l&\equiv&\frac{3 M}{2}\pm\frac{1}{2} \sqrt{9 M^{2}-8 Q^{2}-Q^{2}
\epsilon ^{2}}\ ,\\
\tilde{\epsilon}_{\pm}&\equiv&
\frac{1}{\sqrt{2}Q}\sqrt{5M^{2}\pm4Q^{2}+\sqrt{25M^{2}-24Q^{2}}}\ ,
\end{eqnarray}
and
%
\bea\label{r1novo}
\tilde{\tilde{\epsilon}}_{\pm}&\equiv& \frac{1}{\sqrt{2}Q} \sqrt{ 3M^2-2
Q^2\pm M\sqrt{9M^2-8 Q^2}} \ .
\eea
We note that
\be
\lim_{\epsilon\rightarrow0}r_{s}^\pm=r_{*}=\frac{Q^2}{M}\ ,
\ee
which corresponds to the classical radius of a mass $M$ with charge $Q$, see for example \cite{Qadir:2006va,Belgiorno:2000gq},
and
\be
\lim_{\epsilon\rightarrow0}r_{l}^{\pm}=r_{\gamma}^{\pm}=\frac{3 M}{2}\pm\frac{1}{2} \sqrt{9 M^{2}-8 Q^{2}} \ ,
\ee
which represents the limiting radius at which neutral particles can be in circular motion around a RN naked singularity
\cite{Pugliese:2010ps}.

\begin{figure}
\centering
\begin{tabular}{cc}
\includegraphics[scale=1]{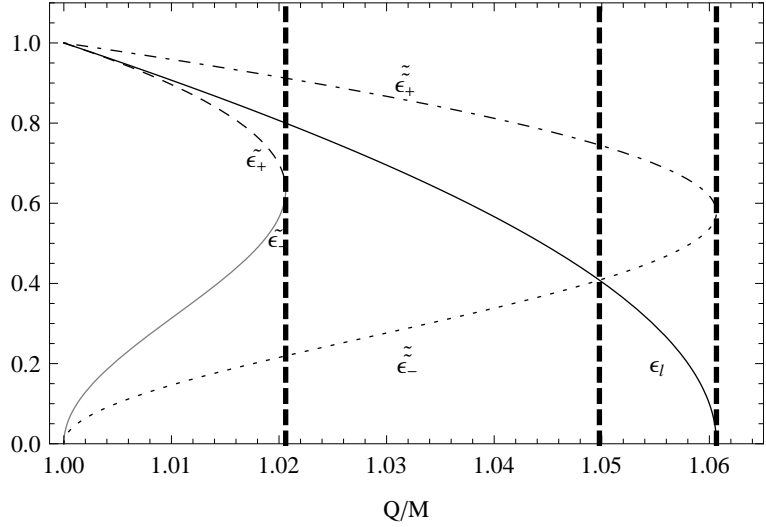}
\end{tabular}
\caption[font={footnotesize,it}]{\footnotesize{The charge parameters
$\epsilon_{l}$ (black solid curve), ${\tilde{\epsilon}}_{-}$ (gray solid curve),
${\tilde{\epsilon}}_{+}$ (dashed curve), $\tilde{ \tilde{\epsilon}}_{-}$ (dotted curve), and $\tilde{\tilde{\epsilon}}_{+}$ (dotdashed curve)
as functions of the charge--to--mass ratio of the RN naked singularity. The special lines  $Q/M=5/\left(2\sqrt{6}\right)\approx 1.02$,
$Q/M=3\sqrt{6}/7\approx 1.05$, and $Q/M=\sqrt{9/8}\approx 1.06$ are also plotted.}}
 \label{Plotettmp}
\end{figure}

The behavior of the charge parameters defined above is depicted in
Fig. \il\ref{Plotettmp} in terms of the ratio $Q/M>1$.
It follows from Fig. \il\ref{Plotettmp} that it is necessary to consider the following intervals:
\bea
Q/M&\in&(1,5/(2\sqrt{6})],\\
Q/M&\in&( 5/(2\sqrt{6}),(3\sqrt{6})/7],\\
Q/M&\in&( (3\sqrt{6})/7,\sqrt{9/8}],\\
Q/M&\in&[\sqrt{9/8}, \infty).
\eea
Our approach consists in analyzing the conditions for the existence of circular orbits by using the expressions for the angular momentum,
Eq.\il(\ref{ECHEEUNAL}), and the energy,  Eq.\il(\ref{Lagesp}), of the particle together with the expressions for the velocity obtained
in Sec.  \ref{BHBHTR}.
We consider separately the case $\epsilon>0$ in Secs.\il\ref{eme1} and \ref{eme1a},
and $\epsilon<0$ in Secs. \ref{bicioc} and \ref{bicioca}.
In the Appendix \ref{noten} we present equivalent results by using the alternative
method of the proper linear velocity of test particles in an orthonormal frame as formulated in Sec.  \ref{BHBHTR}.
%

\subsection{Case $\epsilon>1$}
\label{eme1}
\begin{figure}
\centering
\begin{tabular}{cc}
\includegraphics[scale=.7]{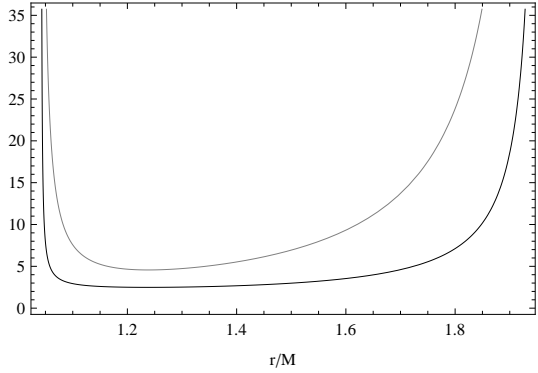}
\end{tabular}
\caption[font={footnotesize,it}]{\footnotesize{The case $\epsilon>1$.
The energy (black curve) and angular momentum (gray curve)  for a  test particle with charge--to--mass ratio $\epsilon=2$ in a RN
naked singularity with   $Q=1.06M$. Circular orbits
exist  in the interval $r_{\gamma}^-<r<r_{\gamma}^+$, where $r_\gamma^-=1.04196M$ and $r_\gamma^+ =1.95804M$.}} \label{pelHa}
\end{figure}

For $\epsilon>0$ the condition
(\ref{14}) implies in general that $r>r_{*}\equiv Q^{2}/M$.
Imposing this constraint on Eqs.(\ref{ECHEEUNAL}) and (\ref{Lagesp}),
we obtain the following results for timelike orbits.
For $\epsilon>1$ and  $M<Q<\sqrt{9/8}M$ circular orbits
exist with angular momentum $L=L^{+}$  in the interval $r_\gamma^-<r<r_\gamma^+ $,
 while for $Q\geq\sqrt{9/8}M$ no circular orbits exist (see Fig. \il\ref{pelHa}).
Clearly, the energy and angular momentum of circular orbits diverge as $r$ approaches the limiting orbits at $r_\gamma^\pm $.
This means that charged test particles located in   the region $r_\gamma^-<r<r_\gamma^+ $
need to acquire an infinite amount of energy to reach the orbits at $r_\gamma^\pm $.
The energy of the states is always positive. A hypothetical accretion disk would consist in this case
of a charged ring of inner radius $r_\gamma^-$ and outer radius $r_\gamma^+$, surrounded by a disk of neutral particles.
The boundary $r=r_\gamma^+$ in this case would be a lightlike hypersurface.

Since for $\epsilon Q>0$ the Coulomb interaction is repulsive, the situation characterized by the values
 for $Q\geq\sqrt{9/8}M$  and  $\epsilon>1$ corresponds to a repulsive electromagnetic effect that cannot be  balanced
 by the attractive gravitational interaction.
We note that  the case  $Q\geq\sqrt{9/8}M$ and  $\epsilon>1$ could be associated  to the realistic configuration of a positive ion  or a positron
in the background of a RN naked singularity.
%
\subsection{Case $0<\epsilon<1$}
\label{eme1a}
It turns out that in this case it is necessary to consider separately each of the four different regions for the ratio $Q/M$
that follow from Fig. \il\ref{Plotettmp}.
Moreover, in each region of $Q/M$ it is also necessary to consider the value of $\epsilon$ for each
of the zones determined by the charge parameters $\epsilon_l$, $\tilde{\epsilon}_\pm$, and $\tilde{\tilde{\epsilon}}_\pm$, as shown
in Fig. \il\ref{Plotettmp}. We analyzed  all the resulting cases in detail and found the values of the energy and angular momentum
of charged test particles in all the intervals where circular motion is
allowed.  We summarize the results as follows.

There is always a minimum radius $r_{min}$ at which circular motion is allowed.
We found that either $r_{min}=r_{s}^+$ or $r_{min}=r_\gamma^-$. Usually, at the radius $r_{s}^+$ the test particle acquires a zero angular
momentum so that a static observer at infinity would consider the particle as being at rest. Furthermore, at the radius $r_\gamma^-$ the energy
 of the test
particle diverges, indicating that the hypersurface $r=r_\gamma^-$ is lightlike. In the simplest case, circular orbits are allowed
in the  infinite interval $[r_{min},\infty)$ so that, at any given radius greater than $r_{min}$, it is always possible to have
a charged test particle moving on a circular trajectory. Sometimes, inside the infinite interval $[r_{min},\infty)$, there exists
a lightlike hypersurface situated at $r_\gamma^+>r_{min}$.

\begin{figure}
\centering
\begin{tabular}{cc}
\includegraphics[scale=.5]{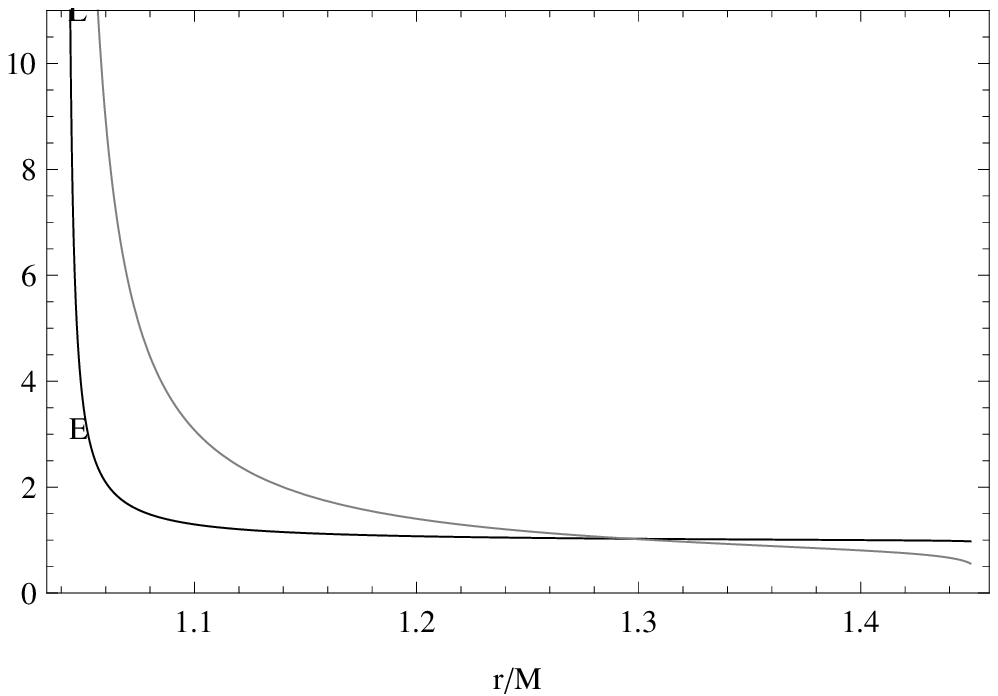}&
\includegraphics[scale=.5]{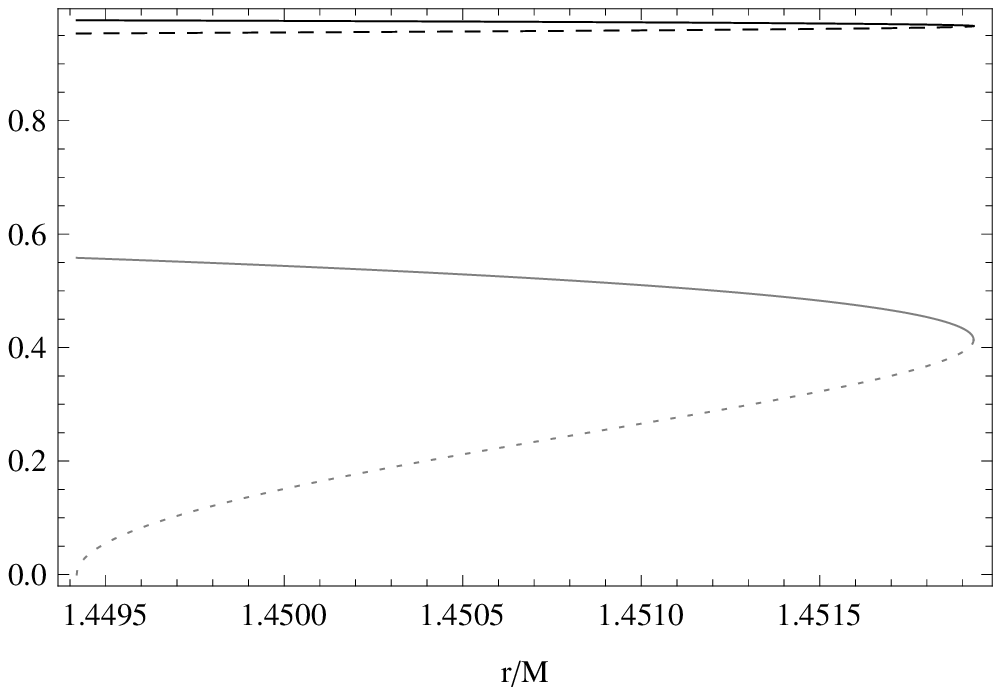}\\
\includegraphics[scale=.5]{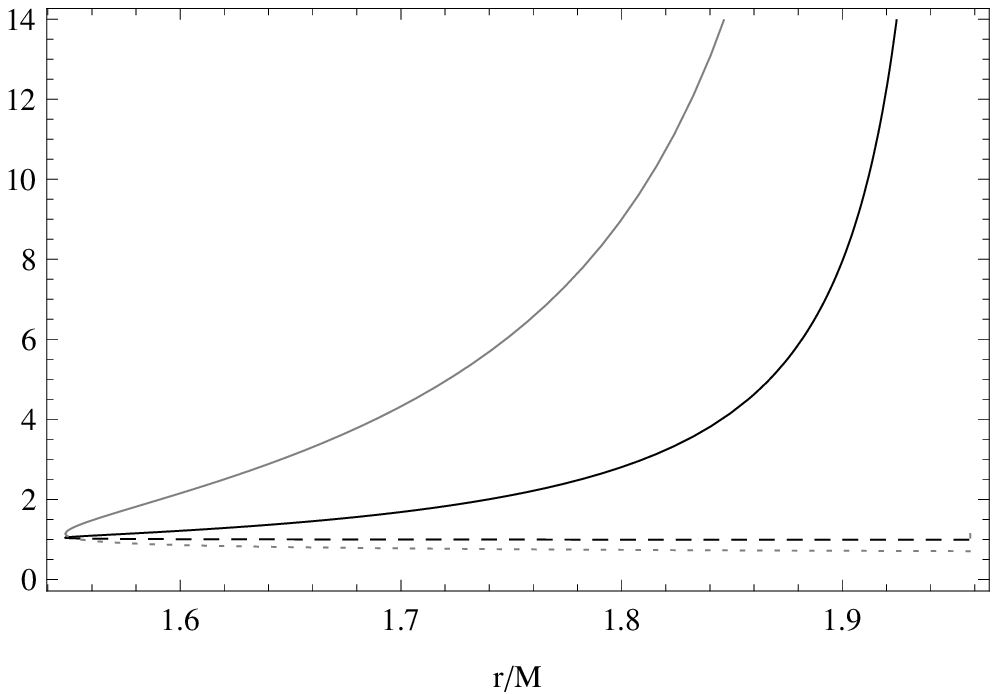}&
\includegraphics[scale=.5]{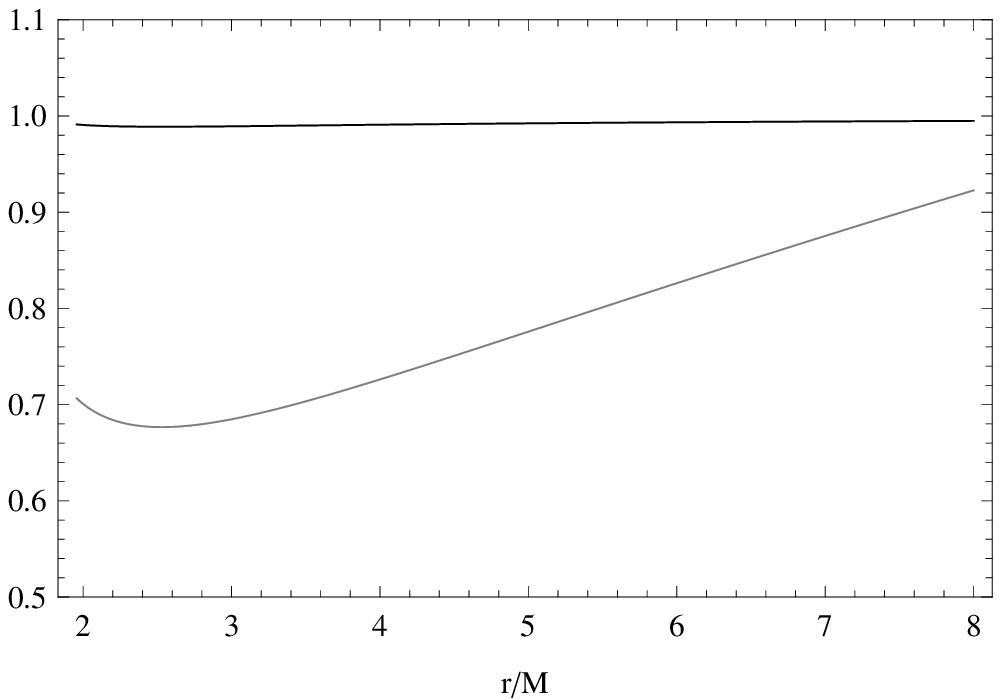}\\
\end{tabular}
\caption[font={footnotesize,it}]{\footnotesize{Case: $M<Q\leq 5/(2\sqrt{6})M$ and  $\tilde{\epsilon}_{+}<\epsilon\leq \epsilon_{l}$.
Parameter choice: $Q= 1.01M$ and $\epsilon =0.902$. Then $\epsilon_{l}=0.907$, $\tilde{\epsilon}_{+}=0.8963$, $r_{s}^+ =1.44942M$, $r_\gamma^-=1.04196M$,
$r_\gamma^+ =1.95804M$, $r_l^- =1.45192M$, and $r_l^+  =1.548077M$.
Circular orbits exist with angular momentum
$L=L^{+}$ (gray curves) and energy  $E=E^{+}$ (black curves) in $r_\gamma^-<r<r_{s}^{+}$ (upper left plot);
$L=L^{\pm}$ in  $r_{s}^{+}\leq r<r_l^-$ (upper right plot) and  $r_l^+ \leq r<r_\gamma^+ $ (bottom left plot);
$L=L^{-}$  in $r\geq r_\gamma^+ $ (bottom right plot).
}}
\label{Mercedes}
\end{figure}

Another possible structure is that of a finite region filled with charged particles within the spatial
interval $(r_{min}=r_\gamma^-,r_{max}=r_\gamma^+)$. This region is usually
surrounded by an empty finite region in which no motion is allowed. Outside the empty region, we find a zone of allowed circular
motion in which either only neutral particles or neutral and charged particles can exist in circular motion.
Clearly, this spatial configuration formed by two separated regions in which circular motion is allowed, could be used
to build with test particles an accretion disk of disconnected rings. A particular example of this case is illustrated in Fig. \ref{Mercedes}

%
\subsection{Case $\epsilon<-1$}
\label{bicioc}
%
The contribution of the electromagnetic interaction in this case is always attractive. Hence,
the only repulsive force to balance the attractive effects of the gravitational and Coulomb interactions can be
generated only by the RN naked singularity. This case therefore can be compared with the neutral test particle motion as studied in \cite{Pugliese:2010ps,Pugliese:2010he}.
Then, it is convenient, as in the case of a neutral test particle, to consider the two regions
 $Q>\sqrt{9/8}M$ and $M<Q\leq\sqrt{9/8}M$ separately.

\begin{figure}
\centering
\begin{tabular}{cc}
\includegraphics[scale=1]{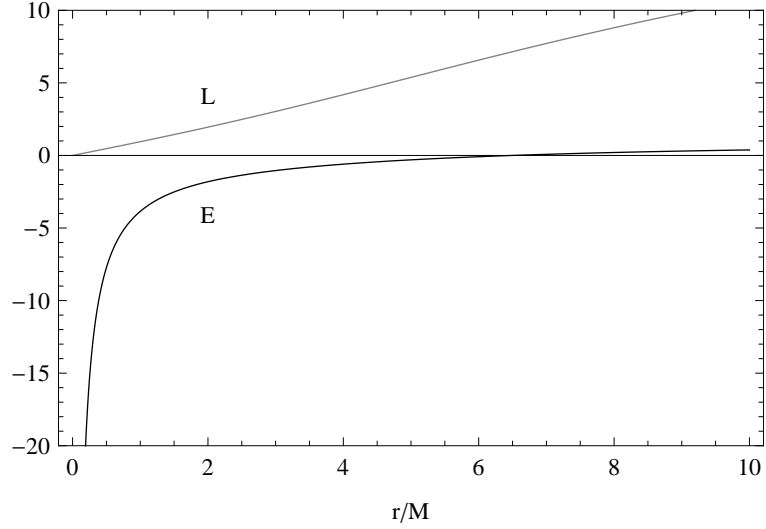}
\end{tabular}
\caption[font={footnotesize,it}]{\footnotesize{Case: $\epsilon<-1$ and  $Q>\sqrt{9/8}M$. Parameter choice: $Q= 2M$ and $\epsilon =-2$.
Circular orbits exist with angular momentum $L=L^{+}$ (gray curve) and energy  $E=E^{+}$ (black curve).
}}
\label{Faria}
\end{figure}

For $\epsilon<-1$ and for $Q>\sqrt{9/8}M$ circular orbits with $L=L^{+}$
always exist for $r>0$ (in fact, however, one has to consider also the limit $r>r_*$ for the existence of timelike trajectories).
This case is illustrated in Fig. \il\ref{Faria} where  the presence of orbits with negative energy states is evident.

\begin{figure}
\centering
\begin{tabular}{cc}
\includegraphics[scale=.7]{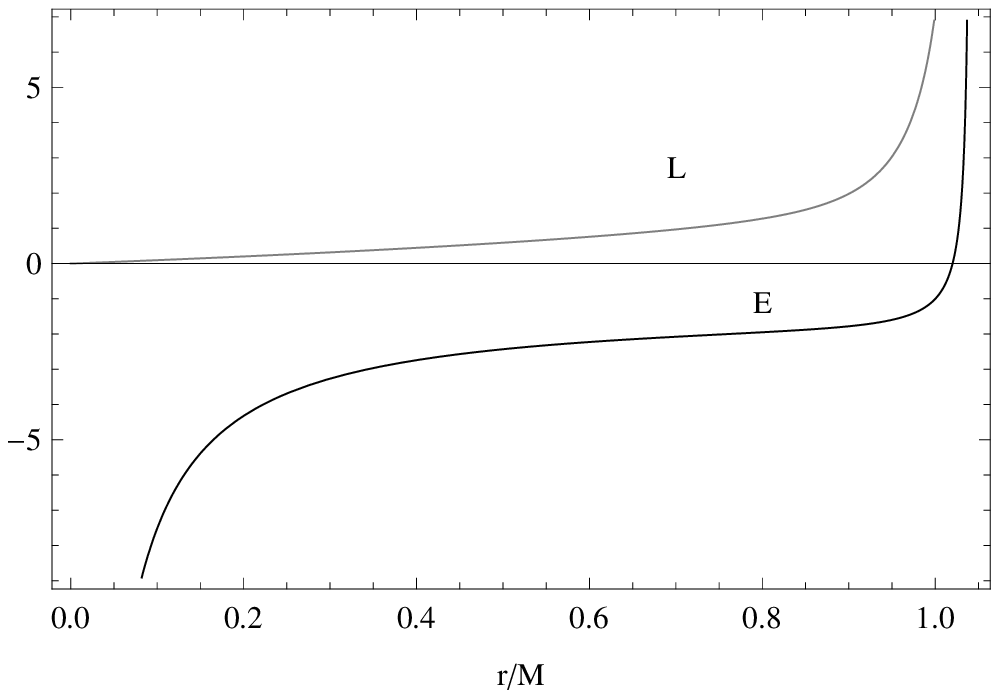}
\includegraphics[scale=.7]{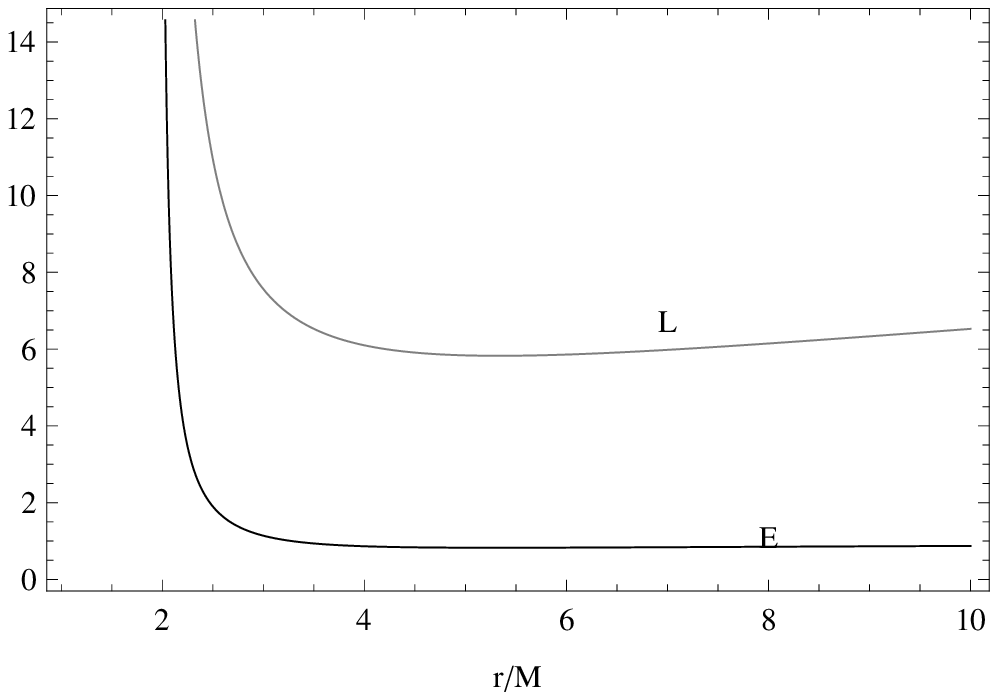}
\end{tabular}
\caption[font={footnotesize,it}]{\footnotesize{Case: $\epsilon<-1$ and   $M<Q\leq\sqrt{9/8}M$.
Parameter choice: $Q= 1.01M$ and $\epsilon =-2$. Then, $r_\gamma^-=1.04196M$ and $r_\gamma^+ =1.95804M$.
Circular orbits exist with angular momentum $L=L^{+}$ (gray curve) and energy  $E=E^{+}$ (black curve) in $0<r<r_\gamma^-$ and $r>r_\gamma^+ $.
}} \label{Antonioc}
\end{figure}

For $M<Q\leq\sqrt{9/8}M$ circular orbits exist with
$L=L^{+}$ in $0<r<r_\gamma^-$ and $r>r_\gamma^+ $ (see Fig. \il\ref{Antonioc}).
We note that for neutral test particles in the region $M<Q\leq\sqrt{9/8}M$, (stable) circular orbits are possible
for $r>r_{*}=Q^2/M$. At $r=r_*$, the angular momentum of the particle vanishes \cite{Pugliese:2010ps}.
On the contrary, charged test particles with $\epsilon<-1$ can move along circular orbits also  in the region $(0,r_{*}]$.
The value of the energy on circular orbits increases  as $r$ approaches  $r=0$.
However, the angular momentum, as seen by an observer located at infinity, decreases as the radius of the orbit decreases.
In  the region $M<Q\leq\sqrt{9/8}M$, two limiting orbits appear at $r_\gamma^\pm $, as in  the  neutral particle case \cite{Pugliese:2010ps}.

\subsection{Case $-1<\epsilon<0$}
\label{bicioca}

For this range of the ratio $\epsilon$, it is also convenient to analyze separately the two cases  $Q>\sqrt{9/8}M$ and $M<Q\leq\sqrt{9/8}M$.
In each case it is necessary to analyze the explicit value of $\epsilon$ with respect to the ratio $M/Q$.
Several cases arise in which we must find the regions where circular motion is allowed and the value of the angular momentum and energy
of the rotating charged test particles.

We summarize the results in the following manner. There are two different configurations for the regions in which circular motion
of charged test particles is allowed. The first one arises in the case $Q>\sqrt{9/8}M$, and consists in
a continuous region that extends from a minimum radius $r_{min}$ to infinity, in principle.
The explicit value of the minimum radius depends on the value of $\epsilon$ and can be either $r_s^-$, $r_s^+$, or
$r_{min}=Q^2/(2M)$. In general, we find that particles standing on the minimum radius are characterized by $L=0$, i. e., they are
static with respect to a non-rotating observer located at infinity.

\begin{figure}
\centering
\begin{tabular}{cc}
\includegraphics[scale=.7]{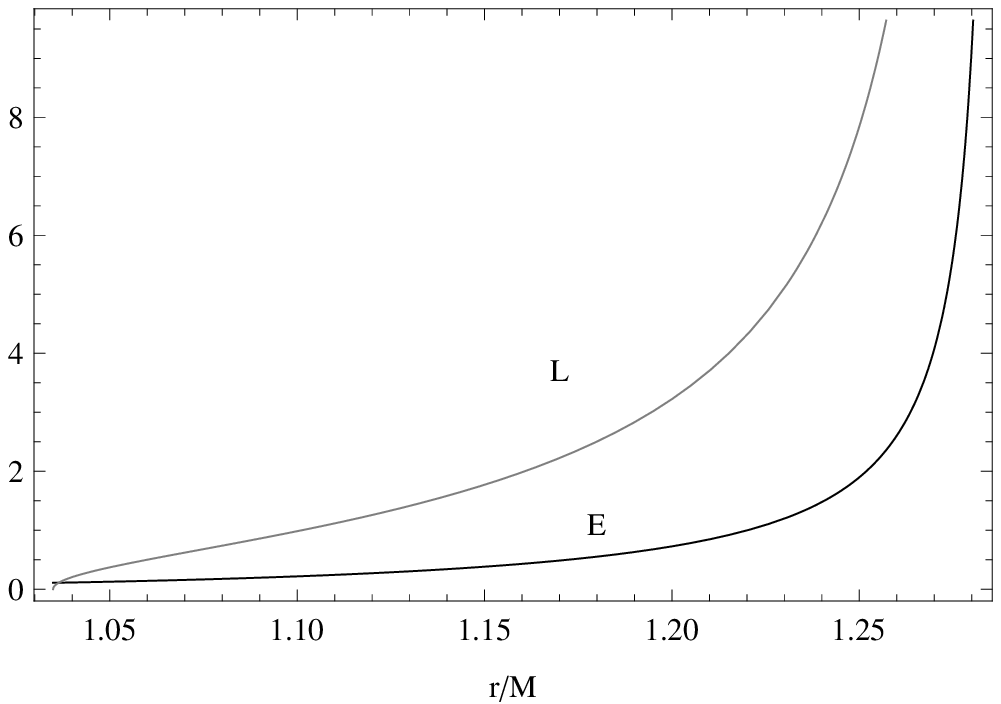}
\includegraphics[scale=.7]{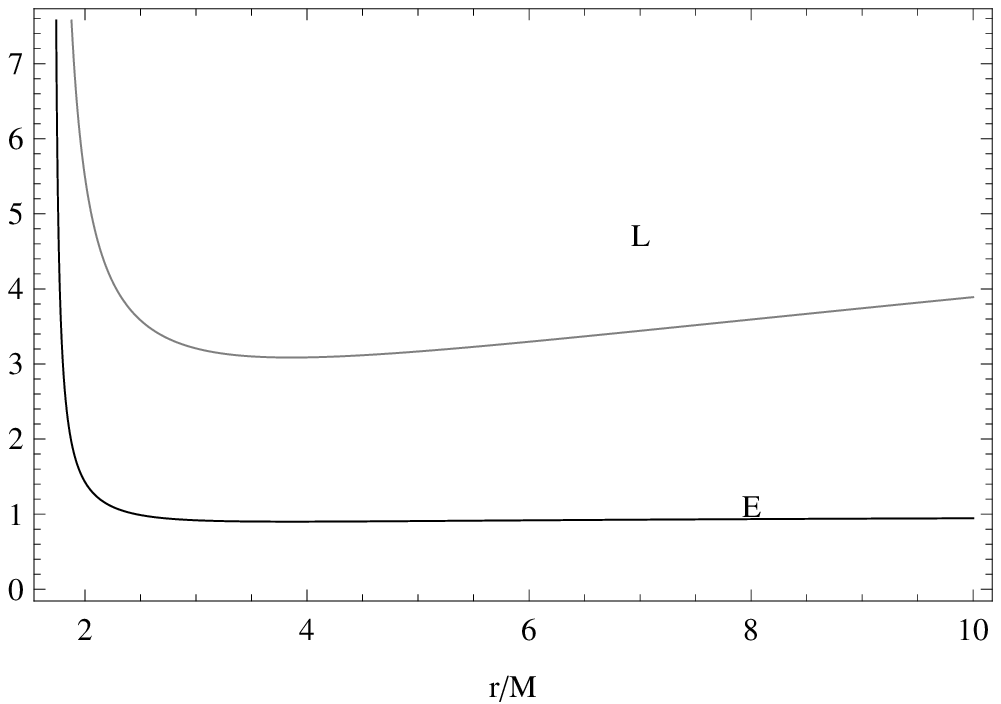}
\end{tabular}
\caption[font={footnotesize,it}]{\footnotesize{Case: $M<Q\leq\sqrt{9/8}M$ and  $-M/Q<\epsilon<0$.
Parameter choice: $Q= 1.05M$ and $\epsilon=-0.2$.
Then   $r_\gamma^-=1.28787M$, $r_\gamma^+ =1.71213M$, and $r_{s}^-=1.03487M$.
Circular orbits exist with angular momentum
$L=L^{+}$ (gray curve) and energy  $E=E^{+}$ (black curve) in $r_{s}^- <r<r_\gamma^-$   (left plot) and in $r>r_\gamma^+ $ (right  plot).
For $r=r_{s}^-$, $L=0$.
}}
\label{Grotta}
\end{figure}

The second configuration appears for  $M<Q\leq\sqrt{9/8}M$. It also extends from $r_{min}$ to infinity, but inside it there is
a forbidden region delimited by the radii $r_\gamma^-$ and $r_\gamma^+$. The configuration is therefore composed of two
disconnected regions. At the minimum radius, test particles are characterized by $L=0$. On the boundaries ($r_\gamma^\pm$)
of the interior forbidden region only photons can stand on circular orbits. A particular example of this case is presented in
Fig. \ref{Grotta}.

%

\subsection{Stability}

\begin{figure}
\centering
\begin{tabular}{cc}
\includegraphics[scale=1]{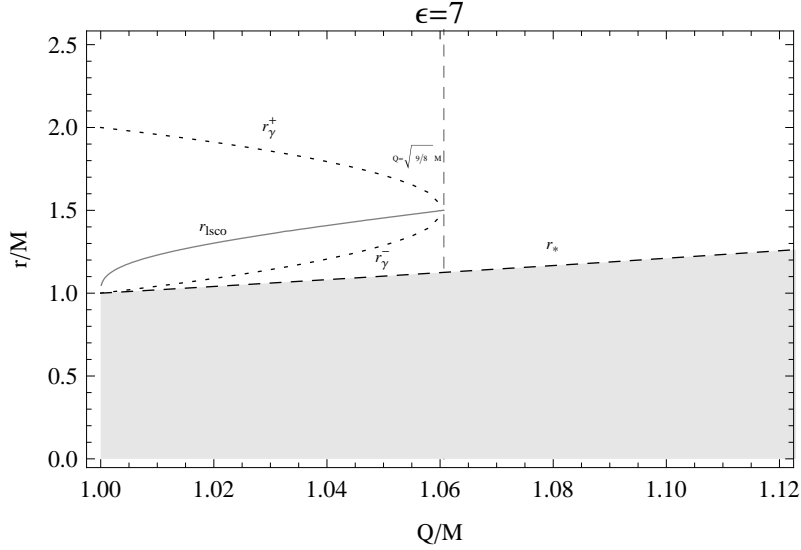}
\end{tabular}
\caption[font={footnotesize,it}]{\footnotesize{The radius of the last stable circular orbit
$r_{\ti{lsco}}$ (gray curve) of a charged particle with ratio $\epsilon=+7$ in a RN
naked singularity with ratio $Q/M \in [1,1.2].$
The radii $r_*=Q^2/M$ and $r=r_\gamma^{\pm}\equiv[3M\pm\sqrt{9M^2-8Q^2}]/2$ are also plotted.
Circular orbits exist only in the interval $1<Q/M<\sqrt{9/8}$.
The shaded region is forbidden for timelike particles.
Stable orbits are located in the region $r>r_{\ti{lsco}}$.}}
\label{PlotVns11}
\end{figure}

To explore the stability properties of the circular motion of charged test
particles in a RN naked singularity, it is necessary to
investigate the equation (\ref{Lagespep}) or, equivalently,
Eqs.(\ref{nube}), (\ref{nuvola}), and (\ref{Carmara}),
considering the different values for $\epsilon$ and $Q/M>1$.
We can distinguish  two different cases, $|\epsilon|>1$ and  $0<|\epsilon|<1$.
Let us consider the case $|\epsilon|>1$.
In particular, as it was shown in Sec.  \ref{eme1}, for $\epsilon>1$ and  $M<Q<\sqrt{9/8}M$ circular orbits
exist with $L=L^{+}$  in the interval  $r_\gamma^-<r<r_\gamma^+ $ whereas no circular orbits exist for
$\epsilon>1$ and  $Q>\sqrt{9/8}M$. For this particular case,  a numerical analysis of
condition (\ref{Lagespep}) leads to the conclusion that a circular orbit is stable only if its radius $r_0$
satisfies the condition $r_0> r_{\ti{lsco}}$, where $r_{\ti{lsco}}$ is depicted in Fig. \il\ref{PlotVns11}.
We see that in general the radius of the last stable circular orbit is located inside the interval $(r_\gamma^-,r_\gamma^+)$.
It then follows that the only stable region is determined by the  interval $r_{\ti{lsco}}<r<r_\gamma^+$.

\begin{figure}
\centering
\begin{tabular}{cc}
\includegraphics[scale=.71]{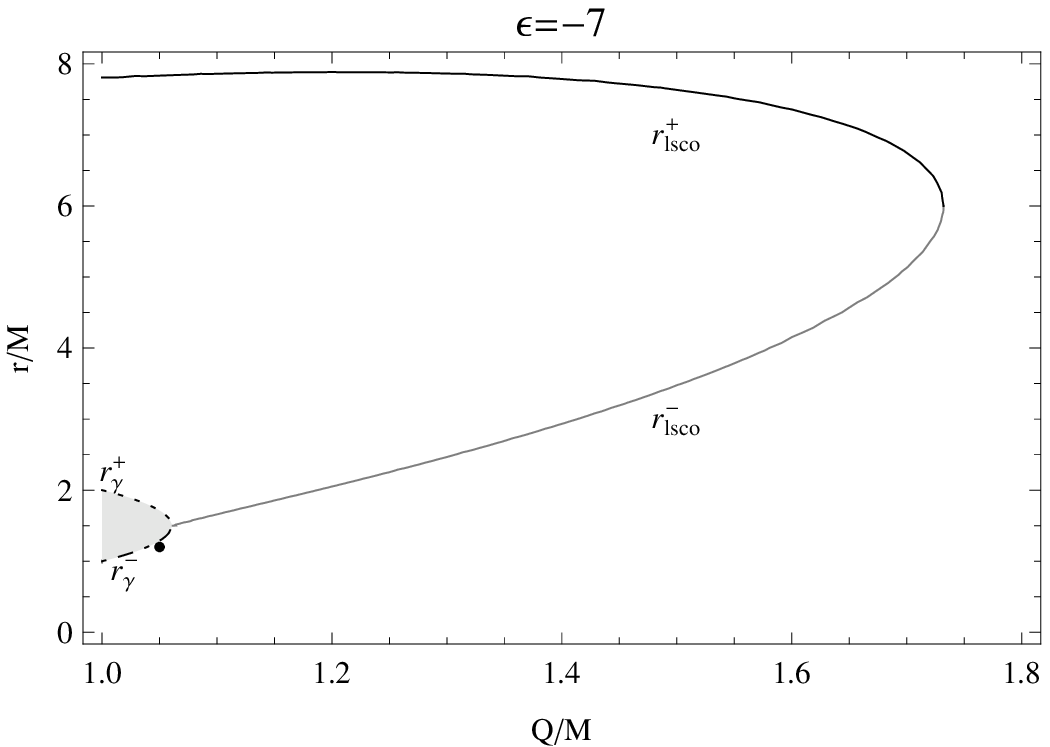}
\includegraphics[scale=.71]{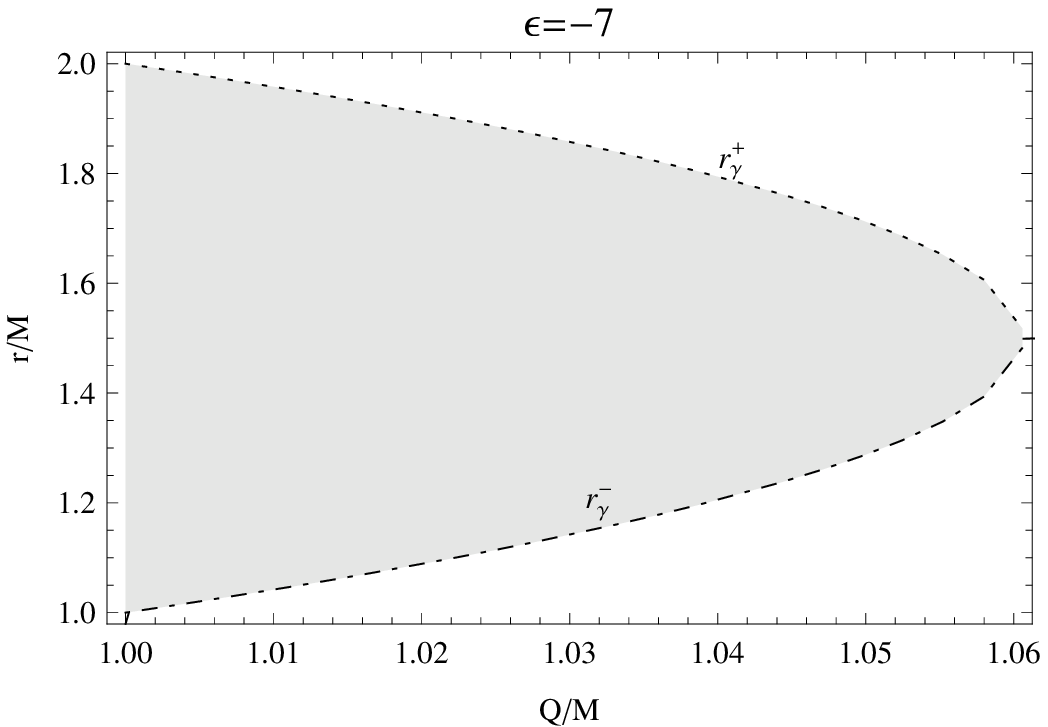}
\end{tabular}
\caption[font={footnotesize,it}]{\footnotesize{
The radius of the last stable circular orbit
$r_{\ti{lsco}}^\pm$ (black curves) of a charged particle with ratio $\epsilon=-7$ in a RN
naked singularity with ratio $Q/M \in [1,1.8].$ The radii $r_*$, and $r_\gamma^\pm$ are also
plotted for comparison. In the shaded region no circular orbits can exist. Stable circular orbits are situated outside the region with boundaries
$r_{\ti{lsco}}^+$, $r_{\ti{lsco}}^-$, and the vertical axis $Q/M=1$.
}}
\label{NSm7}
\end{figure}

Consider now the case $\epsilon<-1$. The numerical investigation of the condition (\ref{Lagespep}) for the last
stable circular orbit shows that in this case there are two solutions $r_{\ti{lsco}}^\pm$ such that
$r_{\ti{lsco}}^-\leq r_{\ti{lsco}}^+$, where the equality is valid for $Q/M\approx 1.72$. Moreover, for $Q/M=\sqrt{9/8}$
we obtain that $r_{\ti{lsco}}^-=r_\gamma^- = r_\gamma^+$.
This situation is illustrated in Fig. \il\ref{NSm7}.
Stable orbits corresponds to points located outside the region delimited by the curves $r=r_{\ti{lsco}}^+$, $r=r_{\ti{lsco}}^-$,
and the axis $Q/M=1$.
On the other hand, we found in Sec.  \ref{bicioc} that for $\epsilon<-1$ and $1<Q/M\leq\sqrt{9/8}$ circular orbits exist
in the interval $0<r<r_\gamma^-$  and $r>r_\gamma^+ $. It then follows that the region of stability corresponds in this case
to two disconnected zones determined by $0<r<r_\gamma^-$ and $r>r_{\ti{lsco}}^+$. Moreover, we established in
 Sec.  \ref{bicioc} that for $\epsilon<-1$ and $\sqrt{9/8}<Q/M$  circular
orbits always exist for $r>0$. Consequently, in the interval $\sqrt{9/8}<Q/M\lesssim 1.72$, the stable circular orbits are located
in the two disconnected regions
defined by $0<r<r_{\ti{lsco}}^-$  and $r>r_{\ti{lsco}}^+$. Finally, for $Q/M\gtrsim 1.72$ all the circular orbits are stable
(see Fig. \il\ref{NSm7}).

\begin{figure}
\centering
\begin{tabular}{cc}
\includegraphics[scale=1]{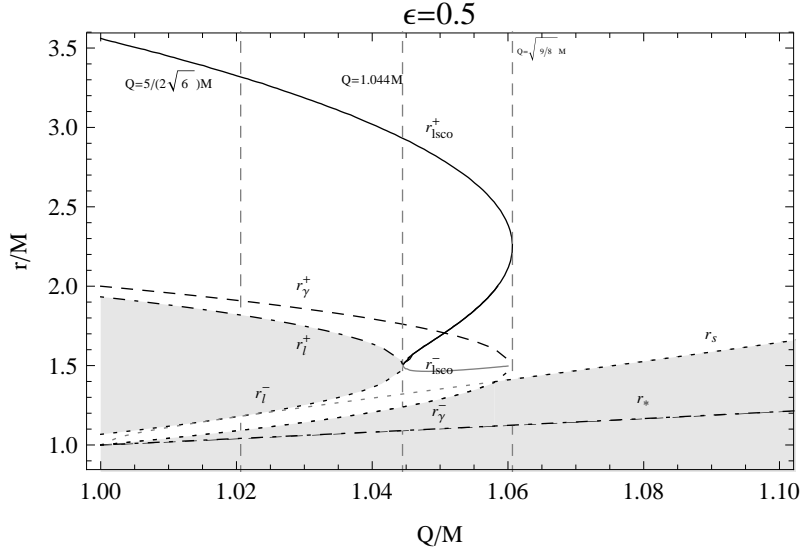}
\end{tabular}
\caption[font={footnotesize,it}]{\footnotesize{
The radius of the last stable circular orbit
$r_{\ti{lsco}}^\pm$ (black and grey curves) of a charged particle with ratio $\epsilon=0.5$ in a RN
naked singularity with ratio $Q/M \in [1,1.1].$
Also plotted: $r_{\gamma}^{\pm}\equiv[3M\pm\sqrt{(9M^2-8Q^2)}]/2$,
$r_{s}^{+}\equiv\frac{Q^{2}}{\epsilon^{2}Q^{2}-M^{2}}\left[\sqrt{\epsilon^2(\epsilon^{2}-1)(M^{2}-Q^{2})} M(\epsilon^{2}-1)\right]$,
$r^{\pm}_l \equiv\frac{3 M}{2}\pm\frac{1}{2}\sqrt{9 M^{2}-8 Q^{2}-Q^{2} \epsilon ^{2}} $, and
$r_{*}=Q^2/M$.
Regions of stability are: for $Q>\sqrt{9/8}M$ in $r>r_{s}$, for  $(3 \sqrt{6}/7)M<Q<\sqrt{9/8}M$ exist
stable orbits in $r_{\gamma}^-<r$, for
$(5/(2\sqrt{6}))M<Q<(3 \sqrt{6}/7)M$ exist stable orbits in
$r_{\gamma}^-<r$.
For $M<Q<(5/(2\sqrt{6}))M$  stable orbits are located in
$r>r^+_{lsco}$.}}
\label{NSp05}
\end{figure}

\begin{figure}
\centering
\begin{tabular}{c}
\includegraphics[scale=1]{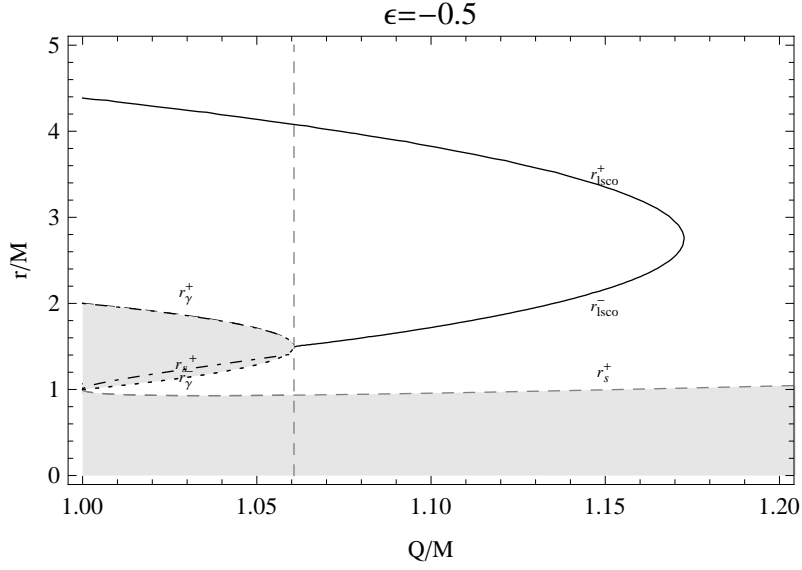}
\end{tabular}
\caption[font={footnotesize,it}]{\footnotesize{
The radius of the last stable circular orbit
$r_{\ti{lsco}}^\pm$ (black and grey curves) of a charged particle with ratio $\epsilon=-0.5$ in a RN
naked singularity with ratio $Q/M \in [1,1.2].$
Also plotted: $r_{\gamma}^{\pm}\equiv[3M\pm\sqrt{(9M^2-8Q^2)}]/2$,
$r_{s}^{+}\equiv\frac{Q^{2}}{\epsilon^{2}Q^{2}-M^{2}}\left[\sqrt{\epsilon^2(\epsilon^{2}-1)(M^{2}-Q^{2})} M(\epsilon^{2}-1)\right]$,
$r^{\pm}_l \equiv\frac{3 M}{2}\pm\frac{1}{2}\sqrt{9 M^{2}-8 Q^{2}-Q^{2} \epsilon ^{2}} $, and
$r_{*}=Q^2/M$.
Shaded regions are forbidden.
Regions of stability are: for
$Q>(\sqrt{9/8})M$ stable circular orbits exist in $r_{s}^+<r<r_{lsco}^-$, and $r>r_{lsco}^+$.  For
$M<Q<(\sqrt{9/8})M$ stable circular orbits exist  in $r_{s}^+<r<r_{\gamma}^-$, and
$r>r_{\gamma}^+$.
}}\label{NSm05}
\end{figure}

The case $0<|\epsilon|<1$  is much more complex, and needs to be described for different subcases following the classification of orbital regions
traced in Sec.  \il\ref{eme1a} for the case $0<\epsilon<1$, and in Sec.  \il\ref{bicioca} for the case $-1<\epsilon<0$.
The results for the specific ratio $\epsilon=0.5$ are given in Fig. \il\ref{NSp05} and for $\epsilon=-0.5$ in Fig. \il\ref{NSm05}.
In general, we find that the results are similar to those obtained for the case $\epsilon<-1$. Indeed, the zone of stability
consists of either one connected region or two disconnected regions. The explicit value of the radii that determine the
boundaries of the stability regions depend on the particular values of the ratio $Q/M$.

\clearpage

\section{Conclusions}
\label{milka}
In this work,  we explored  the motion of charged test particles along circular orbits in the spacetime described by the
Reissner--Nordstr\"om (RN) metric. We performed a very detailed discussion of all the regions of the spacetime where
circular orbits are allowed, using as parameters the charge--to--mass ratio $Q/M$ of the source of gravity and
the charge--to--mass ratio $\epsilon=q/\mu$ of the test particle. Depending on the value of $Q/M$, two major cases
must be considered: The black hole case, $|Q/M|\leq 1$, and the naked singularity case, $|Q/M|> 1$.
Moreover, we found out that the two cases $|\epsilon|\leq 1$ and $|\epsilon|>1$ must also be investigated separately.
Whereas the investigation of the motion of charged test particles with $|\epsilon|>1$ can be carried out in a relatively
simple manner, the case with $|\epsilon|\leq 1$ is much more complex, because it is necessary to consider various
subcases which depend on the explicit value of $\epsilon$ in this interval.

To perform the analysis of circular motion of charged test particles in this gravitational field we use two different methods.
The first one consists in using constants of motion to reduce the equations of motion to a single first--order differential
equation for a particle moving in an effective potential. The properties of this effective potential are then used
to find the conditions under which circular motion is possible. The second approach uses a local orthonormal frame to
introduce a   ``local proper linear velocity" for the test particle. The conditions for this velocity to be timelike
are then used to determine the regions of space where circular orbits are allowed. The results of both methods are
equivalent and, in fact, for the sake of simplicity  it is sometimes convenient to use a combination of both approaches.
In this work, we analyzed in detail the conditions for the existence of circular orbits and found all the solutions for
all the regions of space in the case of black holes and naked singularities.

To formulate the main results of this work in a plausible manner, let us suppose that an accretion disk around a RN gravitational
source can be made of test particles moving along circular orbits \cite{Kovacs:2010xm}. Then, in the case of black hole we find
two different types of accretion disks made of charged test particles. The first type consists of a disk that begins
at a minimum radius $R$ and can extend to infinity, in principle. In the second possible configuration, we find a circular ring
of charged particles with radii $(r_{int},r_{ext})$, surrounded by the disk, i. e., with $r_{ext}<R$.
For certain choices of the parameter $\epsilon$ the exterior disk might be composed only of neutral particles.
A study of the stability of circular orbits shows that the second structure of a ring plus a disk is highly unstable.
This means that test particles in stable circular motion  around RN black holes can be put together to form only
a single disk that can, in principle, extend to infinity.

In the case of RN naked singularities we find the same two types of accretion disks. The explicit
values of the radii $r_{min}$, $r_{ext}$, and $R$ depend on the values of the ratios $\epsilon$ and $Q/M$, and
differ significantly from the case of black holes. In fact, we find that the case of naked singularities offers
a much richer combination of values of the charge--to--mass ratios for which it is possible to find a structure
composed of an interior ring plus an exterior disk. A study of the stability of this specific situation
shows that for certain quite general combinations of the parameters the configuration is stable. This
result implies that test particles in stable circular motion around RN naked singularities can be put together to form
either a single disk that can extend, in principle, to infinity or a configuration of an interior ring with an exterior disk.
This is the main difference between black holes and naked singularities from the viewpoint of these hypothetical
accretion disks made of test particles.

The question arises whether it is possible to generalize these results to the case of more realistic accretion disks
around more general gravitational sources, taking into account, for instance,
the rotation of the central body, \cite{LoraClavijo:2010ih,Vogt:2004qx}.
It seems reasonable to expect that in the case of Kerr and Kerr-Newman naked singularities, regions can be found where
stable circular motion is not allowed so that an accretion disk around such an object would exhibit a discontinuous
structure. Indeed, some preliminary calculations of circular geodesics in the field of rotating compact objects support this
expectation. Thus, we can conjecture that the discontinuities in the accretion disks around naked singularities are
a consequence of the  intensity of the repulsive gravity effects that characterize these speculative objects.
Furthermore, it was recently proposed that static compact objects with quadrupole moment can be interpreted as describing
the exterior gravitational field of naked singularities \cite{quev11a,quev11b}. It would be interesting to test the
above conjecture in this relatively simple case in which rotation is absent. If the conjecture turns out to be true,
it would give us the possibility of distinguishing between black holes and naked singularities
by observing their accretion disks.

\section*{Acknowledgments}
Daniela Pugliese and Hernando Quevedo would like to thank the ICRANet for support. We would like to thank Andrea Geralico for helpful comments and discussions. One of us (DP) gratefully acknowledges financial support from the A. Della Riccia Foundation.
This work was supported in part by DGAPA-UNAM, grant No. IN106110.


\appendix
\section{Velocity of test particles in a RN naked singularity}
\label{noten}
In this Appendix we explore charged test particles in circular motion in a RN naked singularity by using
the tetrad formalism, as developed in Sec.  \ref{BHBHTR} for the black hole analysis.
In Sec.  \ref{NSNSTRE}, we studied the timelike circular motion in the naked singularity case by analyzing directly
the existence conditions for the energy,  Eq.\il(\ref{Lagesp}), and the angular momentum,  Eq.\il(\ref{ECHEEUNAL}).
Here we use the formalism of ``local proper linear velocity" as measured by an observer attached to an orthonormal frame.
The results are equivalent to those obtained by using the expressions for the energy and angular momentum.

In Sec.  \il\ref{BHBHTR}, we showed that the linear velocity of a test particle in a RN spacetime can be written
as
\begin{equation}\label{vel}
\nu_{\epsilon}^{\pm}=\nu_{g} \left[\Lambda\pm\sqrt{\Lambda^{2}-1+(\epsilon/\epsilon_0)^{2}}\right]^{1/2}\ ,
\end{equation}
where
\begin{equation}\label{notvel}
\Lambda=1-\frac{\nu_{g} ^{2}}{2}\left(\frac{\epsilon}{\epsilon_0}\right)^{2}\ , \quad \nu_g= \sqrt{\frac{Mr-Q^2}{\Delta}}\ ,
\quad \epsilon_0 = \frac{Mr-Q^2}{Q\sqrt{\Delta}} \ .
\end{equation}
Then, the conditions for the existence of timelike velocities are
\begin{eqnarray}
\label{cond1}
\Lambda^{2}-1+(\epsilon/\epsilon_{0})^{2} &\geq& 0 \ ,\\
\label{cond2}
\Lambda\pm\sqrt{\Lambda^{2}-1+(\epsilon/\epsilon_{0})^{2}}&\geq& 0 \ ,\\
\label{cond3}
(\nu_{\epsilon}^{\pm})^{2}&<&1\ .
\end{eqnarray}
We first note that, in the case of a naked singularity, these conditions can be satisfied only for
$r\geq Q^{2}/M$.

\begin{figure}
\centering
\begin{tabular}{cc}
\includegraphics[scale=.7]{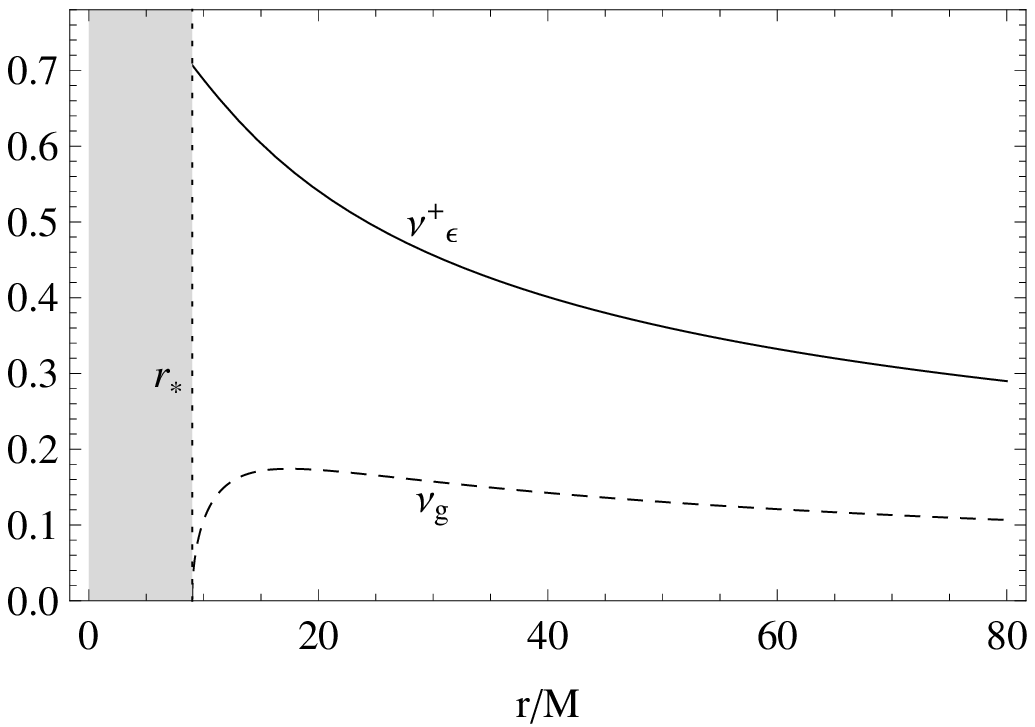}
&
\includegraphics[scale=.7]{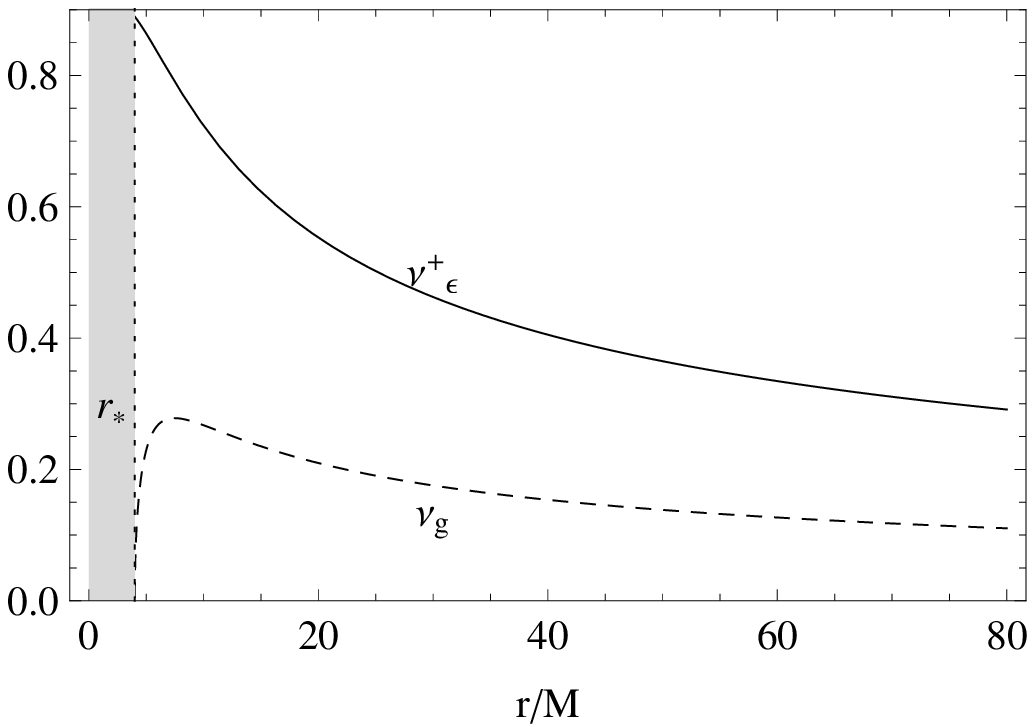}\\(a)&(b)\\
\includegraphics[scale=.7]{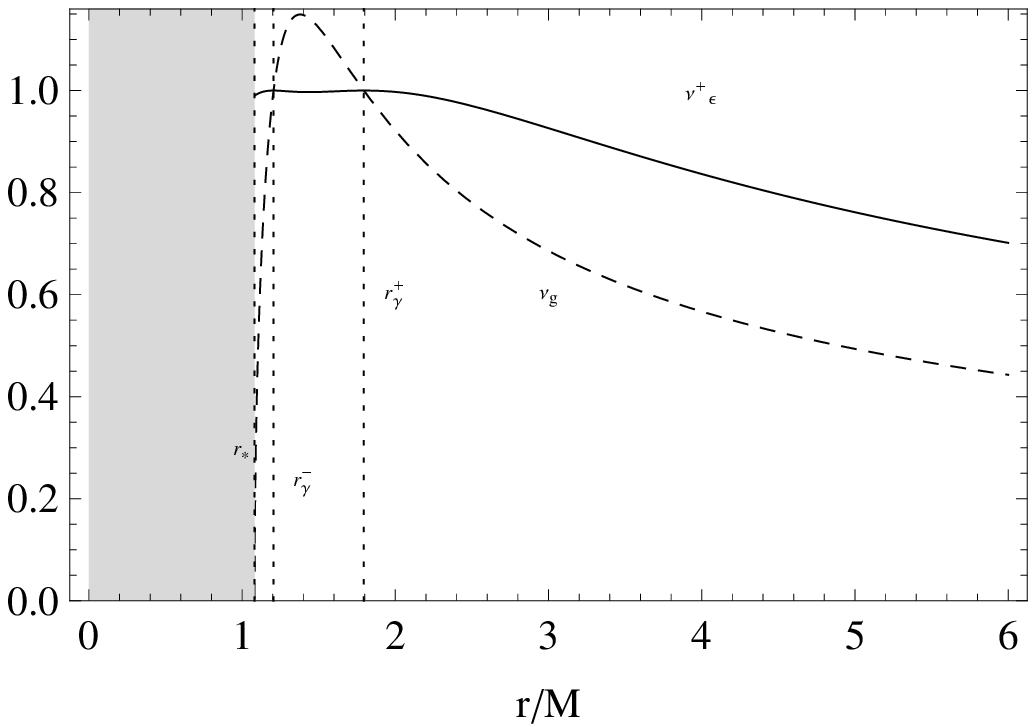}&
\includegraphics[scale=.7]{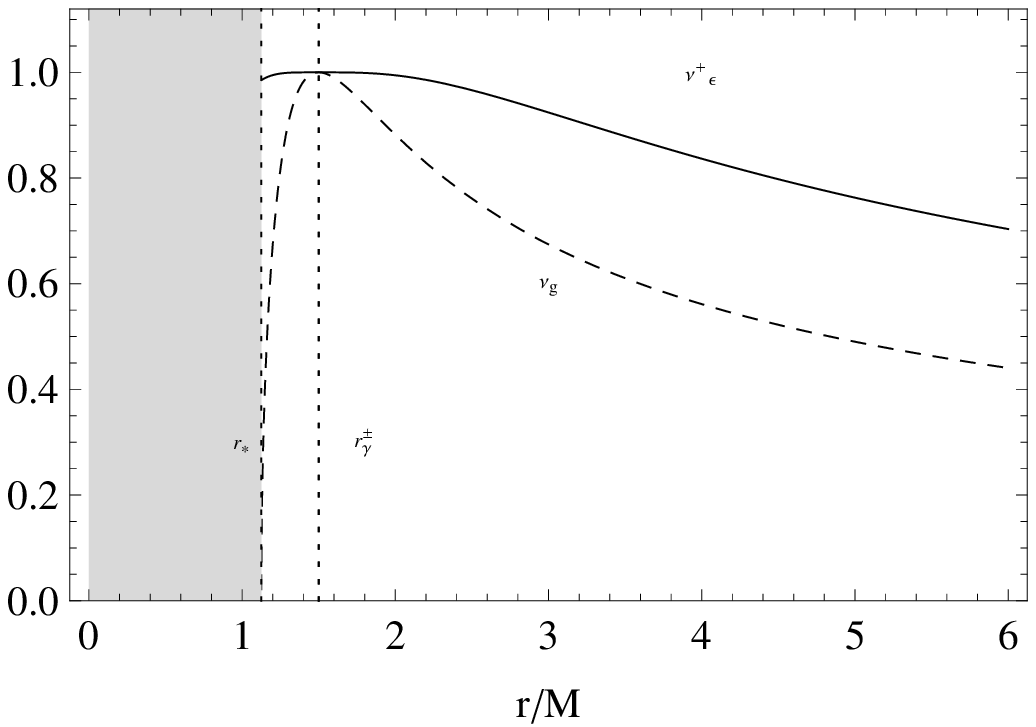}\\(c)&(d)
\end{tabular}
\caption[font={footnotesize,it}]{\footnotesize{The positive solution
of the linear velocity $\nu_{\epsilon}^{+}$ is plotted as a function
of the radial distance $r/M$ for different values of the ratios
$Q/M$ and $\epsilon$. The geodesic velocity $\nu_{g} $ is also shown
(dashed curve). Shaded region is forbidden. In (\emph{a})  the
parameter choice is $Q/M=3$ and $\epsilon=2$, with $r_{*}\equiv
Q^{2}/M=9M$.
In (\emph{b}) the parameter choice is $Q/M=2$ and
$\epsilon=3$, with $r_{*}\equiv Q^{2}/M=4M$. In (\emph{c})  the
parameter choice is $Q/M=1.04$ and $\epsilon=2$, with $r_{*}\equiv
Q^{2}/M\approx1.08M$,
$r_\gamma^{+}\equiv[3M+\sqrt{9M^{2}-8Q^{2}}]/2\approx1.79 M$, and
$r_\gamma^{-} \equiv[3M-\sqrt{9M^{2}-8Q^{2}}]/2\approx1.201 M$. In
(\emph{d}) the parameter choice is $Q/M=\sqrt{9/8}$ and
$\epsilon=2$, with $r_{*}\equiv Q^{2}/M=(9/8)M$,
$r_{\gamma}^{\pm}\equiv[3M\pm\sqrt{9M^{2}-8Q^{2}}]/2=(3/2) M$. }
} \label{Sax11b}
\end{figure}

For $\epsilon>1$  and $\epsilon<-1$ the
solutions are the geodesic velocities $\nu=\pm\nu_{\epsilon}^{+}$. In fact, in this case, condition (\ref{cond2})
with the minus sign is no more satisfied.
On the other hand, conditions (\ref{cond1}), (\ref{cond2}), and (\ref{cond3}) imply
that circular timelike orbits exist for $Q/M>\sqrt{9/8}$ in the entire range $r>Q^{2}/M$.
For $1<Q/M<\sqrt{9/8}$ circular orbits are possible in $r>Q^{2}/M$ and
$r\neq r_\gamma^\pm \equiv[3M\pm\sqrt{9M^{2}-8Q^{2}}]/2$.
Finally, for $Q/M=\sqrt{9/8}$ timelike circular orbits exist for all $r>Q^{2}/M$,
except at $r=(3/2) M$. Moreover, the radii  $r= r_\gamma^\pm $ correspond  to
photon orbits in the RN spacetime (see Fig.  \ref{Sax11b}).

Consider now the  case $|\epsilon|<1$. It is useful to introduce here the following notations:
\begin{eqnarray}
r^{\pm}_l&\equiv&\frac{3 M}{2}\pm\frac{1}{2} \sqrt{9 M^{2}-8 Q^{2}-Q^{2}
\epsilon ^{2}},\\
\tilde{\epsilon}_{\pm}&\equiv&
\frac{1}{\sqrt{2}Q}\sqrt{5M^{2}\pm4Q^{2}+\sqrt{25M^{2}-24Q^{2}}},
\end{eqnarray}
and
\bea\label{r1}
r_{s}^\pm &\equiv&
\frac{Q^{2}}{\epsilon^{2}Q^{2}-M^{2}}
\left[M(\epsilon^{2}-1)\pm \sqrt{\epsilon^2(\epsilon^2-1)(M^{2}-Q^{2})}\right]\ .
\eea

First, consider the case $0<\epsilon<1$. For $\epsilon>0$ condition
(\ref{14}) implies that $r>Q^{2}/M$. Applying this constraint on conditions (\ref{cond1})
and (\ref{cond2}), we obtain the following results for timelike geodesics.

\begin{figure}
\centering
\begin{tabular}{cc}
\includegraphics[scale=.7]{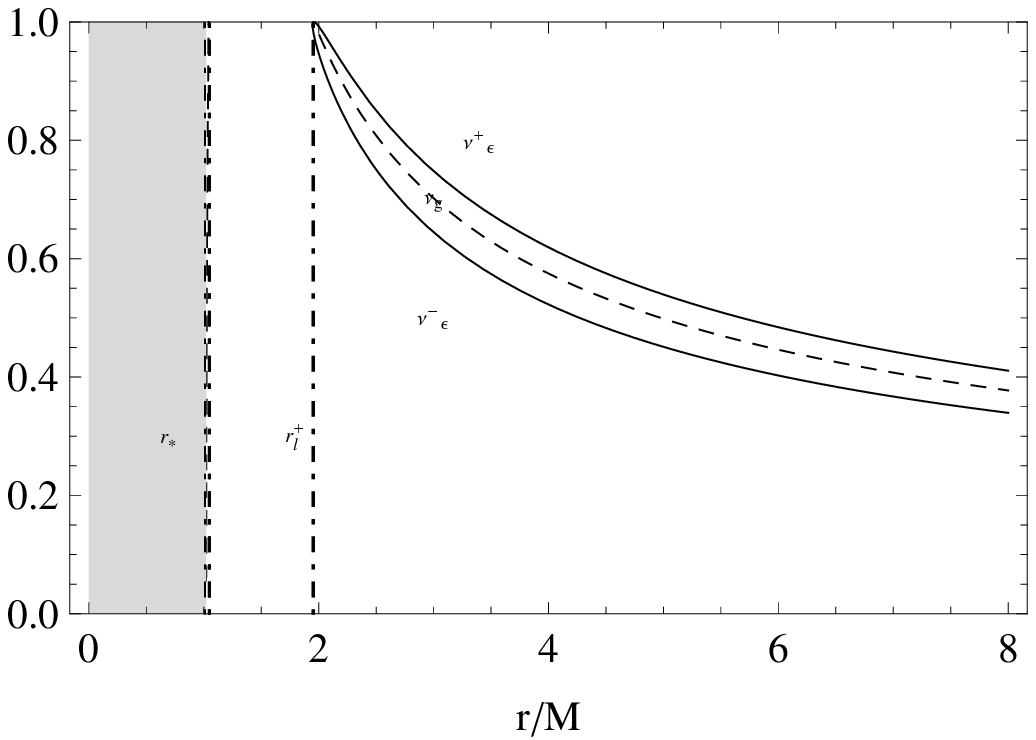}&
\includegraphics[scale=.7]{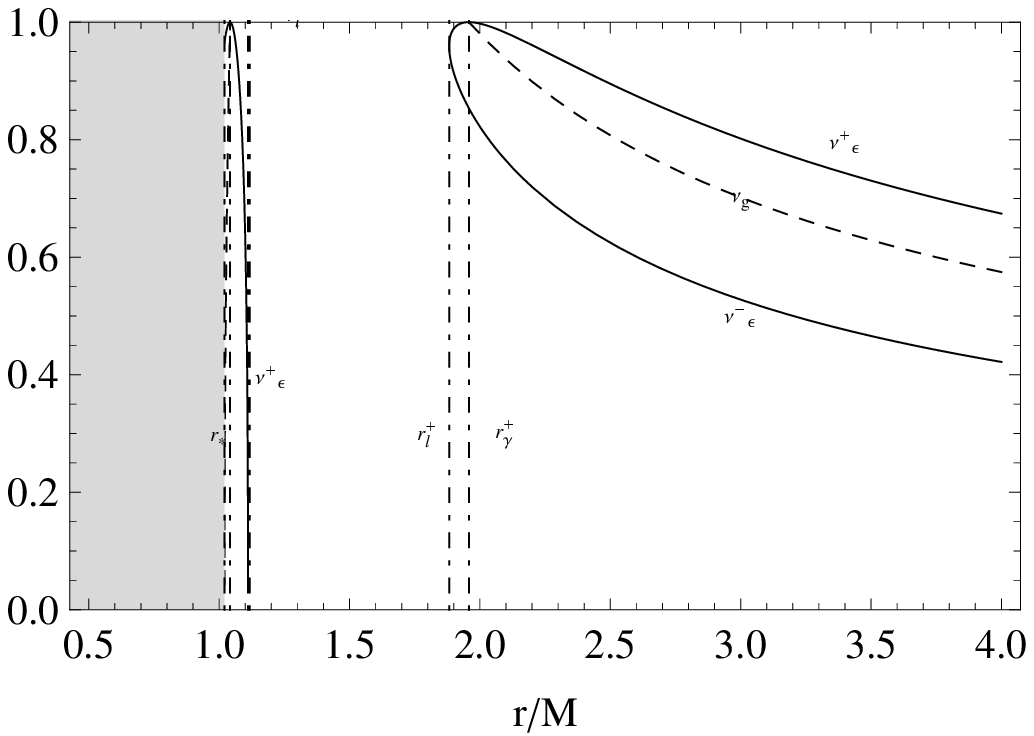}\\(a)&(b)\\
\includegraphics[scale=.7]{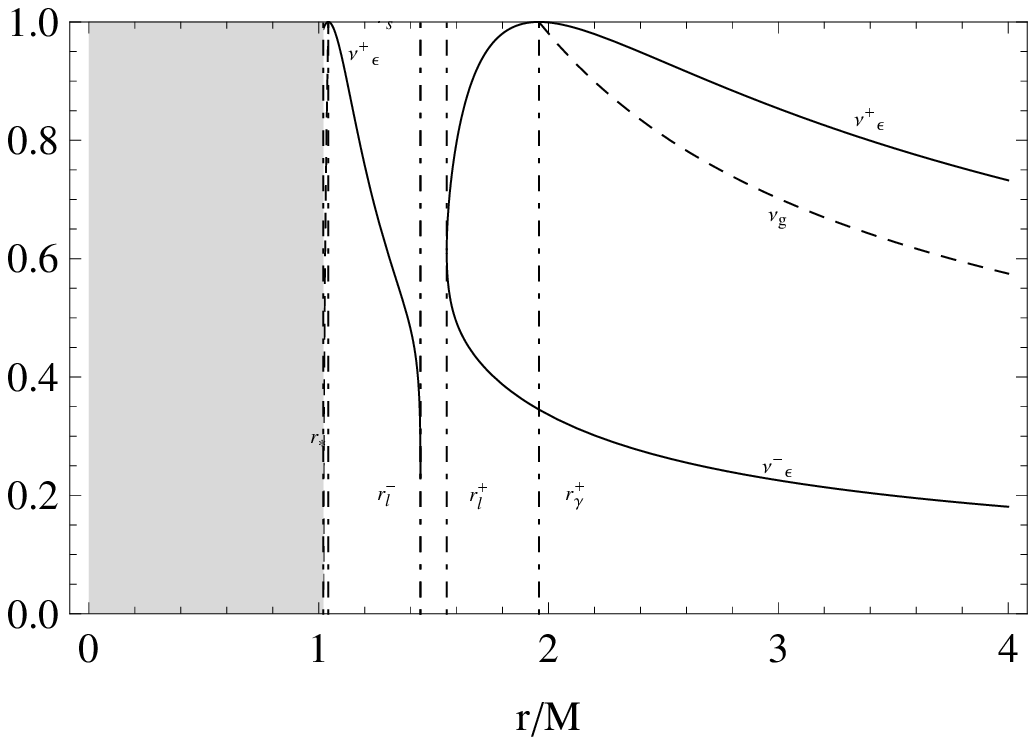}
&
\includegraphics[scale=.7]{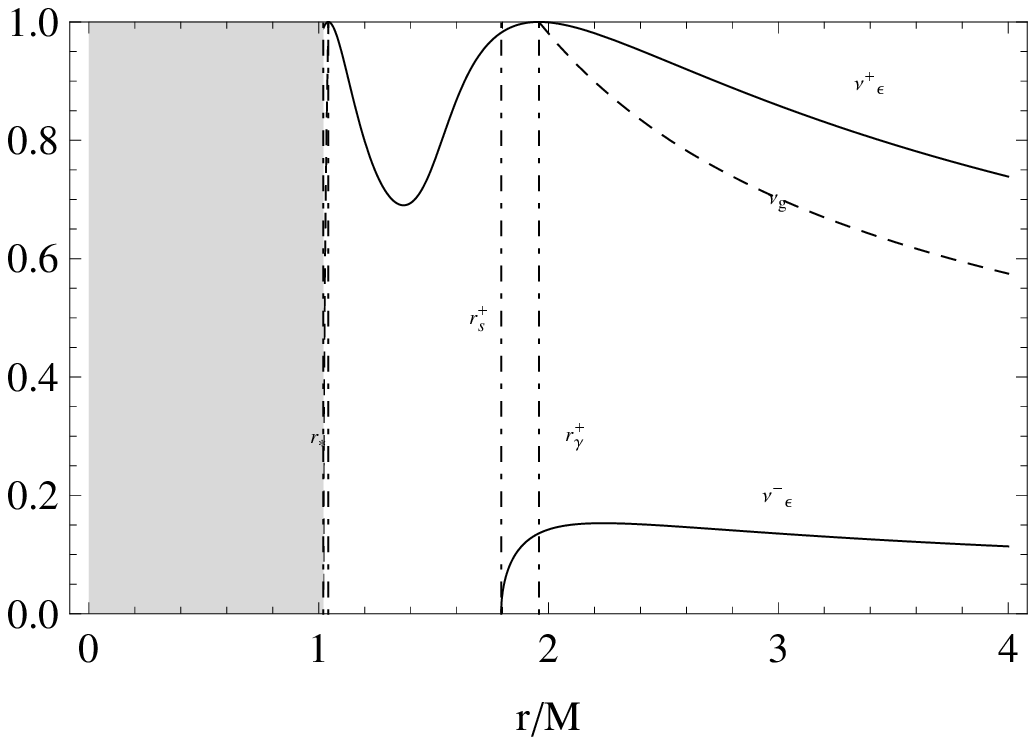}\\(c)&(d)\\
\end{tabular}
\caption[font={footnotesize,it}]{\footnotesize{The positive solution
of the linear velocity $\nu_{\epsilon}$ is plotted as a function of
the radial distance $r/M$ for $Q/M=1.01$ and different values of the
ratio $\epsilon$. In this case $r_{\gamma}^{+}=1.96M$,  $r_{\gamma}^{-}=1.042M$
with $r_{*}\equiv Q^{2}/M=1.02M$, $\tilde{\epsilon}_{-}=0.31$,
$\tilde{\epsilon}_{+}=0.9$, $\epsilon_{l}\approx0.91$.
The geodesic velocity $\nu_{g} $ is also shown (dashed curve). Shaded
region is forbidden. In (\emph{a}) the parameter choice is
$\epsilon=0.2$ with $r_{s}^{+}=1.05M$, and $r_{l}^{+}=1.95M$,
$r_{l}^{-}=1.05M$. In (\emph{b}) the parameter choice is
$\epsilon=0.5$. Here $r_{s}^{+}=1.11M$, and $r_{l}^{+}=1.88M$,
$r_{l}^{-}=1.12M$. In (\emph{c})  the parameter choice is
$\epsilon=0.9$. Here $r_{s}^{+}=1.11M$, and $r_{l}^{+}=1.88M$,
$r_{l}^{-}=1.12M$. In (\emph{d}) the parameter choice is
$\epsilon=0.95$. Here $r_{s}^{+}=1.79M$, and $r_{l}^{\pm}$ do not exist.
} }
\label{Sax4}
\end{figure}

\begin{enumerate}
\item
For $1<Q/M\leq5/(2\sqrt{6})$  the following subcases occur:
\\
\begin{description}
\item[a)]
$0<\epsilon<\tilde{\epsilon}_{-}$   : Fig.  \il\ref{Sax4}a

The velocity $\nu=\pm\nu_{\epsilon}^{+}$ exists in   the range
$Q^{2}/M<r\leq r_{l}^{-}$ and $r\geq r_{l}^{+}$ with $r\neq
r_{\gamma}^{\pm}$, $\nu=\pm\nu_{\epsilon}^{-}$ exists in   the range
$r_{s}^+<r\leq r_{l}^{-}$ and $r\geq r_{l}^{+}$.
\\
\item[b)]  $\tilde{\epsilon}_{-}\le\epsilon \leq\tilde{\epsilon}_{+}$    : Fig.  \il\ref{Sax4}b

The velocity $\nu=\pm\nu_{\epsilon}^{+}$ exists in   the range
$Q^{2}/M<r< r_{l}^{-}$ and $r\geq r_{l}^{+}$ with $r\neq
r_{\gamma}^{\pm}$, $\nu=\pm\nu_{\epsilon}^{-}$ exists in   the range
$r\geq r_{l}^{+}$.
\\
\item[c)]
 $\tilde{\epsilon}_{+}<\epsilon<\epsilon_{l}$,  : Fig.  \il\ref{Sax4}.

The velocity $\nu=\pm\nu_{\epsilon}^{+}$ exists in   the range
$Q^{2}/M<r\leq r_{l}^{-}$ and $r\geq r_{l}^{+}$ with $r\neq
r_{\gamma}^{\pm}$, $\nu=\pm\nu_{\epsilon}^{-}$ exists in   the range
$r_{s}^{+}<r\leq r_{l}^{-}$ and $r\geq r_{l}^{+}$.
\\
\item[d)] $\epsilon_{l} \le\epsilon <1$   : Fig.  \il\ref{Sax4}d

The solutions are the geodesic velocities
$\nu=\pm\nu_{\epsilon}^{+}$ in the range $r>Q^{2}/M$ with $r\neq
r_{\gamma}^{\pm}$. The solution $\nu=\pm\nu_{\epsilon}^{-}$ exists for
$\epsilon_{l} \le\epsilon <M/Q$ in the range $r>r_{s}^{+}$.
\end{description}
\item
For $5/(2\sqrt{6})<Q/M<\sqrt{9/8}$  the following subcases occur:
\begin{description}
\item[a)] $0<\epsilon<\epsilon_{l}$  : Fig.  \il\ref{Sax4}b

The velocity $\nu=\pm\nu_{\epsilon}^{+}$ exists in   the range
$Q^{2}/M<r\leq r_{l}^{-}$ and $r> r_{l}^{+}$ with $r\neq
r_{\gamma}^{\pm}$, $\nu=\pm\nu_{\epsilon}^{-}$ exists in   the range
$r_{s}^{+}<r\leq r_{l}^{-}$ and $r\geq r_{l}^{+}$.
\\
\item[b)] $\epsilon_{l}\le\epsilon <1$    : Fig.  \il\ref{Sax4}a

The velocity $\nu=\pm\nu_{\epsilon}^{+}$ exists in   the range
$r>Q^{2}/M$, $\nu=\pm\nu_{\epsilon}^{-}$ exists in   the range
$r> r_{s}^{+}$.

\end{description}
\item $Q/M\geq\sqrt{9/8}$  : Figs.\il\ref{Sax2} and \ref{Sax2bc}

The velocity $\nu=\pm\nu_{\epsilon}^{+}$ exists in   the range
$r>Q^{2}/M$ for $Q/M>\sqrt{9/8}$ whereas for $Q/M=\sqrt{9/8}$
this is a solution in  $r/M> 9/8$ with $r/M\neq 3/2$,
$\nu=\pm\nu_{\epsilon}^{-}$ exists for
$0<\epsilon<M/Q$ in the range $r>r_{s}^{+}$.
\end{enumerate}


\begin{figure}
\centering
\begin{tabular}{cc}
\includegraphics[scale=.7]{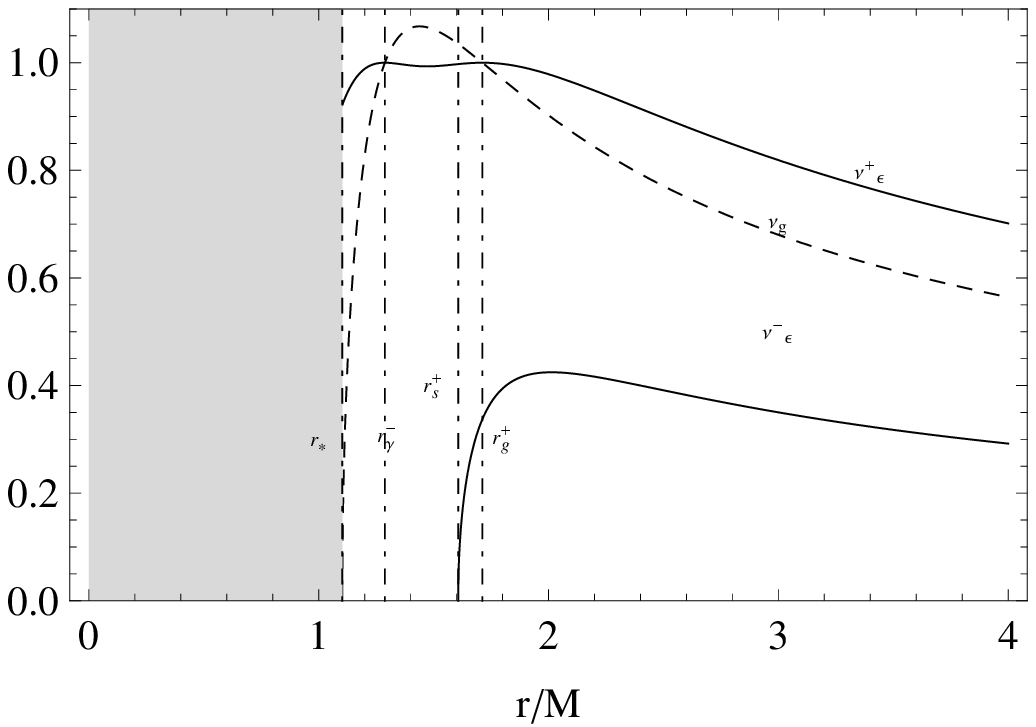}
&
\includegraphics[scale=.7]{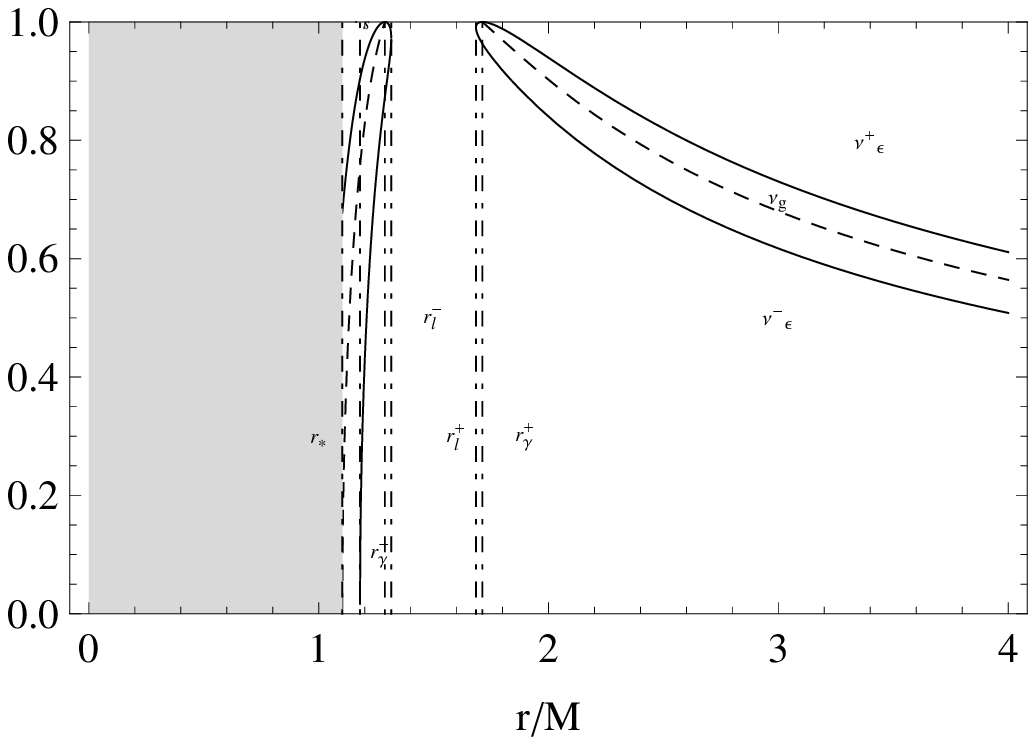}\\(a)&(b)\\
\end{tabular}
\caption[font={footnotesize,it}]{\footnotesize{The positive solution
of the linear velocity $\nu_{\epsilon}$ is plotted as a function of
the radial distance $r/M$ for $Q/M=1.05$ and different values of the
ratio $\epsilon$. In this case $r_{\gamma}^{+}=1.71M$, $r_{\gamma}^{-}=1.29M$
with $r_{*}\equiv Q^{2}/M\approx1.102M$, $\epsilon_{l}\approx0.40$.
The geodesic velocity $\nu_{g} $ is also shown (dashed curve). Shaded
region is forbidden. In (\emph{a}) the parameter choice is
$\epsilon=0.7$. Here $r_{s}^{+}=1.61M$, and $r_{l}^{\pm}$ are not defined.
In (\emph{b}) the parameter choice is $\epsilon=0.2$. Here
$r_{s}^{+}=1.2M$, and $r_{l}^{+}=1.68M$,  $r_{l}^{-}=1.32M$. } }
\label{Sax5}
\end{figure}
\begin{figure}
\centering
\begin{tabular}{cc}
\includegraphics[scale=.7]{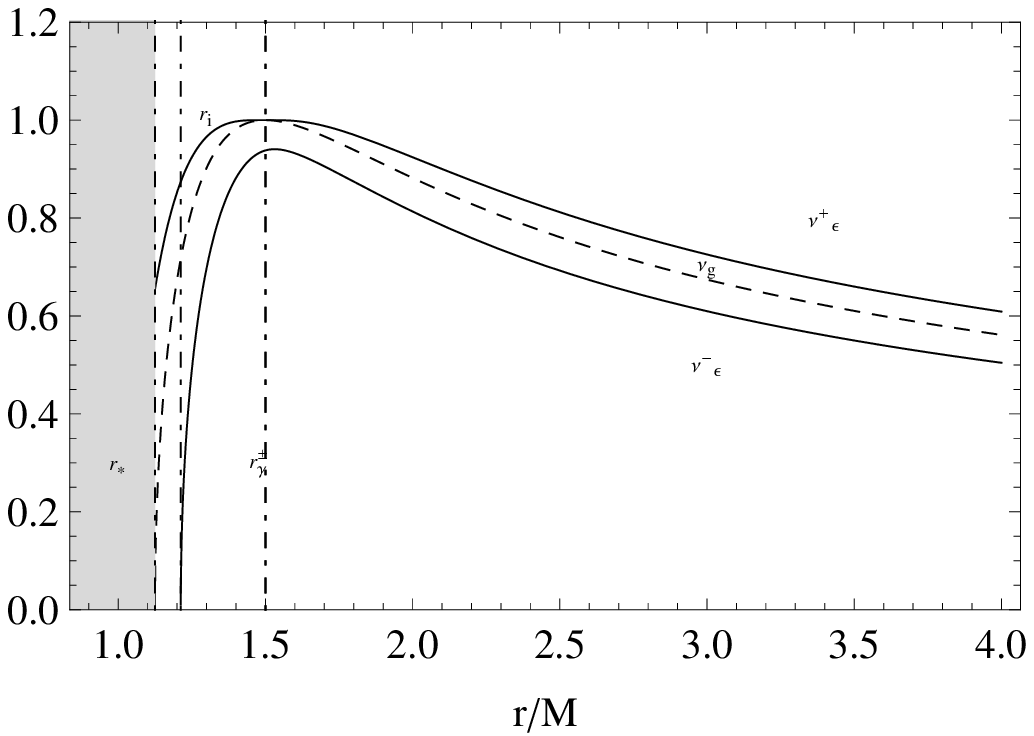}
&
\includegraphics[scale=.7]{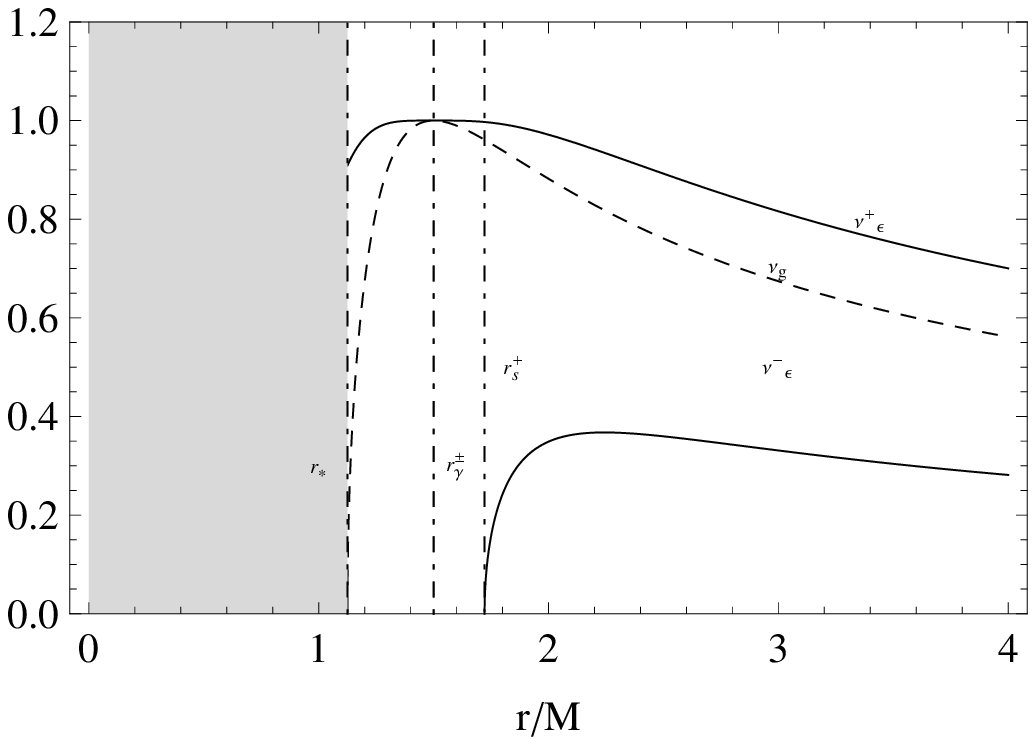}\\(a)&(b)\\
\end{tabular}
\caption[font={footnotesize,it}]{\footnotesize{The positive solution
of the linear velocity $\nu_{\epsilon}$ is plotted as a function of
the radial distance $r/M$ for $Q/M=\sqrt{9/8}$ and different values
of the ratio $\epsilon$. In this case $r_{\gamma}^{+}=r_{\gamma}^{-}=3/2M$
with $r_{*}\equiv 9/8M$, $\epsilon_{l}=0$. The geodesic velocity
$\nu_{g} $ is also shown (dashed curve). Shaded region is forbidden.
In (\emph{a}) the parameter choice is $\epsilon=0.2$ with
$r_{s}^{+}=1.2M$. In (\emph{b}) the parameter choice is $\epsilon=0.7$
with $r_{s}^{+}=1.72M$. } } \label{Sax2}
\end{figure}
\begin{figure}
\centering
\begin{tabular}{cc}
\includegraphics[scale=.7]{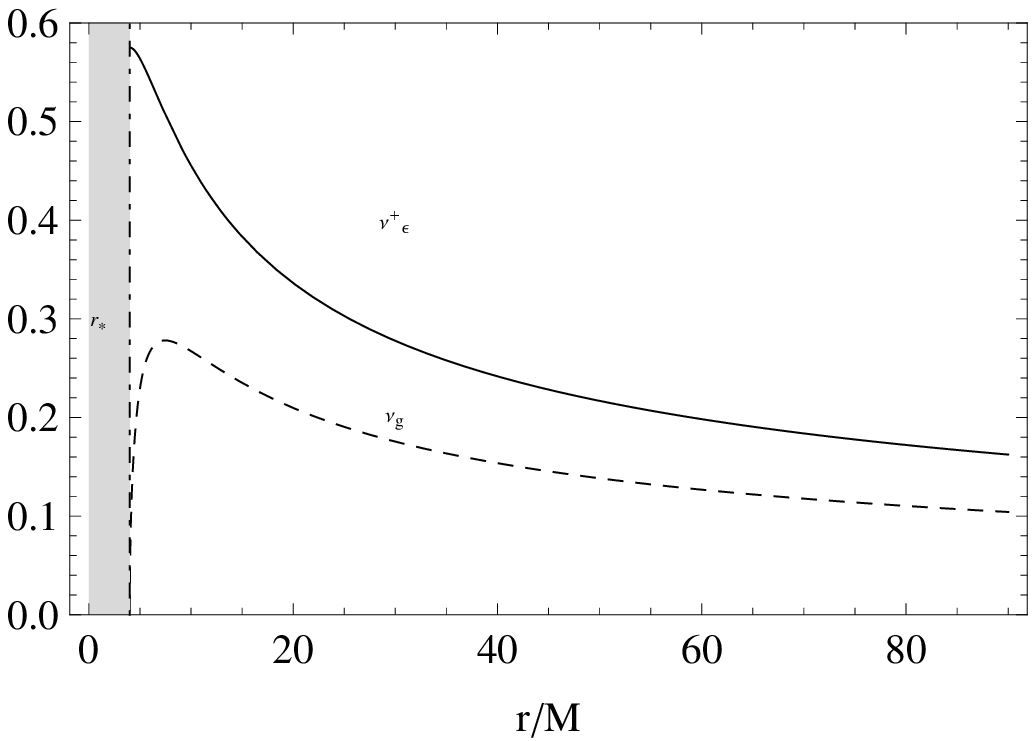}
&
\includegraphics[scale=.7]{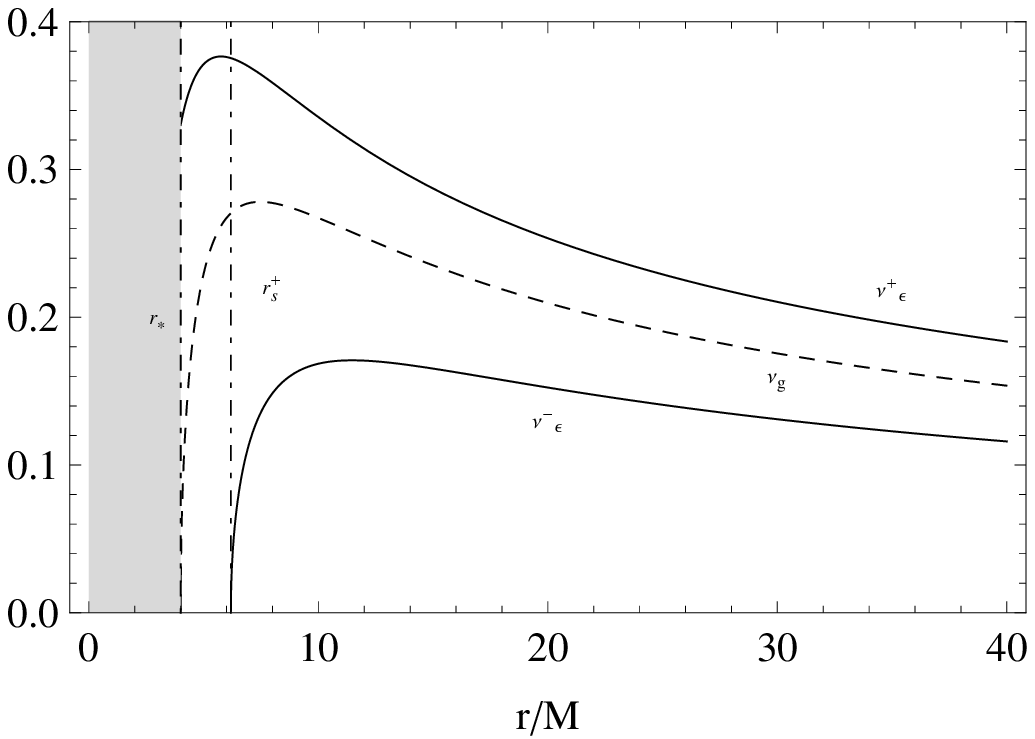}\\(a)&(b)\\
\end{tabular}
\caption[font={footnotesize,it}]{\footnotesize{The positive solution
of the linear velocity $\nu_{\epsilon}$ is plotted as a function of
the radial distance $r/M$ for $Q/M=2$ and different values of the
ratio $\epsilon$. In this case  $r_{*}\equiv Q^{2}/M\approx4M$. The
geodesic velocity $\nu_{g} $ is also shown (dashed curve). Shaded
region is forbidden. In (\emph{a}) the parameter choice is
$\epsilon=0.7$ with $r_{s}^{+}=1.48M$. In (\emph{b}) the parameter
choice is $\epsilon=0.2$ with $r_{s}^{+}=6.2M$. } }
\label{Sax2bc}
\end{figure}

\begin{figure}
\centering
\begin{tabular}{cc}
\includegraphics[scale=.7]{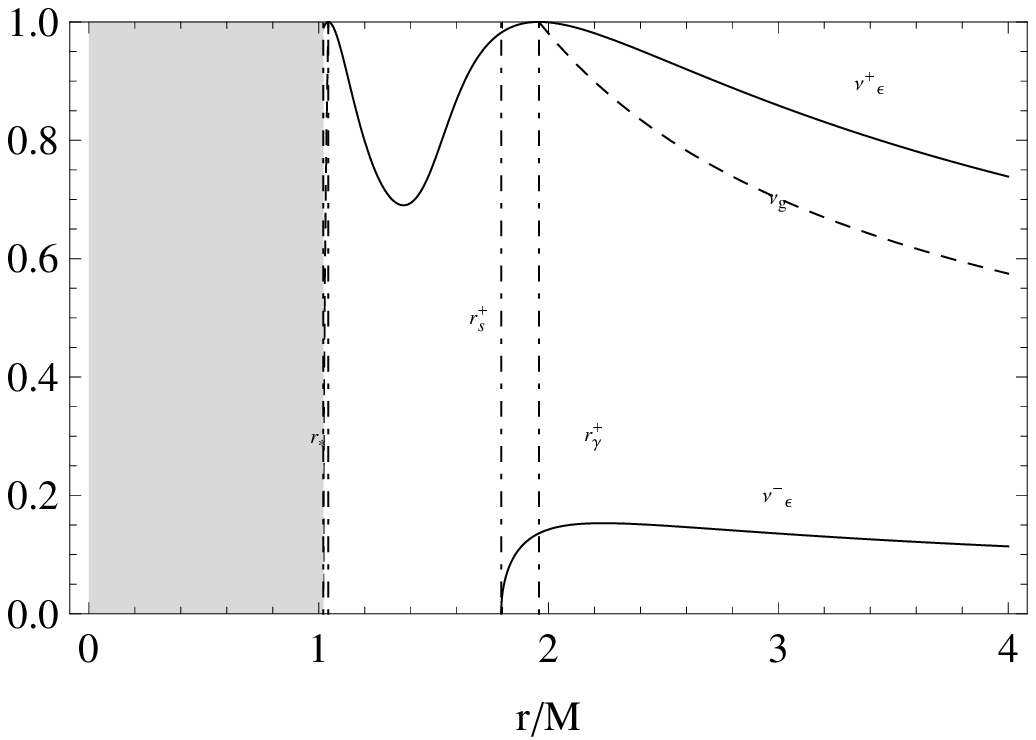}&
\includegraphics[scale=.7]{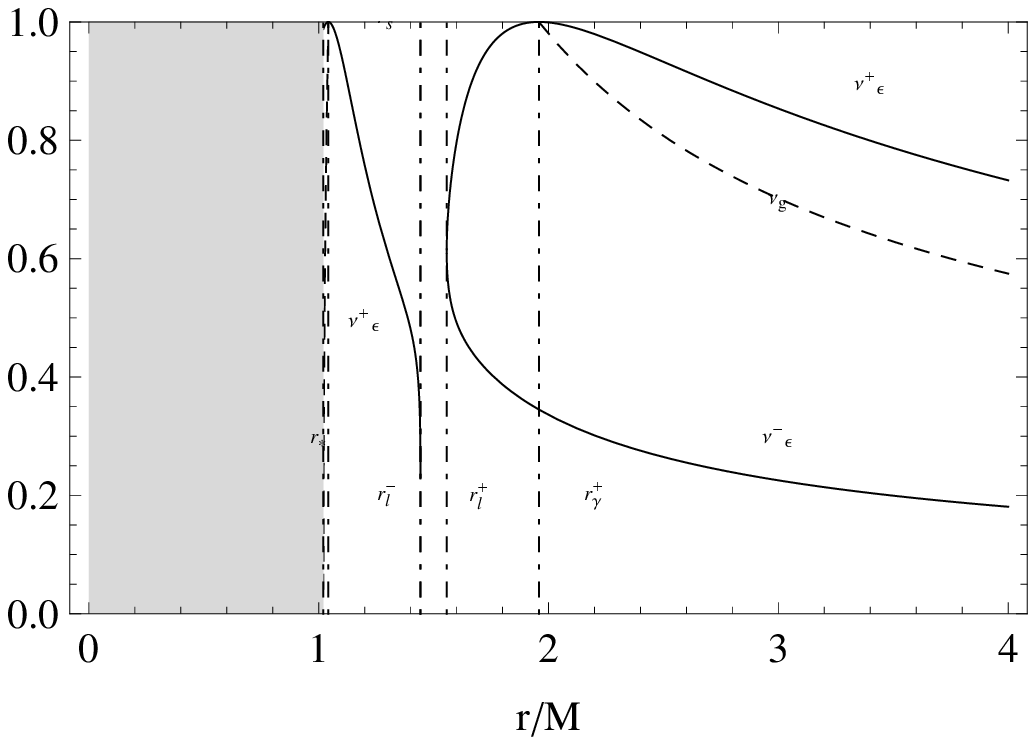}\\(a)&(b)\\
\includegraphics[scale=.7]{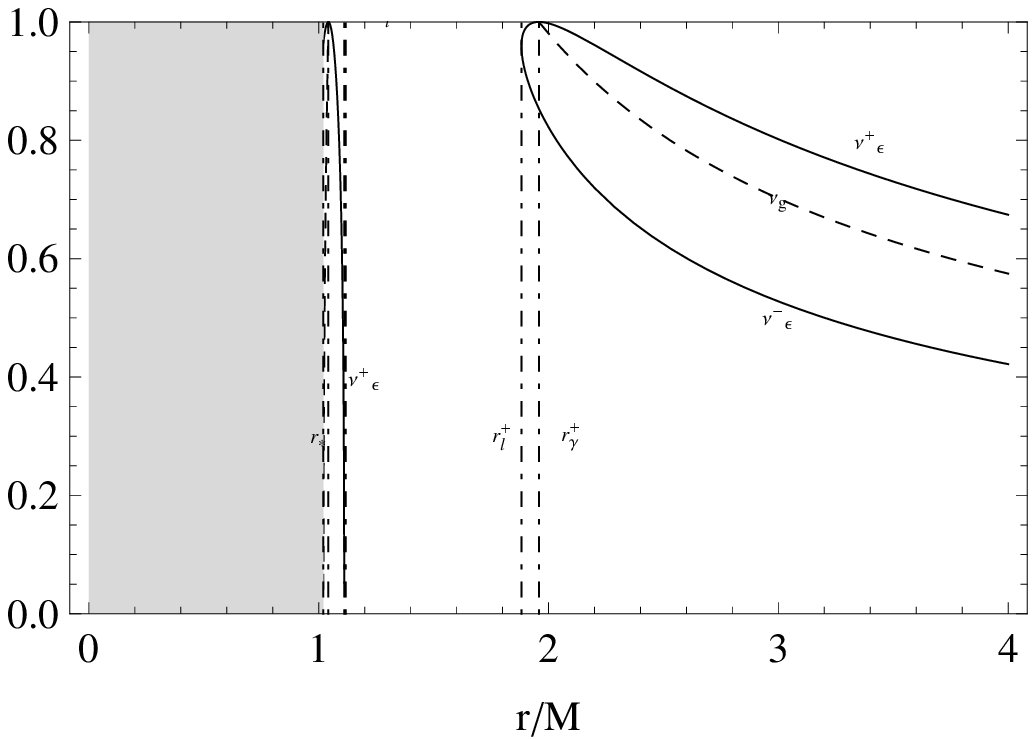}
&
\includegraphics[scale=.7]{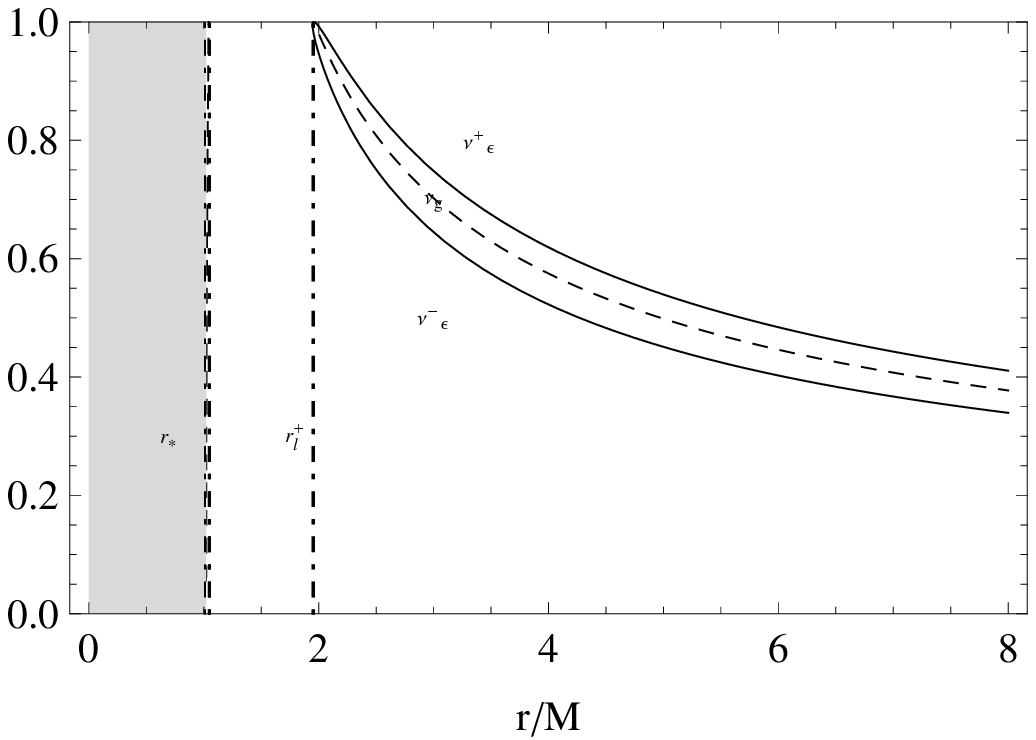}\\(c)&(d)\\
\end{tabular}
\caption[font={footnotesize,it}]{\footnotesize{The positive solution
of the linear velocity $\nu_{\epsilon}$ is plotted as a function of
the radial distance $r/M$ for $Q/M=1.01$ and different values of the
ratio $\epsilon$. In this case $r_{\gamma}^{+}=1.96M$,  $r_{\gamma}^{-}=1.042M$
 $r_{*}\equiv Q^{2}/M=1.02M$, $\tilde{\epsilon}_{-}=0.31$,
$\tilde{\epsilon}_{+}=0.9$, and $\epsilon_{l}\approx0.91$.
The geodesic velocity $\nu_{g} $ is also shown (dashed curve). Shaded
region is forbidden. In (\emph{a}) the parameter choice is
$\epsilon=-0.95$. Here $r_{s}^{+}=1.79M$, and $r_{l}^{\pm}$ do not
exist.  In (\emph{b})  the parameter choice is $\epsilon=-0.9$. Here
$r_{s}^{+}=1.11M$, $r_{l}^{+}=1.88M$, and $r_{l}^{-}=1.12M$. In
(\emph{c}) the parameter choice is $\epsilon=-0.5$. Here
$r_{s}^{+}=1.11M$, $r_{l}^{+}=1.88M$, and $r_{l}^{-}=1.12M$. In
(\emph{d}) the parameter choice is $\epsilon=-0.2$. Here
$r_{s}^{+}=1.05M$, $r_{l}^{+}=1.95M$, and $r_{l}^{-}=1.05M$. }
} \label{Sax4n}
\end{figure}

The results for $-1<\epsilon<0$  are summarized below.
\begin{enumerate}
\item
For $1<Q/M\leq5/(2\sqrt{6})$  the following subcases occur:
\\
\begin{description}
\item[a)]
For $-1<\epsilon\leq-\epsilon_{l}$, the velocity
$\nu=\pm\nu_{\epsilon}^{+}$ exists in   the range $r>Q^{2}/M$
with $r\neq r_{\gamma}^{\pm}$, $\nu=\pm\nu_{\epsilon}^{-}$ exists for
$-(M/Q)<\epsilon\leq-\epsilon_{l}$ in   the range $r>r_{s}^{+}$ (see
Fig.  \il\ref{Sax4n}a).
\\
\item[b)]
For $-\epsilon_{l}<\epsilon<-\tilde{\epsilon}_{+}$, the solution
is $\nu=\pm\nu_{\epsilon}^{+}$ in the range
$Q^{2}/M<r\leq r_{l}^{-}$ and $r\geq r_{l}^{+}$ with $r\neq
r_{\gamma}^{\pm}$, $\nu=\pm\nu_{\epsilon}^{-}$ exists in   the range
$r_{s}^{+}<r\leq r_{l}^{-}$ and $r\geq r_{l}^{+}$ (see Fig.  \il\ref{Sax4n}b).
\\
\item[c)]
For  $-\tilde{\epsilon}_{+}\le\epsilon \leq-\tilde{\epsilon}_{-}$,
the velocity $\nu=\pm\nu_{\epsilon}^{-}$ exists in   the range
$r\geq r_{l}^{+}$. $\nu=\pm\nu_{\epsilon}^{+}$ exists  for
$-\tilde{\epsilon}_{+}<\epsilon<-\tilde{\epsilon}_{-}$ in   the
range $(Q^{2}/M)<r<r_{s}^{+}$,  and $r\geq r_{l}^{+}$ with $r\neq
r_{\gamma}^{\pm}$,  and  for $\epsilon=-\tilde{\epsilon}^{\pm}$ the
velocity $\nu_{\epsilon}^{+}$ exists for $Q^{2}/M<r<r_{l}^{-}$ and
$r\geq r_{l}^{+}$ with $r\neq r_{\gamma}^{\pm}$. Finally, for
$Q=5/(2\sqrt{6})M$ and $\epsilon=-\tilde{\epsilon}_{+}$,
$\nu_{\epsilon}^{+}$ exists  for $(Q^{2}/M)<r<r_{l}^{-}$,  and
$r\geq r_{l}^{+}$ (see Fig.  \il\ref{Sax4n}c).
\\
\item[d)]
For  $-\tilde{\epsilon}_{-} <\epsilon<0$, the solutions are the
geodesic velocities $\nu=\pm\nu_{\epsilon}^{+}$ in the range $
(Q^{2}/M)<r\leq r_{l}^{-}$  and $r\geq r_{l}^{+}$ with $r\neq
r_{\gamma}^{\pm}$. The solution $\nu=\pm\nu_{\epsilon}^{-}$ exists in $
r_{s}^{+}<r\leq r_{l}^{-}$  and $r\geq r_{l}^{+}$ (see Fig.  \il\ref{Sax4n}d).
\end{description}
\item
For $5/(2\sqrt{6})<Q/M<\sqrt{9/8}$  the following subcases occur:
\begin{description}
\item[a)]
For  $-1<\epsilon\leq-\epsilon_{l}$, the velocity
$\nu=\pm\nu_{\epsilon}^{+}$ exists in   the range $r>Q^{2}/M$
with $r\neq r_{\gamma}^{\pm}$, $\nu=\pm\nu_{\epsilon}^{-}$ exists for
$-(M/Q)<\epsilon\leq-\epsilon_{l}$ in   the range $r>r_{s}^{+}$
(see Fig.  \il\ref{Sax4n}b).
\\
\item[b)]
For $-\epsilon_{l}\le\epsilon <0$, the velocity
$\nu=\pm\nu_{\epsilon}^{+}$ exists in   the range $Q^{2}/M<r\leq
r_{l}^{-}$  and $r\geq r_{l}^{+}$, $r\neq r_{\gamma}^{\pm}$,
$\nu=\pm\nu_{\epsilon}^{-}$ exists in   the range $ r_{s}^{+}<r\leq
r_{l}^{-}$  and $r\geq r_{l}^{+}$ (see Fig.  \il\ref{Sax4n}a).
\end{description}
\item
For $Q/M\geq\sqrt{9/8}$ the velocity $\nu=\pm\nu_{\epsilon}^{-}$
exists for $-(M/Q)<\epsilon<0$ in   the range $r>r_{s}^{+}$.
$\nu=\pm\nu_{\epsilon}^{+}$ is a solution for $Q/M>\sqrt{9/8}$ and
$-1<\epsilon<0$  in $r>Q^{2}/M$ whereas  for $Q/M=\sqrt{9/8}$ this
is a solution in $r/M> 9/8$ with $r/M\neq 3/2$  (see Figs.\il\ref{Sax2n} and \ref{Sax2bcn}).
\end{enumerate}

\begin{figure}
\centering
\begin{tabular}{cc}
\includegraphics[scale=.7]{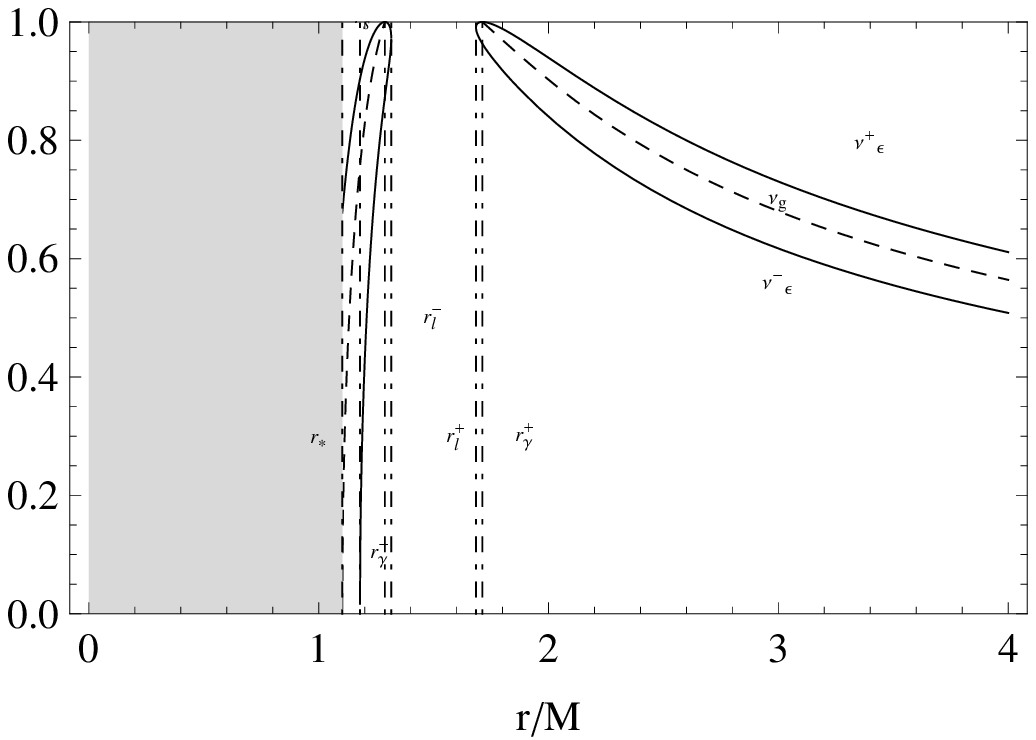}
&
\includegraphics[scale=.7]{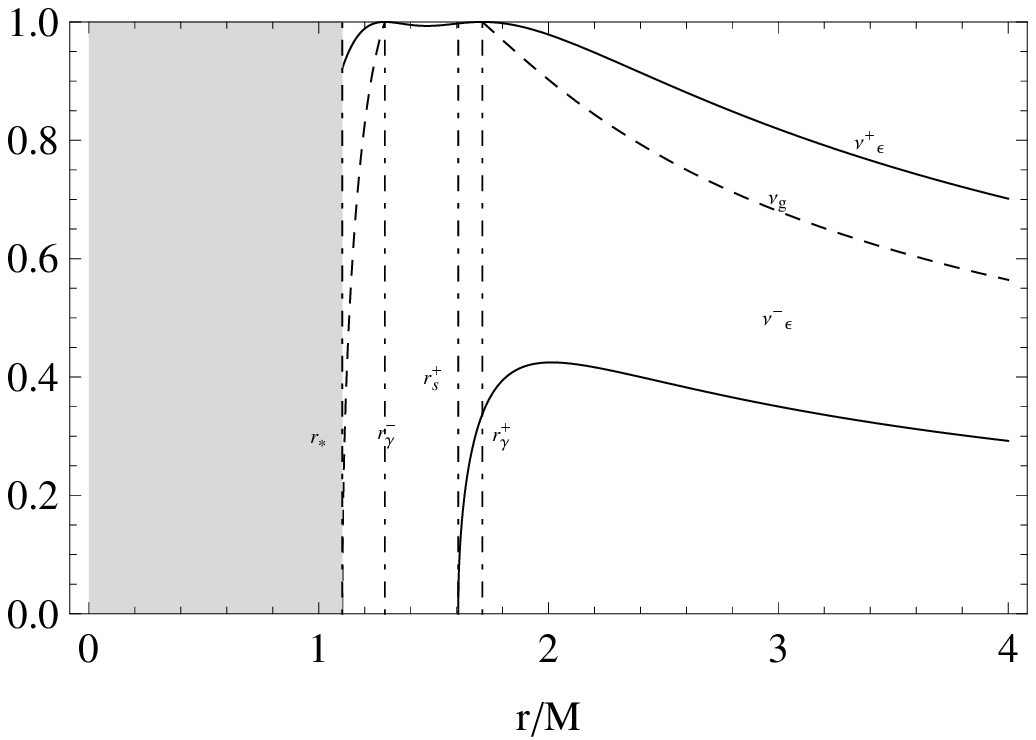}\\(a)&(b)\\
\end{tabular}
\caption[font={footnotesize,it}]{\footnotesize{The positive solution
of the linear velocity $\nu_{\epsilon}$ is plotted as a function of
the radial distance $r/M$ for $Q/M=1.05$ and different values of the
ratio $\epsilon$. In this case $r_{\gamma}^{+}=1.71M$, $r_{\gamma}^{-}=1.29M$,
 $r_{*}\equiv Q^{2}/M\approx1.102M$, and $\epsilon_{l}\approx0.40$.
The geodesic velocity $\nu_{g} $ is also shown (dashed curve). Shaded
region is forbidden. In (\emph{a}) the parameter choice is
$\epsilon=-0.2$. Here $r_{s}^{+}=1.2M$, $r_{l}^{+}=1.68M$, and
$r_{l}^{-}=1.32M$. In (\emph{b}) the parameter choice is
$\epsilon=-0.7$. Here $r_{s}^{+}=1.61M$, and $r_{l}^{\pm}$ are not defined.
} } \label{Sax5n}
\end{figure}

\begin{figure}
\centering
\begin{tabular}{cc}
\includegraphics[scale=.7]{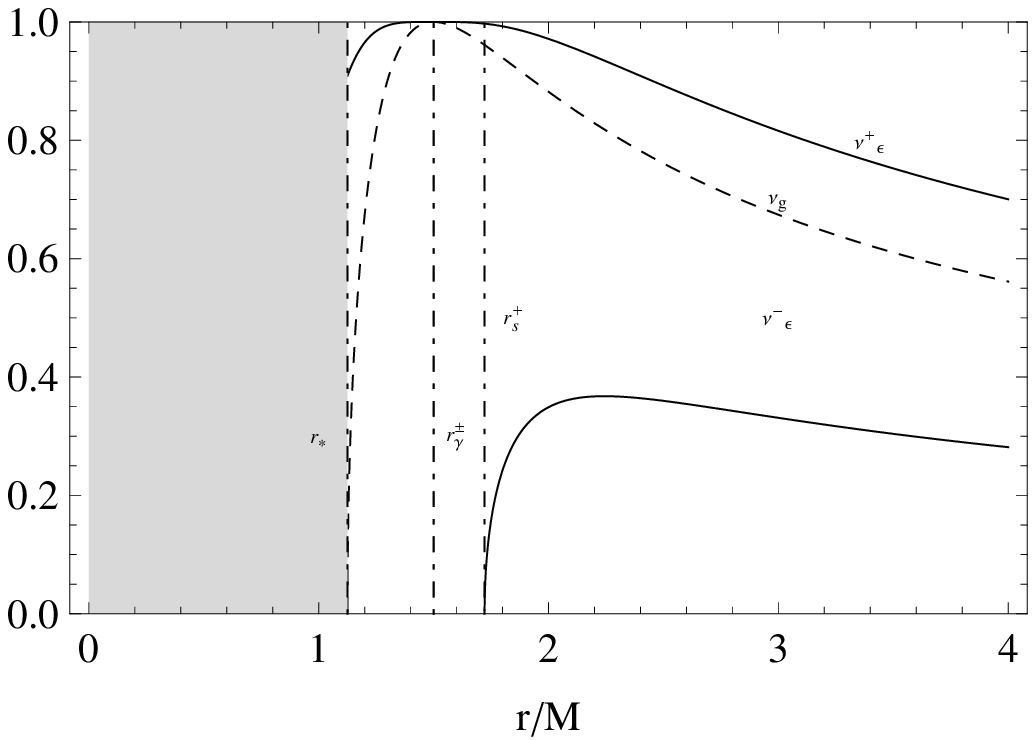}
&
\includegraphics[scale=.7]{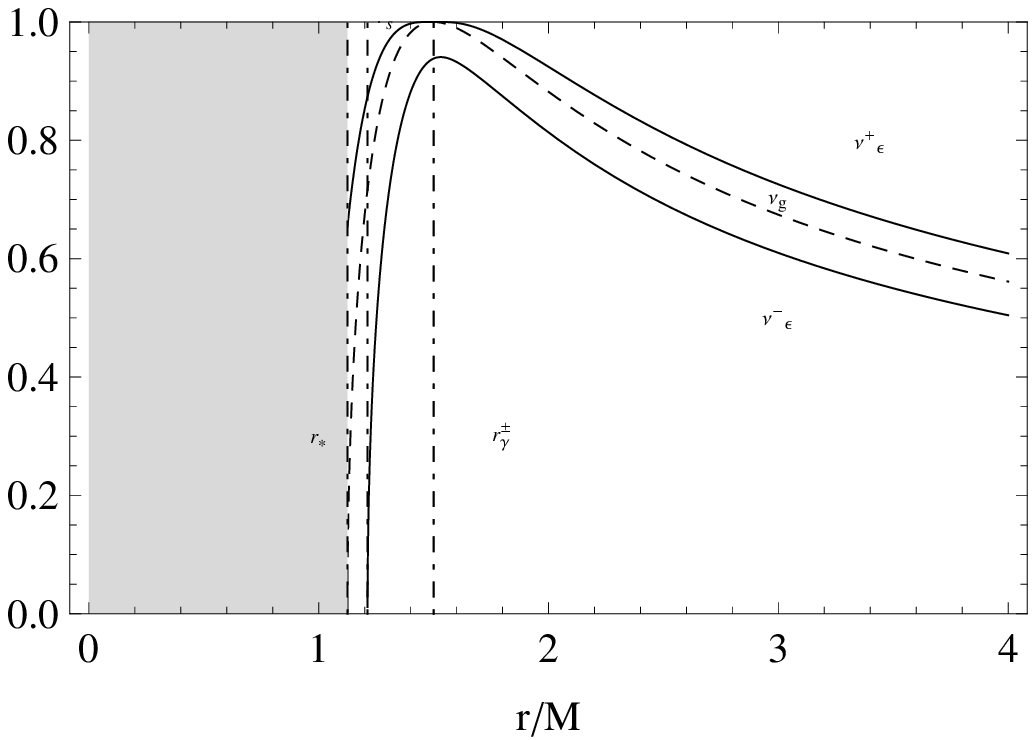}\\(a)&(b)\\
\end{tabular}
\caption[font={footnotesize,it}]{\footnotesize{The positive solution
of the linear velocity $\nu_{\epsilon}$ is plotted as a function of
the radial distance $r/M$ for $Q/M=\sqrt{9/8}$ and different values
of the ratio $\epsilon$. In this case $r_{\gamma}^{+}=r_{\gamma}^{-}=3/2M$,
$r_{*}\equiv 9/8M$, and $\epsilon_{l}=0$. The geodesic velocity
$\nu_{g} $ is also shown (dashed curve). Shaded region is forbidden.
In (\emph{a}) the parameter choice is $\epsilon=-0.7$ with
$r_{s}^{+}=1.72M$. In (\emph{b}) the parameter choice is $\epsilon=-0.2$
with $r_{s}^{+}=1.2M$. } } \label{Sax2n}
\end{figure}

\begin{figure}
\begin{tabular}{cc}
\includegraphics[scale=.7]{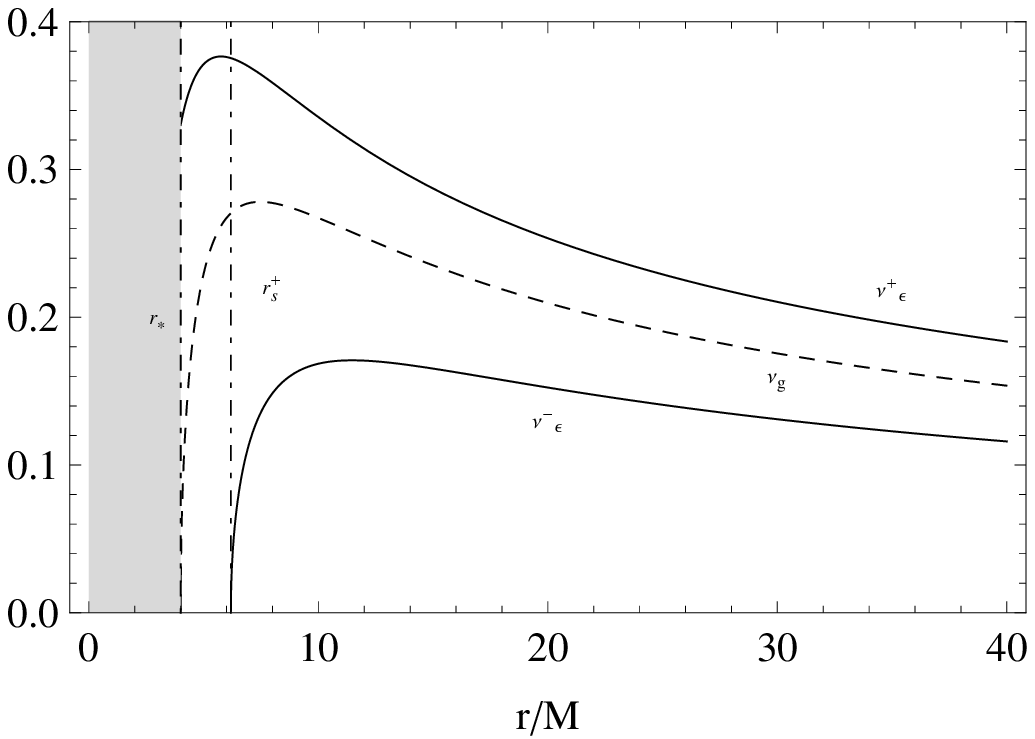}
&
\includegraphics[scale=.7]{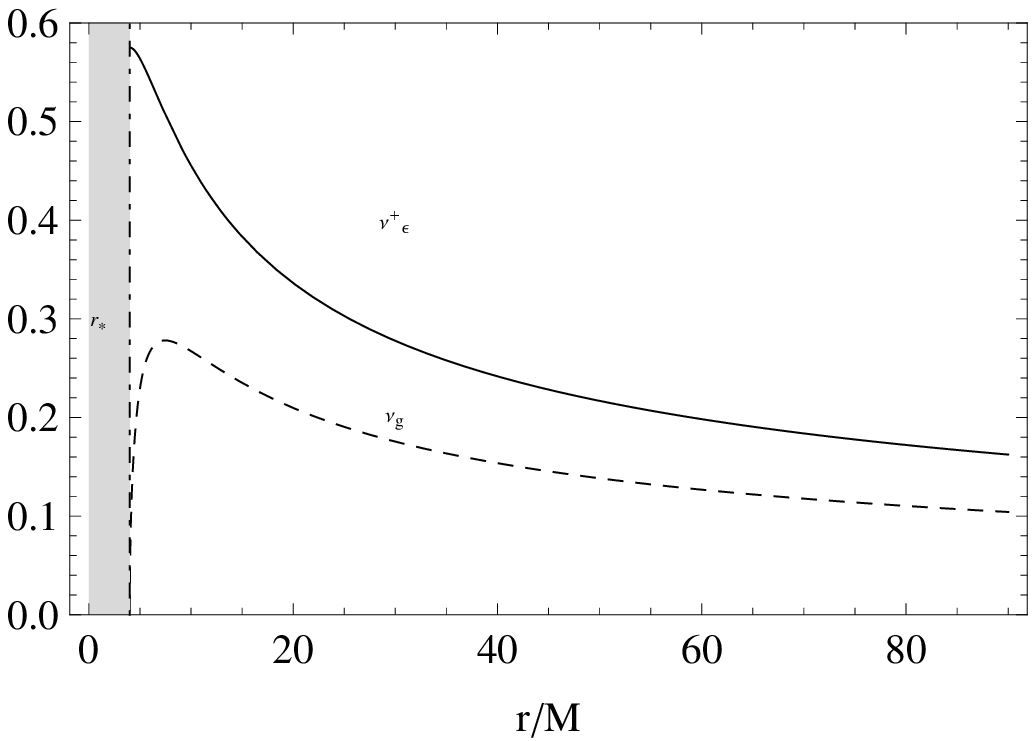}\\(a)&(b)\\
\end{tabular}
\caption[font={footnotesize,it}]{\footnotesize{The positive solution
of the linear velocity $\nu_{\epsilon}$ is plotted as a function of
the radial distance $r/M$ for $Q/M=2$ and different values of the
ratio $\epsilon$. In this case  $r_{*}\equiv Q^{2}/M\approx4M$. The
geodesic velocity $\nu_{g} $ is also shown (dashed curve). Shaded
region is forbidden. In (\emph{a}) the parameter choice is
$\epsilon=-0.2$. Here $r_{s}^{+}=6.2M$.  In (\emph{b}) the parameter
choice is $\epsilon=-0.7$. Here $r_{s}^{+}=1.48M$. } }
\label{Sax2bcn}
\end{figure}

\end{document}